\documentclass[11pt]{article}

\usepackage[preprint]{acl}

\usepackage{times}
\usepackage{latexsym}

\usepackage[T1]{fontenc}

\usepackage[utf8]{inputenc}

\usepackage{microtype}

\usepackage{inconsolata}

\usepackage{graphicx}
\usepackage{amsmath}
\usepackage{amssymb}
\usepackage[ruled,vlined]{algorithm2e}
\usepackage{adjustbox}
\usepackage{booktabs}
\usepackage{caption}
\usepackage{enumitem}
\usepackage{float}
\usepackage{longtable}
\usepackage{linguex}
\usepackage{listings}
\usepackage{multirow}
\usepackage{natbib}
\usepackage{soul}
\usepackage{subcaption}
\usepackage{relsize}
\usepackage{tabularray}
\usepackage{todonotes}
\usepackage{xcolor}
\usepackage{comment}

\lstdefinestyle{promptbox}{
  basicstyle=\ttfamily\small,
  frame=single,
  breaklines=true,
  columns=fullflexible,
  tabsize=2
}

\title{Concentrated Causal Influence, Distributed Diagnostic Information: Cognitive Impairment Decisions in Large Audio-Language Models}

\author{Xiutian Zhao\textsuperscript{1}, 
Luqi Sun\textsuperscript{1}, 
Bj\"orn Schuller\textsuperscript{2}, 
Berrak Sisman\textsuperscript{1}
\\
\textsuperscript{1} Center for Language and Speech Processing (CLSP), Johns Hopkins University, USA \\
\textsuperscript{2} Group on Language, Audio \& Music (GLAM), Imperial College London, UK
}

\begin{document}
\maketitle
\begin{abstract}
Large audio-language models (LALMs) can assess cognitive impairment (CI) from speech, yet their answers reveal neither which internal components control the decision nor how much CI information is accessible in their representations. We investigate both questions through CI-contrast neurons (CICNs), selected by class-contrastive activation frequency in decoder MLPs. Across eight open LALMs and two corpora (Pitt and PROCESS-2), we causally test CICNs against size- and layer-matched random masks and activity-matched controls. In three of the five models that meet our eligibility criteria, ablation of small CICN sets causes large, class-directed prediction shifts; speech-identified masks also control transcript-only decisions, and three of four primary Pitt masks retain specificity on PROCESS-2. Position-restricted interventions locate the influence before the final answer position, most strongly after the speech or transcript evidence under letter-labeled options, largely as a decision-direction shift. Beyond causal localization, supervised linear readouts of frozen CICN activations achieve 0.92--0.97 AUC within Pitt, versus 0.67--0.84 for the models' direct answers. Matched random and all-neuron readouts reach comparable performance; top-ranked CICNs show only an exploratory advantage at very small unit budgets. These results distinguish concentrated causal control over output decisions from broadly distributed diagnostic information for CI detection in LALMs.

\end{abstract}

\section{Introduction}
Cognitive impairment (CI), ranging from mild cognitive impairment (MCI) to dementia, is a growing health concern as populations age \cite{scheltens2021alzheimer, https://doi.org/10.1002/alz.71345}. The number of people living with dementia worldwide is projected to nearly triple by 2050 \cite{nichols2022estimation}. Speech is a promising signal for detecting cognitive decline: it is non-invasive, inexpensive to collect, and reflects decline in both what is said and how it is said \cite{doi:10.3233/JAD-200888, 10767227}.

Large audio-language models (LALMs) have begun to be applied to cognitive impairment detection \cite{huang2026generalizablecognitiveimpairmentdetection}, leveraging their capabilities to process lexical content, syntax, pauses, disfluencies, and timing. For health assessment, the key interpretability question is therefore not only \emph{what} an LALM predicts, but \emph{which internal computations} transform speech and transcript evidence into that prediction.

We study CI rather than dementia alone. Dementia involves cognitive decline that interferes with independence in everyday activities \cite{https://doi.org/10.1016/j.jalz.2011.03.005}, whereas CI is broader and includes MCI, where functional independence is largely preserved \cite{https://doi.org/10.1111/j.1365-2796.2004.01388.x, https://doi.org/10.1016/j.jalz.2011.03.008}. Our probe asks whether a speaker is cognitively healthy or cognitively impaired: speakers with dementia define the CI class used for neuron identification and evaluation, while MCI is held out to test whether the identified interventions transfer to milder impairment.

Cookie Theft picture descriptions \cite{goodglass1972assessment} provide a controlled setting for this question. Transcripts are strong predictors of CI, while speech contributes timing, expressiveness, and disfluency information \cite{doi:10.3233/JAD-150520, yuan20_interspeech, 10.3389/fcomp.2020.624488, rohanian21_interspeech}; benchmark corpora may also contain shortcuts such as interviewer turns, transcription conventions, and recording quality \cite{luz2020adress, 10767227, sun2026cleanerspeechweakergeneralization}. Output accuracy alone reveals neither which sources a model uses nor how its internal representations control the decision.

Neuron-level interpretability allows us to distinguish \emph{information available in a model} from \emph{causal influence over its decisions}. Activation statistics can identify units selective for languages and concepts, while intervention tests their functional importance \cite{bau2017network, meng2022locating, gurnee2024universal, tang-etal-2024-language}. Multimodal neuron studies have largely focused on modality- or domain-specific units \cite{huo-etal-2024-mmneuron, huang2024minerminingunderlyingpattern}; their role in clinical speech decisions remains largely unexplored. We ask whether control over CI decisions is concentrated in a small neuron set, whether that control is shared across speech and transcript inputs, and how it relates to the distribution of diagnostic information.

We study eight open LALMs on DementiaBank Pitt \cite{10.1001/archneur.1994.00540180063015} and PROCESS-2 \cite{10889017, pahar2026process2benchmarkspeechcorpus}, under speech-only (\texttt{A}), transcript-only (\texttt{T}), and joint (\texttt{AT}) input and, for Pitt, joint input with pauses and disfluencies rendered into the transcript (\texttt{AT-chat}). We identify \emph{CI-contrast neurons} (CICNs): decoder MLP neurons with the largest activation-frequency differences between recordings that an LALM correctly and confidently classifies as cognitively healthy (NC) and cognitively impaired (CI), yielding an NC and a CI mask per model and condition. Our central test is whether these masks exert more causal influence over the CI decision than neuron sets with the same size and layer distribution. We then examine whether intervention changes diagnostic discrimination or primarily shifts decision bias, and separately test how efficiently CICN activations support diagnostic readouts. Our analyses combine matched-control ablations, cross-condition and cross-corpus transfer, and position-restricted interventions, supplemented by steering, speaker-attribute controls, and token-level analyses.

The central finding is that \emph{concentrated decision control coexists with distributed diagnostic information}. Five of the eight models meet our eligibility criteria for neuron identification (\S\ref{sec:setup}); in three of these five, ablating small CICN sets produces class-directed prediction shifts beyond matched controls. These effects transfer across speech- and transcript-based decisions and, for several masks, to an independently collected corpus. Position-restricted interventions place their influence primarily before the final answer position, with the strongest localized effects at post-content question and option positions under letter-labeled options. The highest-ranked CICNs show an exploratory readout advantage at small feature budgets, while larger MLP feature sets contain redundant CI information.

Our contributions are: (1) to the best of our knowledge, the first causal neuron-level analysis of CI decisions in LALMs, with matched random masks and activity-matched neuron sets as specificity baselines; (2) evidence of cross-condition and cross-corpus portability of the identified interventions; (3) position-level characterization of their influence on the prompted decision; and (4) an empirical distinction between \emph{causal concentration} and \emph{informational concentration}.

\section{Related Work}
\label{sec:related}

\paragraph{Neuron-Level Interpretability.}
Identifying units that respond selectively to a concept is a long-standing theme, from Network Dissection for CNN units \cite{bau2017network, bau2020understanding} to partial concept selectivity of neurons in LLMs \cite{voita-etal-2024-neurons, yu-ananiadou-2024-neuron, gurnee2024universal, tang-etal-2024-language}, including units localized by causal tracing \cite{meng2022locating, alkhamissi-etal-2025-llm}.
Such units are typically validated by intervention \cite{lee-etal-2025-large, zhao-etal-2026-discovering, wang2026do}.
In multimodal models \cite{wang2023large}, neuron-level analyses have mainly focused on modality- or domain-specific units \cite{huo-etal-2024-mmneuron, huang2024minerminingunderlyingpattern, fang2024towards, xu2025deciphering}, and audio work labels acoustic units with natural-language descriptions \cite{wu2024andaudionetworkdissection} or probes speech transformer layers for phonetic, speaker and prosodic content \cite{10031189, Akman2025Improving, 11434723}.
Whether units found this way are also privileged predictive features is less clear. Sparse-autoencoder features are often comparable to difference-of-means directions or dense probes \cite{wu2025axbench, pmlr-v267-kantamneni25a}, while fine-tuning selected neurons has produced both gains over random-neuron baselines \cite{gurgurov-etal-2025-language} and parity with them \cite{mondal-etal-2025-language}.

\paragraph{Interpreting CI Detection Models.}
Speech-based CI detection has evolved from hand-crafted linguistic and acoustic features to pretrained encoders, LLMs, and multimodal systems that combine speech and transcript evidence \cite{doi:10.3233/JAD-150520, balagopalan20_interspeech, luz24_interspeech, rohanian21_interspeech}. LALMs now place both modalities within a single model \cite{xu2025qwen25omnitechnicalreport, kimiteam2025kimiaudiotechnicalreport, sun2026cleanerspeechweakergeneralization}, making it possible to ask not only which input cues are predictive, but where those cues are integrated internally.
Interpretability of CI detectors has primarily remained at the input level, using feature importance \cite{doi:10.3233/JAD-150520, 10.3389/fnagi.2021.635945}, saliency or attention over transcript tokens \cite{karlekar-etal-2018-detecting, 9769980}, and post-hoc attribution such as SHAP \cite{doi:10.3233/JAD-200888, 10767227}. These methods say which inputs a prediction correlates with, not which internal computations produce it. The closest neuron-level work studies a text-only LLM \cite{he2026activationguidedneuroninterventioninduce}.
No prior work causally localizes the CI decision inside models that process speech directly, or tests whether its control is shared across speech and transcript input.

\section{Methods}
\label{sec:method}

We pose CI detection as a binary selection question, record decoder MLP activations during correct predictions, identify CICNs by the largest between-class differences in activation frequency, and intervene on these masks against size- and layer-matched random controls. We repeat the analysis across four input conditions (\S\ref{subsec:conditions}) to test whether causal influence is specific to, or shared across, speech and transcript pathways.

\subsection{Probe Task}
\label{subsec:probe}

Each item is one participant's description of the Cookie Theft picture. The model sees a fixed prompt skeleton (Listing~\ref{lst:cicn-prompt}) that states the setting without clinical framing, says that both options are equally likely, and offers ``The speaker is cognitively healthy'' (NC) and ``The speaker is cognitively impaired'' (CI). Every item is run under both option orders, which counterbalances letter and position preferences \cite{zheng2024large, zhao-etal-2024-measuring}: a preference that adds a constant to one letter's logit cancels exactly in the order-averaged log-odds and therefore in the prediction, which is its sign; other forms are reduced but not removed.

\paragraph{Prediction from first-step logits.}
For each order $o$ we take the first-step logits of the two option letters and renormalize them, giving $P_o(c)$; the item-level probability is $P(c)=\tfrac12\sum_o P_o(c)$ and the prediction its argmax. We also log the margin $P(y)-P(\bar y)$ under each order and the first-step log-odds $\ell=\tfrac12\sum_o\big(z_o(\text{CI})-z_o(\text{NC})\big)$, where $z_o(c)$ is the logit of the letter that stands for $c$. Decoding is greedy and deterministic (Appendix~\ref{appendix:determinism}).

\subsection{Input Conditions}
\label{subsec:conditions}

We use four input conditions: transcript only (\texttt{T}), speech only (\texttt{A}), speech with a clean transcript (\texttt{AT}, the main setting) and speech with an annotated transcript (\texttt{AT-chat}).
The \emph{clean} transcript is orthographic, with all transcription codes removed.
The \emph{annotated} transcript keeps the same words, adds the fillers and word fragments, and renders pauses, retracings and unintelligible speech as readable text by a fixed convention (Appendix~\ref{appendix:render}); transcriber judgments about errors are dropped.
\texttt{AT-chat} tests whether explicitly rendering disfluency and timing cues in the transcript changes the joint-input decision.
In every condition,
the model receives only the participant's transcript, and on Pitt only the participant's speech, since investigator prompts are more frequent for impaired speakers and would leak the label; PROCESS-2 recordings have no timings and are passed whole, after excluding those whose transcript shows another speaker (Appendix~\ref{appendix:data}). For the joint conditions we also record activations over the audio-token and the transcript-token spans of the prompt (Appendix~\ref{appendix:impl}).

\subsection{Activation Statistics and CICN Identification}
\label{subsec:identification}

\paragraph{Logged activations.}
All studied decoders use gated MLPs \cite{shazeer2020gluvariantsimprovetransformer}, $\mathrm{MLP}(x)=W_{\text{down}}\big(\sigma(W_{\text{gate}}x)\odot W_{\text{up}}x\big)$, where the gate nonlinearity $\sigma$ is SiLU (SwiGLU) in every model except Gemma-4-12B, which uses GELU. We hook the gate nonlinearity and log $a_{l,n,t}$, the $n$-th coordinate of $\sigma(W_{\text{gate}}x_t)$ at layer $l$ and prompt position $t$. Only the prefill pass is recorded, i.e., the forward pass that produces the first-step logits.
For the mixture-of-experts decoder of Qwen3-Omni-30B, each unit of each expert is a neuron (Appendix~\ref{appendix:impl}).

\paragraph{Identification pool.}
For each class $c$, condition and span, we pool only identification recordings that the unmodified model classifies correctly under \emph{both} option orders with margin at least $m=0.1$ under each order. From these eligible recordings we draw a balanced pool of $N=15$ per class with a fixed seed, in every cell that reaches identification (\S\ref{subsec:eligibility}). Restricting to confident correct decisions defines class profiles around the model's stable decision regime.

\paragraph{Activation probability.}

Over the pooled recordings of class $c$, we define each neuron's activation probability as the fraction of positions at which its gate activation is positive:
\begin{equation}
P^{(c)}_{l,n}
=
\frac{1}{T_c}
\sum_t
\mathbb{I}\!\left(a^{(c)}_{l,n,t}>0\right).
\end{equation}
This statistic is used only for neuron selection; functional importance is established by intervention. We also retain the class-mean activation $\mu^{(c)}_{l,n}$ used for steering.

\paragraph{Selector.}
Our selector is the contrastive activation margin (ConAct; \citealp{zhao-etal-2026-finding}). With two classes it reduces to
\begin{equation}
\begin{aligned}
\mathrm{ConAct}^{(c)}_{l,n}
&=
\big|P^{(\mathrm{NC})}_{l,n}
      -P^{(\mathrm{CI})}_{l,n}\big| \\
&\quad {}\times
\mathbb{I}\!\left[
c=\arg\max_{c'} P^{(c')}_{l,n}
\right].
\end{aligned}
\end{equation}
Thus, each neuron can enter only the mask corresponding to the class for which it fires more often. Neurons that fire about equally often for both classes score near zero and enter neither mask.
For every class and condition, the mask $\mathcal{I}^{(c)}=\{\mathcal{I}^{(c)}_l\}_l$ is the top $r=0.5\%$ of all decoder MLP neurons of the model, ranked globally across layers.

\subsection{Interventions}
\label{subsec:intervention}

Let $g_{l,t}\in\mathbb{R}^{D_l}$ be the gate output at layer $l$ and position $t$. We patch each MLP's forward pass so that only the masked coordinates change. Interventions apply at every position of the prefill, except in the position-restricted ablations of Section~\ref{subsec:mechanism_method}.

\paragraph{Ablation (causal influence).}
We zero the masked coordinates:
$\tilde g_{l,t,n}=0 \quad\text{for } n\in\mathcal{I}^{(c)}_l .$
If the mask supports class $c$, ablating it should move predictions \emph{away} from $c$: removing NC neurons raises $P(\text{CI})$, and removing CI neurons lowers it.

\subsection{Specificity Baselines}
\label{subsec:controls}

With two classes, any neuron set that moves the decision boundary looks ``class-specific'' in the direction it pushes. Specificity must therefore be shown against references outside the class contrast.

\paragraph{Matched random masks.}
For every real mask, we draw $K$ random masks with the same number of neurons in every layer, sampled from the neurons outside the real mask (seeds $1..K$); $K=20$ throughout (Appendix~\ref{appendix:stats} Table~\ref{tab:overview}).

\paragraph{Activity-matched sets.}
ConAct favors neurons with particular activation statistics, which uniform draws do not reproduce. For the eight Pitt masks with the largest own-condition effects we therefore also draw 20 sets with the same per-layer counts, from outside the mask and without the class label, matched per layer on activation probability and mean gate activation, and ablate them like the random masks (Appendix~\ref{appendix:controlmasks}).

\subsection{Where CICNs Act}
\label{subsec:mechanism_method}

\paragraph{Position-restricted ablation.}
To locate where CICNs act, we ablate them during the prefill only at a subset of positions: the audio span, the transcript span, the final prompt position (where the answer is predicted), all positions but the final one, and the prompt template outside the content, split into the part \emph{before} the content, which under causal attention cannot see the recording, and the part \emph{after} it (question and options). Each variant is compared with 20 matched random masks restricted in the same way.

\subsection{Exploitation Tests}
\label{subsec:exploit_method}

We test how efficiently CICNs support CI detection relative to the model's own answer, the same procedure on matched random neurons, and a CICN-independent dense reference (Appendix~\ref{appendix:exploit}).
The test uses a class-balanced ridge \emph{readout} of mean neuron activations, trained on the identification split, which includes the 30 recordings used for mask selection (the unit-budget analysis excludes them), and compared with a readout using all neurons. Because large feature sets can make readout performance saturate, we also evaluate fixed budgets of 8--2{,}048 units, comparing the top-ranked CICNs with layer-matched random neurons.

\subsection{Metrics and Statistical Testing}
\label{subsec:metrics}
\paragraph{Intervention effects.}
Two co-primary metrics, always reported together, compare each intervention with the unmodified model on the same items:
(1) $\Delta P(\text{CI})$, the change in mean $P(\text{CI})$;
(2) $\Delta\ell$, the mean signed per-item change in first-step log-odds of CI vs.\ NC, which does not saturate for confident decisions.
For interpretability, we additionally report $\text{flip}_{\text{own}}$, the fraction of baseline-correct examples of the mask's class whose prediction flips under ablation; this measure is descriptive and does not enter the significance criterion.

\paragraph{Real vs.\ random.}
The unit of inference is the mask: each real mask is compared with its matched random masks under the corresponding intervention. A mask supports the claim only if it differs from the random-mask distribution on \emph{both} co-primary metrics, using an intersection--union test with
$p_{\mathrm{conj}}=\max(p_{\Delta P},p_{\Delta\ell})$.
Multiplicity across masks is controlled with the Benjamini--Hochberg false discovery rate \cite{https://doi.org/10.1111/j.2517-6161.1995.tb02031.x} within each experiment family, pooling models; a mask passes when $q<0.05$ and the effect lies in the predicted direction.
Appendix~\ref{appendix:stats} specifies the statistics, corrections, and robustness checks.

\section{Experiment Setup}
\label{sec:setup}

\paragraph{Models.}
We study eight open LALMs that take speech input and produce text:
Qwen2.5-Omni-7B \cite{xu2025qwen25omnitechnicalreport},
Qwen3-Omni-30B \cite{xu2025qwen3omnitechnicalreport},
Kimi-Audio \cite{kimiteam2025kimiaudiotechnicalreport},
MiniCPM-o-4.5 \cite{cui2026minicpmo45realtimefullduplex},
Gemma-4-12B \cite{gemmateam2026gemma4},
Audio-Flamingo-Next \cite{ghosh2026audioflamingonextnextgeneration},
Voxtral-Small-24B \cite{liu2025voxtral}, and
MOSS-Audio-8B \cite{yang2026mossaudiotechnicalreport}.
Table~\ref{tab:checkpoints} gives the checkpoints, and Table~\ref{tab:models} their decoder sizes and mask sizes.
All models receive the same 16~kHz mono waveform, and no model's audio-length limit binds (Appendix~\ref{appendix:models}).

\paragraph{Datasets.}
We evaluate on the Cookie Theft task from two English cognitive-impairment corpora: DementiaBank Pitt \cite{10.1001/archneur.1994.00540180063015} and PROCESS-2 \cite{10889017,pahar2026process2benchmarkspeechcorpus}. In both corpora, cognitively healthy participants define NC and dementia/Alzheimer's disease defines CI, while MCI and other diagnoses are held out for transfer analyses. Identification and evaluation participants are disjoint. Pitt serves as the primary identification corpus, while PROCESS-2 differs substantially in collection and transcription regime and therefore provides an independent cross-corpus transfer setting. Because PROCESS-2 transcription style is associated with diagnosis, we normalize the transcripts, stratify the split by style, and verify effects within each style. Dataset sizes, exclusions, preprocessing, transcript handling, and partition details are provided in Appendix~\ref{appendix:data}.

\paragraph{Baseline performance and analysis eligibility.}
\label{subsec:eligibility}
The eight LALMs span substantially different zero-shot operating regimes (Tables~\ref{tab:pilot_full} and~\ref{tab:pilot_full_p2}), with Pitt UAR ranging from 0.49 to 0.75 and substantial variation in class preference across models and input conditions. Adding audio can shift the decision toward either class depending on the model and corpus (Appendix~\ref{appendix:pilot}). We restrict neuron identification to model--condition cells with sufficiently reliable baseline behavior: \texttt{T} must reach at least 0.65 UAR, audio-containing conditions must have a bootstrap 95\% UAR interval excluding 0.5, and at least $N=15$ identification recordings per class must satisfy the both-orders confidence criterion. This yields twelve Pitt cells from five models and two PROCESS-2 transcript cells for within-corpus identification; PROCESS-2 audio and joint-input conditions additionally serve as independent transfer targets. The UAR criteria are screened on the evaluation split and the pool criterion on identification recordings; both are reproduced on the identification split for every reported Pitt cell, and the four strongest Pitt effects remain significant when the correction is taken over all 32 Pitt model--condition cells (Appendix~\ref{appendix:impl}).

\section{Results}
\label{sec:results}

\subsection{Concentrated Causal Control}
\label{subsec:ablation}

\begin{figure*}[h]
\centering
\includegraphics[width=\textwidth]{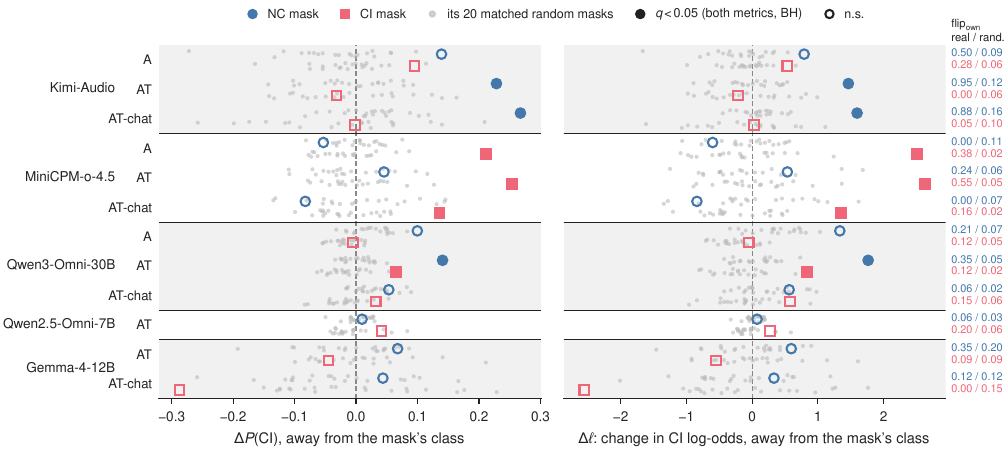}
\caption{Ablating CICN masks on Pitt in their identification condition (24 masks). Each row shows the NC mask (blue circle), CI mask (red square), and matched random masks (gray). Left: $\Delta P(\text{CI})$; right: $\Delta\ell$. Both axes are signed so that positive values are shifts away from the mask's class (toward CI for NC masks, toward NC for CI masks), the direction expected under ablation. Filled markers pass the predefined significance test (Section~\ref{subsec:metrics}); hollow markers do not. Right margin: $\text{flip}_{\text{own}}$ of the real mask / random-mask mean (descriptive).}
\label{fig:ablation}
\end{figure*}

Figure~\ref{fig:ablation} shows concentrated causal control in three models: ablating 0.5\% of decoder MLP neurons shifts CI decisions substantially more than size- and layer-matched random masks.
Seven of the 24 masks pass on both co-primary metrics ($q<0.05$), all in the predicted direction. Four have especially large effects: Qwen3-Omni-30B \texttt{AT} NC, MiniCPM-o-4.5 \texttt{A} CI and \texttt{AT} CI, and Kimi-Audio \texttt{AT-chat} NC ($|\Delta\ell|=1.6$--$2.6$ per item, versus $|\Delta\ell|\le0.17$ on average for matched random masks; $q\le0.002$). These four serve as the \emph{primary} masks for follow-up analyses.
These shifts substantially alter class predictions: ablating Kimi-Audio's \texttt{AT-chat} NC mask flips 88\% of baseline-correct NC items to CI, compared with 16\% for random.

\paragraph{Beyond activation statistics.}
Uniform random masks do not control for how active the selected neurons are. We therefore also compare the eight strongest Pitt masks with activity-matched neuron sets. Six of eight, including all four primary masks, pass against this stronger null in the predicted direction ($q=0.040$; Table~\ref{tab:robust}; Appendix~\ref{appendix:controlmasks}). 

\paragraph{Scope and robustness.}
Control is class-asymmetric and model-specific: Kimi-Audio shows the strongest effects for NC masks, MiniCPM-o-4.5 for CI masks, and Qwen3-Omni-30B most clearly for NC masks.
Effect sizes vary with mask and identification-pool size, supporting an interpretation in terms of neuron sets with disproportionate influence rather than individually indispensable units (Appendix~\ref{appendix:sensitivity}). All four primary masks retain significant effects under the tested paraphrased prompt and numeric option labels (Appendix~\ref{appendix:prompt_robust}).

\paragraph{Task selectivity and steering.}
\label{subsec:steering}
Relative to a speaker-gender control, primary CICN masks leave gender predictions unchanged and mostly remain within matched-random log-odds effects, whereas gender masks strongly affect gender decisions (Appendix~\ref{appendix:attribute}).
Class-mean steering changes CI decisions monotonically but does not outperform norm-matched random directions after correction in the masks' own conditions; the evidence for selective causal influence therefore rests on ablation (Appendices~\ref{appendix:steering} and~\ref{appendix:steer_transfer}).

\subsection{Decision Bias versus Diagnostic Discrimination}
\label{subsec:bias}
We separate changes in decision bias from changes in diagnostic ranking using offsets, rank correlations, and AUC.

\begin{table}[t]
\centering
\small
\resizebox{\columnwidth}{!}{%
\begin{tabular}{@{}lll rrr rr r@{}}
\toprule
& & & \multicolumn{3}{c}{$\Delta$AUC} & \multicolumn{2}{c}{Spearman $\rho$} & offset \\ \cmidrule(lr){4-6}\cmidrule(lr){7-8}
Model & Cond. & Mask & real & random & $q$ & real & random & share \\ \midrule
Kimi-Audio & \texttt{AT-chat} & NC & \textbf{$-$0.059} & +0.002{\scriptsize$\pm$0.019} & \textbf{0.009} & 0.925 & 0.968 & 0.91 \\
MiniCPM-o-4.5 & \texttt{A} & CI & \textbf{$-$0.040} & $-$0.003{\scriptsize$\pm$0.010} & \textbf{0.003} & 0.953 & 0.989 & 0.89 \\
MiniCPM-o-4.5 & \texttt{AT} & CI & +0.001 & +0.003{\scriptsize$\pm$0.009} & 0.95 & 0.964 & 0.986 & 0.89 \\
Qwen3-Omni-30B & \texttt{AT} & NC & 0.000 & 0.000{\scriptsize$\pm$0.009} & 0.95 & 0.935 & 0.981 & 0.61 \\ \midrule
Kimi-Audio & \texttt{A} & NC & $-$0.023 & +0.004{\scriptsize$\pm$0.017} & 0.16 & 0.896 & 0.967 & 0.69 \\
Kimi-Audio & \texttt{AT} & NC & $-$0.028 & 0.000{\scriptsize$\pm$0.016} & 0.13 & 0.930 & 0.978 & 0.82 \\
MiniCPM-o-4.5 & \texttt{AT-chat} & CI & +0.004 & $-$0.002{\scriptsize$\pm$0.008} & 0.55 & 0.973 & 0.990 & 0.78 \\
Qwen3-Omni-30B & \texttt{AT} & CI & +0.012 & +0.001{\scriptsize$\pm$0.011} & 0.43 & 0.956 & 0.978 & 0.36 \\ \bottomrule
\end{tabular}%
}
\caption{Discrimination versus intercept shift under CICN ablation on the Pitt evaluation split. The four primary masks are above the divider. $\Delta$AUC is the change in diagnostic AUC after ablation; $\rho$ is the Spearman correlation between unmodified and ablated log-odds; offset share is the fraction of mean squared per-item log-odds change explained by one additive offset.}
\label{tab:bias}
\end{table}

\paragraph{A dominant shift in decision bias.}
For the eight masks in Table~\ref{tab:bias}, 79--100\% of recordings move in the predicted direction (random masks: 37--60\%), including recordings the unmodified model misclassifies.
For the four primary masks, a single additive offset explains 61--91\% of the mean squared per-item log-odds change, while item ordering remains largely preserved (Table~\ref{tab:bias}; decomposition details in Appendix~\ref{appendix:bias}).
The selected neurons therefore exert substantial control over class predictions without comparably large changes in the model's ranking of recordings by impairment.

\paragraph{A secondary discrimination effect.}
The intervention is not a pure intercept shift: each of the eight masks in Table~\ref{tab:bias} reorders items more than all 20 of its random controls, and two primary masks reduce AUC more than their matched random masks ($q\le0.009$; Table~\ref{tab:bias}), although a bootstrap of these changes against zero is not significant after correction (Appendix~\ref{appendix:bias}).
Nevertheless, the dominant effect is a shared decision-direction shift with a smaller item-specific component. This distinction is central: large changes in diagnostic decisions can coexist with substantial preservation of diagnostic discrimination.

\subsection{Control Transfers across Modalities and Corpora}
\label{subsec:attribution}

Having characterized the intervention primarily as a shift in decision bias, we next test whether this control is restricted to the input condition or corpus used to identify the neurons.
All transfer tests apply the identified masks without re-selection.

\begin{table}[t]
\centering
\small
\resizebox{\columnwidth}{!}{%
\begin{tabular}{@{}lll rr rrr@{}}
\toprule
& & & \multicolumn{5}{c}{evaluated under} \\ \cmidrule(l){4-8}
Model & Mask & identified & \texttt{T} & \texttt{T} random & \texttt{A} & \texttt{AT} & \texttt{AT-chat} \\ \midrule
Kimi-Audio & NC & \texttt{A} & \textbf{+0.164}$^{\ast}$ & $-$0.003{\scriptsize$\pm$0.055} & \textit{+0.139} & +0.131 & +0.144 \\
& & \texttt{AT} & \textbf{+0.258}$^{\ast}$ & +0.008{\scriptsize$\pm$0.042} & +0.230 & \textit{+0.228} & +0.146 \\
& & \texttt{AT-chat} & \textbf{+0.246}$^{\ast}$ & +0.036{\scriptsize$\pm$0.044} & +0.266 & +0.240 & \textit{+0.267} \\ \midrule
MiniCPM-o-4.5 & CI & \texttt{A} & \textbf{$-$0.178}$^{\ast}$ & +0.012{\scriptsize$\pm$0.042} & \textit{$-$0.212} & $-$0.202 & $-$0.118 \\
& & \texttt{AT} & \textbf{$-$0.237}$^{\ast}$ & +0.017{\scriptsize$\pm$0.044} & $-$0.221 & \textit{$-$0.253} & $-$0.109 \\
& & \texttt{AT-chat} & \textbf{$-$0.161}$^{\ast}$ & $-$0.006{\scriptsize$\pm$0.056} & $-$0.140 & $-$0.191 & \textit{$-$0.136} \\ \midrule
Qwen3-Omni-30B & NC & \texttt{A} & \textbf{+0.176}$^{\ast}$ & +0.026{\scriptsize$\pm$0.031} & \textit{+0.100} & +0.141 & +0.109 \\
& & \texttt{AT} & \textbf{+0.223}$^{\ast}$ & +0.006{\scriptsize$\pm$0.028} & +0.105 & \textit{+0.141} & +0.092 \\
& & \texttt{AT-chat} & \textbf{+0.126}$^{\ast}$ & +0.019{\scriptsize$\pm$0.030} & +0.041 & +0.072 & \textit{+0.053} \\ \bottomrule
\end{tabular}%
}
\caption{Cross-condition transfer under ablation on Pitt: $\Delta P(\text{CI})$ for masks identified under the row condition and evaluated under the column condition; diagonal in italics. ``\texttt{T} random'' gives the matched-random mean $\pm$ sd under transcript-only evaluation. $^{\ast}$ marks cells that pass the predefined transfer test. Diagonal effects are tested in Figure~\ref{fig:ablation}; other cells without displayed random controls are descriptive.}
\label{tab:transfer}
\end{table}

\paragraph{Across input modalities.}
All nine transfers to transcript-only input shown in Table~\ref{tab:transfer} pass against matched random masks on both co-primary metrics, including the three masks identified from speech alone.
Because speech-only identification contains no transcript and transcript-only evaluation contains no audio, these results show that the same neuron sets influence the prompted decision under both speech and transcript input.
A stricter selection control supports this finding: masks identified specifically from audio positions of the joint input, before any transcript is available, also transfer to transcript-only decisions (Appendix~\ref{appendix:overlap}).
Thus, the observed control is not confined to an acoustic-specific pathway.

\paragraph{Across corpora.}
\label{subsec:crosscorpus}
We apply Pitt-identified masks and their Pitt-matched random controls directly to PROCESS-2.
Three of the four primary masks, and four of the seven masks that pass on Pitt, also pass on PROCESS-2, with comparable shifts across its two transcription styles (Appendix~\ref{appendix:process2}). Both counts include Kimi-Audio's \texttt{AT-chat} NC mask, which is evaluated under PROCESS-2 \texttt{AT} and tested in a separate correction family, because PROCESS-2 has no annotated transcript.
These results establish portability of the causal intervention from older Pitt recordings to a remotely recorded corpus collected under a different regime.

\paragraph{Held-out diagnoses.}
\label{subsec:heldout}
The intervention also extends beyond the NC and dementia-defined CI groups used for identification.
Recordings with held-out diagnoses, including MCI, generally shift in the same direction and by similar amounts, mostly beyond matched random controls on both corpora (Table~\ref{tab:heldout_abl}; Appendix~\ref{appendix:heldout}).
Together, these findings support shared control of the prompted CI decision across modalities, corpora, and diagnostic groups; they do not require identical evidence-processing computations in each setting.

\begin{figure}[t]
\centering
\includegraphics[width=\columnwidth]{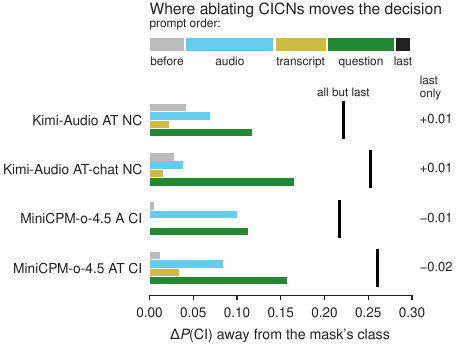}
\caption{Position-restricted ablation on Pitt: change in $P(\text{CI})$ away from the mask's class when ablation is restricted to one set of prefill positions. Each bar is a separate intervention, colored as in the prompt-order strip; the bars are not an additive decomposition. Black tick: all positions but the last; right column: last position only. Values and tests are in Appendix~\ref{appendix:positions}.}
\label{fig:where}
\end{figure}

\subsection{Localizing the Intervention Effects}
\label{subsec:positions}

We next ask where the selected neurons exert their influence during the prompted decision.
Position-restricted ablations provide causal evidence about intervention-sensitive positions.
Position restriction is not implemented for the mixture-of-experts layers of Qwen3-Omni-30B, so these ablations use four masks from the two dense models: the three primary masks of Kimi-Audio and MiniCPM-o-4.5, and Kimi-Audio \texttt{AT} NC, which also passes in its own condition (Appendix~\ref{appendix:positions}).

\paragraph{Influence precedes the final answer position.}
Ablating only the final prompt position has little effect, whereas ablating all earlier positions approximately reproduces the full intervention effect (Figure~\ref{fig:where}).
This distinction also holds under the tested prompt variants (Appendix~\ref{appendix:position_prompts}).
The selected neurons therefore influence the computation leading to the answer primarily through earlier positions.

\paragraph{The strongest effects occur after the content.}
Under the original prompt, the largest localized effects arise at the question and options following the content, with smaller effects from the audio and transcript spans.
Template-only ablation passes against matched random masks for all four tested masks,
after-content-only ablation passes for the three primary masks among them, and last-position-only ablation for none.
Speech-identified masks also affect transcript-only decisions when ablation is restricted to transcript positions, although these effects are smaller than the corresponding all-position effects (Appendix~\ref{appendix:positions}).
These results locate the strongest influence at post-content prompt positions, while retaining a contribution during content processing. This concentration persists when the question and both options are paraphrased and is tied to letter-labeled options, weakening under numeric labels (Appendix~\ref{appendix:position_prompts}).
A layer-wise readout of the answer at the final position, with the depth profile of the masks, is given in Appendix~\ref{appendix:logit_lens}.

\subsection{Distributed Diagnostic Information}
\label{subsec:exploit}
We compare supervised readouts from CICNs, matched random features, and all neurons.

\paragraph{Diagnostic information extends beyond CICNs.}
Within Pitt, whole-mask CICN readouts reach 0.92--0.97 AUC, compared with 0.67--0.84 for the model's own answer (Appendix Figure~\ref{fig:exploit}a).
A supervised lexical baseline on the same split, TF-IDF over transcript uni- and bigrams, reaches 0.91, so the gap to the model's own answer reflects supervised training rather than a property of CICNs (Appendix~\ref{appendix:exploit}).
However, matched-random and all-neuron readouts approach the same performance, leaving small and inconsistent CICN-specific gains (Appendix~\ref{appendix:exploit}).
Thus, high diagnostic readout performance is not unique to the neuron sets with the strongest intervention effects.

\begin{figure}[h]
\centering
\includegraphics[width=\columnwidth]{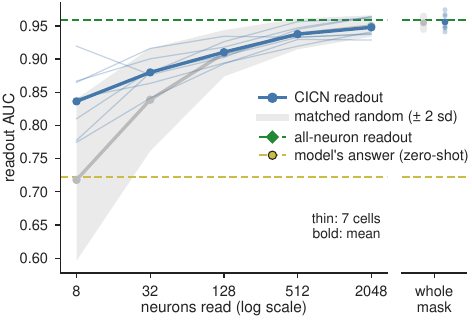}
\caption{Diagnostic readout AUC on Pitt against neuron budget. Bold lines average seven cells; thin lines show individual cells. Blue: top-ranked CICNs; gray: layer-matched random neurons, with mean $\pm$ 2 sd over 20 runs. Dashed lines show the mean all-neuron readout and zero-shot answer. Whole-mask results are shown separately because mask size varies by model. Small-budget differences are descriptive.}
\label{fig:concentration}
\end{figure}

\paragraph{An exploratory efficiency advantage at small budgets.}
At eight units, top-ranked CICNs exceed layer-matched random draws by 0.118 AUC on average across seven cells; at 32 units, the mean advantage is 0.041 (Figure~\ref{fig:concentration}). The advantage is essentially unchanged when every recording of the mask-selection participants is removed from training (Appendix~\ref{appendix:exploit}).
By 128--512 units, random features largely catch up.
No individual budget comparison survives multiplicity correction, so these differences are exploratory evidence that the ranking concentrates informative neurons near the top.
The stronger conclusion is that diagnostic information is broadly recoverable once the feature budget expands.

\section{Conclusion}
We presented, to our knowledge, the first causal neuron-level analysis of CI decisions in LALMs. In three of the five eligible models, CICN ablation produces class-directed decision shifts substantially larger than matched controls, and the same neuron sets influence decisions across input modalities and corpora. 
These interventions primarily change decision bias while largely preserving diagnostic ranking, with the strongest localized effects occurring at post-content prompt positions before the final answer position under letter-labeled options. Meanwhile, diagnostic information extends beyond the selected neurons: top-ranked CICNs show an exploratory readout advantage at very small feature budgets, but larger random feature sets recover comparable information. Together, these findings distinguish concentrated causal control over CI decisions from broadly distributed diagnostic information, showing that substantial changes in a model's predictions can coexist with largely preserved discrimination. This distinction provides a basis for auditing how LALMs translate their internal representations into diagnostic decisions.

\newpage

\section*{Limitations}

\paragraph{Scope of inference.}
CICNs are defined from the model's own confident decisions, so the causal claim concerns the mechanism of the prompted classifier. Recording-property controls show that duration, silence and speech rate explain part of the selection contrast while leaving 58--91\% of the class gap. Token-level analyses relate CICN activity to disfluency, pause and picture-content tokens as correlations only; biomarker interpretation and causal edits of those tokens are outside the present claim (Appendices~\ref{appendix:recording_contrasts} and~\ref{appendix:tokens}).

\paragraph{Model and intervention coverage.}
Our interventions target decoder MLP neurons. Attention, the audio encoder and multimodal adapters remain available for future circuit-level tracing. The predefined analysis-eligibility criteria yield twelve Pitt cells from five models (Appendix~\ref{appendix:impl}). Masks are identified under one prompt template (the four primary masks are re-tested under two prompt variants, Appendix~\ref{appendix:prompt_robust}). Results use two answer options and first-step option logits, and the exploitation experiments instantiate a linear readout. The paper therefore characterizes a specific, reproducible decision interface rather than every possible prompting regime.

\paragraph{Data scope.}
The study uses English Cookie Theft descriptions from two corpora that differ in recording and transcription regime (\S\ref{sec:setup}); the corpus shortcuts we remove or stratify are documented in Appendix~\ref{appendix:data}. PROCESS-2 contributes 20 CI recordings to its evaluation split and 136 held-out MCI recordings, so it serves as an independent transfer environment.

\section*{Ethical Considerations}
This work studies model behavior on access-controlled research corpora containing speech from older adults, including participants with cognitive impairment. Cognitive status and speech recordings are sensitive personal data. We use the datasets only for their intended research purposes, report aggregate results, and do not attempt to identify participants or infer information beyond the provided research labels.

Our experiments do not validate LALMs as diagnostic tools. Cognitive impairment is a heterogeneous clinical construct, and performance on picture-description recordings does not establish clinical validity, reliability across populations, or suitability for real-world screening. In particular, model predictions may reflect dataset artifacts, demographic or recording-related confounds, transcription conventions, and other shortcuts rather than clinically meaningful evidence. Although our intervention analyses help audit internal decision mechanisms, causally influential neurons should not be interpreted as isolated biomarkers or as proof that a model has learned medically valid acoustic or linguistic indicators.

The methods studied here could be used to improve transparency and robustness in speech-based health technologies, but they could also be misused to make unsupported inferences about an individual's cognitive or health status. Any deployment of such systems would require prospective clinical validation, assessment across demographic groups and recording conditions, appropriate consent and privacy protections, human oversight, and compliance with relevant medical-device and data-governance requirements. We therefore frame CI detection solely as a controlled probe for studying multimodal model representations and interventions.

\section*{Acknowledgments}
This work was supported by the National Science Foundation (NSF) under CAREER Award IIS-2533652.

\bibliography{proceedings,preprints,models,datasets}

@article{10.1001/archneur.1994.00540180063015,
    author = {Becker, James T. and Boiler, François and Lopez, Oscar L. and Saxton, Judith and McGonigle, Karen L.},
    title = {The Natural History of Alzheimer's Disease: Description of Study Cohort and Accuracy of Diagnosis},
    journal = {Archives of Neurology},
    volume = {51},
    number = {6},
    pages = {585-594},
    year = {1994},
    month = {06},
    issn = {0003-9942},
    doi = {10.1001/archneur.1994.00540180063015},
    url = {https://doi.org/10.1001/archneur.1994.00540180063015},
    eprint = {https://jamanetwork.com/journals/jamaneurology/articlepdf/592905/archneur_51_6_015.pdf},
}

@INPROCEEDINGS{10889017,
  author={Tao, Fuxiang and Mirheidari, Bahman and Pahar, Madhurananda and Young, Sophie and Xiao, Yao and Elghazaly, Hend and Peters, Fritz and Illingworth, Caitlin and Braun, Dorota and O’Malley, Ronan and Bell, Simon and Blackburn, Daniel and Haider, Fasih and Luz, Saturnino and Christensen, Heidi},
  booktitle={ICASSP 2025 - 2025 IEEE International Conference on Acoustics, Speech and Signal Processing (ICASSP)}, 
  title={Early Dementia Detection Using Multiple Spontaneous Speech Prompts: The PROCESS Challenge}, 
  year={2025},
  volume={},
  number={},
  pages={1-2},
  doi={10.1109/ICASSP49660.2025.10889017}}

@misc{pahar2026process2benchmarkspeechcorpus,
      title={PROCESS-2: A Benchmark Speech Corpus for Early Cognitive Impairment Detection}, 
      author={Madhurananda Pahar and Caitlin H. Illingworth and Bahman Mirheidari and Hend Elghazaly and Fritz Peters and Sophie Young and Wing-Zin Leung and Labhpreet Kaur and Daniel Blackburn and Heidi Christensen},
      year={2026},
      eprint={2605.14888},
      archivePrefix={arXiv},
      primaryClass={cs.SD},
      url={https://arxiv.org/abs/2605.14888}, 
}

@inproceedings{luz2023multilingual,
  title={Multilingual alzheimer’s dementia recognition through spontaneous speech: a signal processing grand challenge},
  author={Luz, Saturnino and Haider, Fasih and Fromm, Davida and Lazarou, Ioulietta and Kompatsiaris, Ioannis and MacWhinney, Brian},
  booktitle={ICASSP 2023-2023 IEEE International Conference on Acoustics, Speech and Signal Processing (ICASSP)},
  pages={1--2},
  year={2023},
  organization={IEEE}
}

@misc{gemmateam2026gemma4,
      title={Gemma 4 Technical Report}, 
      author={{Gemma Team}},
      year={2026},
      eprint={2607.02770},
      archivePrefix={arXiv},
      primaryClass={cs.CL},
      url={https://arxiv.org/abs/2607.02770}, 
}

@misc{kimiteam2025kimiaudiotechnicalreport,
      title={Kimi-Audio Technical Report}, 
      author={{Kimi Team}},
      year={2025},
      eprint={2504.18425},
      archivePrefix={arXiv},
      primaryClass={eess.AS},
      url={https://arxiv.org/abs/2504.18425}, 
}

@misc{cui2026minicpmo45realtimefullduplex,
      title={MiniCPM-o 4.5: Towards Real-Time Full-Duplex Omni-Modal Interaction}, 
      author={Junbo Cui and Bokai Xu and Chongyi Wang and Tianyu Yu and Weiyue Sun and Yingjing Xu and Tianran Wang and Zhihui He and Wenshuo Ma and Tianchi Cai and Jiancheng Gui and Luoyuan Zhang and Xian Sun and Fuwei Huang and Moye Chen and Zhuo Lin and Hanyu Liu and Qingxin Gui and Qingzhe Han and Yuyang Wen and Huiping Liu and Rongkang Wang and Yaqi Zhang and Hongliang Wei and Chi Chen and You Li and Kechen Fang and Jie Zhou and Yuxuan Li and Guoyang Zeng and Chaojun Xiao and Yankai Lin and Xu Han and Maosong Sun and Zhiyuan Liu and Yuan Yao},
      year={2026},
      eprint={2604.27393},
      archivePrefix={arXiv},
      primaryClass={cs.CL},
      url={https://arxiv.org/abs/2604.27393}, 
}

@misc{xu2025qwen25omnitechnicalreport,
      title={Qwen2.5-Omni Technical Report}, 
      author={Jin Xu and Zhifang Guo and Jinzheng He and Hangrui Hu and Ting He and Shuai Bai and Keqin Chen and Jialin Wang and Yang Fan and Kai Dang and Bin Zhang and Xiong Wang and Yunfei Chu and Junyang Lin},
      year={2025},
      eprint={2503.20215},
      archivePrefix={arXiv},
      primaryClass={cs.CL},
      url={https://arxiv.org/abs/2503.20215}, 
}

@misc{xu2025qwen3omnitechnicalreport,
      title={Qwen3-Omni Technical Report}, 
      author={Jin Xu and Zhifang Guo and Hangrui Hu and Yunfei Chu and Xiong Wang and Jinzheng He and Yuxuan Wang and Xian Shi and Ting He and Xinfa Zhu and Yuanjun Lv and Yongqi Wang and Dake Guo and He Wang and Linhan Ma and Pei Zhang and Xinyu Zhang and Hongkun Hao and Zishan Guo and Baosong Yang and Bin Zhang and Ziyang Ma and Xipin Wei and Shuai Bai and Keqin Chen and Xuejing Liu and Peng Wang and Mingkun Yang and Dayiheng Liu and Xingzhang Ren and Bo Zheng and Rui Men and Fan Zhou and Bowen Yu and Jianxin Yang and Le Yu and Jingren Zhou and Junyang Lin},
      year={2025},
      eprint={2509.17765},
      archivePrefix={arXiv},
      primaryClass={cs.CL},
      url={https://arxiv.org/abs/2509.17765}, 
}

@misc{liu2025voxtral,
      title={Voxtral}, 
      author={Alexander H. Liu and Andy Ehrenberg and Andy Lo and Clément Denoix and Corentin Barreau and Guillaume Lample and Jean-Malo Delignon and Khyathi Raghavi Chandu and Patrick von Platen and Pavankumar Reddy Muddireddy and Sanchit Gandhi and Soham Ghosh and Srijan Mishra and Thomas Foubert and Abhinav Rastogi and Adam Yang and Albert Q. Jiang and Alexandre Sablayrolles and Amélie Héliou and Amélie Martin and Anmol Agarwal and Antoine Roux and Arthur Darcet and Arthur Mensch and Baptiste Bout and Baptiste Rozière and Baudouin De Monicault and Chris Bamford and Christian Wallenwein and Christophe Renaudin and Clémence Lanfranchi and Darius Dabert and Devendra Singh Chaplot and Devon Mizelle and Diego de las Casas and Elliot Chane-Sane and Emilien Fugier and Emma Bou Hanna and Gabrielle Berrada and Gauthier Delerce and Gauthier Guinet and Georgii Novikov and Guillaume Martin and Himanshu Jaju and Jan Ludziejewski and Jason Rute and Jean-Hadrien Chabran and Jessica Chudnovsky and Joachim Studnia and Joep Barmentlo and Jonas Amar and Josselin Somerville Roberts and Julien Denize and Karan Saxena and Karmesh Yadav and Kartik Khandelwal and Kush Jain and Lélio Renard Lavaud and Léonard Blier and Lingxiao Zhao and Louis Martin and Lucile Saulnier and Luyu Gao and Marie Pellat and Mathilde Guillaumin and Mathis Felardos and Matthieu Dinot and Maxime Darrin and Maximilian Augustin and Mickaël Seznec and Neha Gupta and Nikhil Raghuraman and Olivier Duchenne and Patricia Wang and Patryk Saffer and Paul Jacob and Paul Wambergue and Paula Kurylowicz and Philomène Chagniot and Pierre Stock and Pravesh Agrawal and Rémi Delacourt and Romain Sauvestre and Roman Soletskyi and Sagar Vaze and Sandeep Subramanian and Saurabh Garg and Shashwat Dalal and Siddharth Gandhi and Sumukh Aithal and Szymon Antoniak and Teven Le Scao and Thibault Schueller and Thibaut Lavril and Thomas Robert and Thomas Wang and Timothée Lacroix and Tom Bewley and Valeriia Nemychnikova and Victor Paltz and Virgile Richard and Wen-Ding Li and William Marshall and Xuanyu Zhang and Yihan Wan and Yunhao Tang},
      year={2025},
      eprint={2507.13264},
      archivePrefix={arXiv},
      primaryClass={cs.SD},
      url={https://arxiv.org/abs/2507.13264}, 
}

@misc{ghosh2026audioflamingonextnextgeneration,
      title={Audio Flamingo Next: Next-Generation Open Audio-Language Models for Speech, Sound, and Music}, 
      author={Sreyan Ghosh and Arushi Goel and Kaousheik Jayakumar and Lasha Koroshinadze and Nishit Anand and Zhifeng Kong and Siddharth Gururani and Sang-gil Lee and Jaehyeon Kim and Aya Aljafari and Chao-Han Huck Yang and Sungwon Kim and Ramani Duraiswami and Dinesh Manocha and Mohammad Shoeybi and Bryan Catanzaro and Ming-Yu Liu and Wei Ping},
      year={2026},
      eprint={2604.10905},
      archivePrefix={arXiv},
      primaryClass={cs.SD},
      url={https://arxiv.org/abs/2604.10905}, 
}

@misc{yang2026mossaudiotechnicalreport,
      title={MOSS-Audio Technical Report}, 
      author={Chen Yang and Chufan Yu and Hanfu Chen and Jie Zhu and Jingqi Chen and Ke Chen and Wenxuan Wang and Yang Wang and Yaozhou Jiang and Yi Jiang and Zhengyuan Lin and Ziqi Chen and Zhaoye Fei and Chenghao Liu and Donghua Yu and Jun Zhan and Kang Yu and Kexin Huang and Liwei Fan and Mingshu Chen and Qinyuan Cheng and Ruixiao Li and Shimin Li and Songlin Wang and Xingjian Zhao and Yang Gao and Yitian Gong and Yiyang Zhang and Zhe Xu and Xipeng Qiu},
      year={2026},
      eprint={2606.01802},
      archivePrefix={arXiv},
      primaryClass={cs.SD},
      url={https://arxiv.org/abs/2606.01802}, 
}

@misc{shazeer2020gluvariantsimprovetransformer,
      title={GLU Variants Improve Transformer}, 
      author={Noam Shazeer},
      year={2020},
      eprint={2002.05202},
      archivePrefix={arXiv},
      primaryClass={cs.LG},
      url={https://arxiv.org/abs/2002.05202}, 
}

@misc{turner2024steeringlanguagemodelsactivation,
      title={Steering Language Models With Activation Engineering}, 
      author={Alexander Matt Turner and Lisa Thiergart and Gavin Leech and David Udell and Juan J. Vazquez and Ulisse Mini and Monte MacDiarmid},
      year={2024},
      eprint={2308.10248},
      archivePrefix={arXiv},
      primaryClass={cs.CL},
      url={https://arxiv.org/abs/2308.10248}, 
}

@misc{huang2024minerminingunderlyingpattern,
      title={MINER: Mining the Underlying Pattern of Modality-Specific Neurons in Multimodal Large Language Models}, 
      author={Kaichen Huang and Jiahao Huo and Yibo Yan and Kun Wang and Yutao Yue and Xuming Hu},
      year={2024},
      eprint={2410.04819},
      archivePrefix={arXiv},
      primaryClass={cs.CL},
      url={https://arxiv.org/abs/2410.04819}, 
}

@misc{he2026activationguidedneuroninterventioninduce,
      title={Activation-Guided Neuron Intervention to Induce Alzheimer's-Related Computational Language Phenotypes in a Large Language Model}, 
      author={Rui He and Ercong Nie and Hong Jiang and Iris E. Sommer and Philipp Homan and Wolfram Hinzen},
      year={2026},
      eprint={2608.03067},
      archivePrefix={arXiv},
      primaryClass={cs.CL},
      url={https://arxiv.org/abs/2608.03067}, 
}

@misc{sun2026cleanerspeechweakergeneralization,
      title={Cleaner Speech, Weaker Generalization: Revisiting Pitt-Derived Benchmarks for Alzheimer's Disease Detection}, 
      author={Luqi Sun and Shreeram Suresh Chandra and Lin Zhang and You-Jin Li and Brian MacWhinney and Yu Tsao and Emily Mower Provost and Berrak Sisman},
      year={2026},
      eprint={2609.00276},
      archivePrefix={arXiv},
      primaryClass={cs.SD},
      url={https://arxiv.org/abs/2609.00276}, 
}

@misc{huang2026generalizablecognitiveimpairmentdetection,
      title={Toward Generalizable Cognitive Impairment Detection with Speech-Based Multimodal Large Language Models}, 
      author={Yingchao Huang and Xin Wang and Yuhan Su and Shanshan Yao},
      year={2026},
      eprint={2607.21496},
      archivePrefix={arXiv},
      primaryClass={eess.SP},
      url={https://arxiv.org/abs/2607.21496}, 
}

@inproceedings{wu2024andaudionetworkdissection,
author = {Wu, Tung-Yu and Lin, Yu-Xiang and Weng, Tsui-Wei},
title = {AND: audio network dissection for interpreting deep acoustic models},
year = {2024},
publisher = {JMLR.org},
booktitle = {Proceedings of the 41st International Conference on Machine Learning},
articleno = {2199},
numpages = {25},
location = {Vienna, Austria},
series = {ICML'24}
}

@inproceedings{xu2025deciphering,
  title={Deciphering Functions of Neurons in Vision-Language Models},
  author={Xu, Jiaqi and Lan, Cuiling and Lu, Yan},
  booktitle={Proceedings of the 33rd ACM International Conference on Multimedia},
  pages={3173--3181},
  year={2025}
}

@INPROCEEDINGS {bau2017network,
author = {Bau, David and Zhou, Bolei and Khosla, Aditya and Oliva, Aude and Torralba, Antonio },
booktitle = { 2017 IEEE Conference on Computer Vision and Pattern Recognition (CVPR) },
title = {{ Network Dissection: Quantifying Interpretability of Deep Visual Representations }},
year = {2017},
volume = {},
ISSN = {1063-6919},
pages = {3319-3327},
doi = {10.1109/CVPR.2017.354},
url = {https://doi.ieeecomputersociety.org/10.1109/CVPR.2017.354},
publisher = {IEEE Computer Society},
address = {Los Alamitos, CA, USA},
month =Jul}

@article{bau2020understanding,
   title={Understanding the role of individual units in a deep neural network},
   volume={117},
   ISSN={1091-6490},
   url={http://dx.doi.org/10.1073/pnas.1907375117},
   DOI={10.1073/pnas.1907375117},
   number={48},
   journal={Proceedings of the National Academy of Sciences},
   publisher={Proceedings of the National Academy of Sciences},
   author={Bau, David and Zhu, Jun-Yan and Strobelt, Hendrik and Lapedriza, Agata and Zhou, Bolei and Torralba, Antonio},
   year={2020},
   month=sep, pages={30071–30078}
}

@inproceedings{tang-etal-2024-language,
    title = "Language-Specific Neurons: The Key to Multilingual Capabilities in Large Language Models",
    author = "Tang, Tianyi  and
      Luo, Wenyang  and
      Huang, Haoyang  and
      Zhang, Dongdong  and
      Wang, Xiaolei  and
      Zhao, Xin  and
      Wei, Furu  and
      Wen, Ji-Rong",
    editor = "Ku, Lun-Wei  and
      Martins, Andre  and
      Srikumar, Vivek",
    booktitle = "Proceedings of the 62nd Annual Meeting of the Association for Computational Linguistics (Volume 1: Long Papers)",
    month = aug,
    year = "2024",
    address = "Bangkok, Thailand",
    publisher = "Association for Computational Linguistics",
    url = "https://aclanthology.org/2024.acl-long.309/",
    doi = "10.18653/v1/2024.acl-long.309",
    pages = "5701--5715"
}

@inproceedings{voita-etal-2024-neurons,
    title = "Neurons in Large Language Models: Dead, N-gram, Positional",
    author = "Voita, Elena  and
      Ferrando, Javier  and
      Nalmpantis, Christoforos",
    editor = "Ku, Lun-Wei  and
      Martins, Andre  and
      Srikumar, Vivek",
    booktitle = "Findings of the Association for Computational Linguistics: ACL 2024",
    month = aug,
    year = "2024",
    address = "Bangkok, Thailand",
    publisher = "Association for Computational Linguistics",
    url = "https://aclanthology.org/2024.findings-acl.75/",
    doi = "10.18653/v1/2024.findings-acl.75",
    pages = "1288--1301"
}

@article{gurnee2024universal,
title={Universal Neurons in {GPT}2 Language Models},
author={Wes Gurnee and Theo Horsley and Zifan Carl Guo and Tara Rezaei Kheirkhah and Qinyi Sun and Will Hathaway and Neel Nanda and Dimitris Bertsimas},
journal={Transactions on Machine Learning Research},
issn={2835-8856},
year={2024},
url={https://openreview.net/forum?id=ZeI104QZ8I},
note={}
}

@inproceedings{huo-etal-2024-mmneuron,
    title = "{MMN}euron: Discovering Neuron-Level Domain-Specific Interpretation in Multimodal Large Language Model",
    author = "Huo, Jiahao  and
      Yan, Yibo  and
      Hu, Boren  and
      Yue, Yutao  and
      Hu, Xuming",
    editor = "Al-Onaizan, Yaser  and
      Bansal, Mohit  and
      Chen, Yun-Nung",
    booktitle = "Proceedings of the 2024 Conference on Empirical Methods in Natural Language Processing",
    month = nov,
    year = "2024",
    address = "Miami, Florida, USA",
    publisher = "Association for Computational Linguistics",
    url = "https://aclanthology.org/2024.emnlp-main.387/",
    doi = "10.18653/v1/2024.emnlp-main.387",
    pages = "6801--6816"
}

@inproceedings{fang2024towards,
author = {Fang, Junfeng and Bi, Zongze and Wang, Ruipeng and Jiang, Houcheng and Gao, Yuan and Wang, Kun and Zhang, An and Shi, Jie and Wang, Xiang and Chua, Tat-Seng},
title = {Towards neuron attributions in multimodal large language models},
year = {2024},
isbn = {9798331314385},
publisher = {Curran Associates Inc.},
address = {Red Hook, NY, USA},
booktitle = {Proceedings of the 38th International Conference on Neural Information Processing Systems},
articleno = {3904},
numpages = {24},
location = {Vancouver, BC, Canada},
series = {NIPS '24}
}

@inproceedings{yu-ananiadou-2024-neuron,
    title = "Neuron-Level Knowledge Attribution in Large Language Models",
    author = "Yu, Zeping  and
      Ananiadou, Sophia",
    editor = "Al-Onaizan, Yaser  and
      Bansal, Mohit  and
      Chen, Yun-Nung",
    booktitle = "Proceedings of the 2024 Conference on Empirical Methods in Natural Language Processing",
    month = nov,
    year = "2024",
    address = "Miami, Florida, USA",
    publisher = "Association for Computational Linguistics",
    url = "https://aclanthology.org/2024.emnlp-main.191/",
    doi = "10.18653/v1/2024.emnlp-main.191",
    pages = "3267--3280"
}

@inproceedings{zhao-etal-2026-finding,
    title = "Finding Culture-Sensitive Neurons in Vision-Language Models",
    author = "Zhao, Xiutian  and
      Choenni, Rochelle  and
      Saxena, Rohit  and
      Titov, Ivan",
    editor = "Demberg, Vera  and
      Inui, Kentaro  and
      Marquez, Llu{\'i}s",
    booktitle = "Proceedings of the 19th Conference of the {E}uropean Chapter of the {A}ssociation for {C}omputational {L}inguistics (Volume 1: Long Papers)",
    month = mar,
    year = "2026",
    address = "Rabat, Morocco",
    publisher = "Association for Computational Linguistics",
    url = "https://aclanthology.org/2026.eacl-long.155/",
    pages = "3366--3381",
    ISBN = "979-8-89176-380-7"
}

@inproceedings{lee-etal-2025-large,
    title = "Do Large Language Models Have ``Emotion Neurons''? Investigating the Existence and Role",
    author = "Lee, Jaewook  and
      Lee, Woojin  and
      Kwon, Oh-Woog  and
      Kim, Harksoo",
    editor = "Che, Wanxiang  and
      Nabende, Joyce  and
      Shutova, Ekaterina  and
      Pilehvar, Mohammad Taher",
    booktitle = "Findings of the Association for Computational Linguistics: ACL 2025",
    month = jul,
    year = "2025",
    address = "Vienna, Austria",
    publisher = "Association for Computational Linguistics",
    url = "https://aclanthology.org/2025.findings-acl.806/",
    doi = "10.18653/v1/2025.findings-acl.806",
    pages = "15617--15639",
    ISBN = "979-8-89176-256-5"
}

@inproceedings{zhao-etal-2024-measuring,
    title = "Measuring the Inconsistency of Large Language Models in Preferential Ranking",
    author = "Zhao, Xiutian  and
      Wang, Ke  and
      Peng, Wei",
    editor = "Li, Sha  and
      Li, Manling  and
      Zhang, Michael JQ  and
      Choi, Eunsol  and
      Geva, Mor  and
      Hase, Peter  and
      Ji, Heng",
    booktitle = "Proceedings of the 1st Workshop on Towards Knowledgeable Language Models (KnowLLM 2024)",
    month = aug,
    year = "2024",
    address = "Bangkok, Thailand",
    publisher = "Association for Computational Linguistics",
    url = "https://aclanthology.org/2024.knowllm-1.14/",
    doi = "10.18653/v1/2024.knowllm-1.14",
    pages = "171--176"
}

@inproceedings{zheng2024large,
title={Large Language Models Are Not Robust Multiple Choice Selectors},
author={Chujie Zheng and Hao Zhou and Fandong Meng and Jie Zhou and Minlie Huang},
booktitle={The Twelfth International Conference on Learning Representations},
year={2024},
url={https://openreview.net/forum?id=shr9PXz7T0}
}

@article{wang2023large,
  title={Large-scale multi-modal pre-trained models: A comprehensive survey},
  author={Wang, Xiao and Chen, Guangyao and Qian, Guangwu and Gao, Pengcheng and Wei, Xiao-Yong and Wang, Yaowei and Tian, Yonghong and Gao, Wen},
  journal={Machine Intelligence Research},
  volume={20},
  number={4},
  pages={447--482},
  year={2023},
  publisher={Springer}
}

@inproceedings{rimsky-etal-2024-steering,
    title = "Steering Llama 2 via Contrastive Activation Addition",
    author = "Rimsky, Nina  and
      Gabrieli, Nick  and
      Schulz, Julian  and
      Tong, Meg  and
      Hubinger, Evan  and
      Turner, Alexander",
    editor = "Ku, Lun-Wei  and
      Martins, Andre  and
      Srikumar, Vivek",
    booktitle = "Proceedings of the 62nd Annual Meeting of the Association for Computational Linguistics (Volume 1: Long Papers)",
    month = aug,
    year = "2024",
    address = "Bangkok, Thailand",
    publisher = "Association for Computational Linguistics",
    url = "https://aclanthology.org/2024.acl-long.828/",
    doi = "10.18653/v1/2024.acl-long.828",
    pages = "15504--15522"
}

@inproceedings{meng2022locating,
author = {Meng, Kevin and Bau, David and Andonian, Alex and Belinkov, Yonatan},
title = {Locating and editing factual associations in GPT},
year = {2022},
isbn = {9781713871088},
publisher = {Curran Associates Inc.},
address = {Red Hook, NY, USA},
booktitle = {Proceedings of the 36th International Conference on Neural Information Processing Systems},
articleno = {1262},
numpages = {14},
location = {New Orleans, LA, USA},
series = {NIPS '22}
}

@ARTICLE{Akman2025Improving,
  author={Akman, Alican and Sun, Qiyang and Schuller, Björn W.},
  journal={IEEE Signal Processing Letters}, 
  title={Improving Audio Explanations Using Audio Language Models}, 
  year={2025},
  volume={32},
  number={},
  pages={741-745},
  doi={10.1109/LSP.2025.3532218}}

@INPROCEEDINGS{10031189,
  author={Singla, Yaman Kumar and Shah, Jui and Chen, Changyou and Shah, Rajiv Ratn},
  booktitle={2022 IEEE International Conference on Data Mining Workshops (ICDMW)}, 
  title={What Do Audio Transformers Hear? Probing Their Representations For Language Delivery \& Structure}, 
  year={2022},
  volume={},
  number={},
  pages={910-925},
  doi={10.1109/ICDMW58026.2022.00120}}

@inproceedings{mean-centring,
title = "Improving Activation Steering in Language Models with Mean-Centring",
author = "Ole Jorgensen and Dylan Cope and Nandi Schoots and Murray Shanahan",
year = "2024",
month = feb,
day = "26",
language = "English",
booktitle = "Responsible Language Models Workshop (ReLM) at AAAI-24",
note = "Responsible Language Models Workshop at AAAI-24, ReLM@AAAI-24 ; Conference date: 26-02-2024 Through 26-02-2024",
url = "https://sites.google.com/vectorinstitute.ai/relm2024/home?authuser=0",
}

@inproceedings{zhao-etal-2026-discovering,
    title = "Discovering and Causally Validating Emotion-Sensitive Neurons in Large Audio-Language Models",
    author = {Zhao, Xiutian  and
      Schuller, Bj{\"o}rn  and
      Sisman, Berrak},
    editor = "Liakata, Maria  and
      Moreira, Viviane P.  and
      Zhang, Jiajun  and
      Jurgens, David",
    booktitle = "Proceedings of the 64th Annual Meeting of the {A}ssociation for {C}omputational {L}inguistics (Volume 1: Long Papers)",
    month = jul,
    year = "2026",
    address = "San Diego, California, United States",
    publisher = "Association for Computational Linguistics",
    url = "https://aclanthology.org/2026.acl-long.687/",
    doi = "10.18653/v1/2026.acl-long.687",
    pages = "15056--15071",
    ISBN = "979-8-89176-390-6"
}

@INPROCEEDINGS{11434723,
  author={Yang, Chih-Kai and Ho, Neo and Lee, Yi-Jyun and Lee, Hung-Yi},
  booktitle={2025 IEEE Automatic Speech Recognition and Understanding Workshop (ASRU)}, 
  title={AudioLens: A Closer Look at Auditory Attribute Perception of Large Audio-Language Models}, 
  year={2025},
  volume={},
  number={},
  pages={1-8},
  doi={10.1109/ASRU65441.2025.11434723}}

@inproceedings{wang2026do,
title={Do {LLM}s {\textquotedblleft}Feel{\textquotedblright}? Emotion Circuits Discovery and Control},
author={Chenxi Wang and Yixuan Zhang and Ruiji Yu and Yufei Zheng and Lang Gao and Zirui Song and Zixiang Xu and Gus Xia and Huishuai Zhang and Dongyan Zhao and Xiuying Chen},
booktitle={Forty-third International Conference on Machine Learning},
year={2026},
url={https://openreview.net/forum?id=a8N0nRG3jA}
}

@book{goodglass1972assessment,
author="Goodglass, Harold and Kaplan, Edith",
title="The assessment of aphasia and related disorders",
publisher="Lea \& Febiger",
year="1972",
URL="https://cir.nii.ac.jp/crid/1971149384814916764"
}

@inproceedings{luz2020adress,
  title     = {{Alzheimer}'s Dementia Recognition through Spontaneous Speech: The {ADReSS} Challenge},
  author    = {Luz, Saturnino and Haider, Fasih and de la Fuente, Sofia and Fromm, Davida and MacWhinney, Brian},
  booktitle = {Proc. Interspeech 2020},
  pages     = {2172--2176},
  year      = {2020}
}

@article{doi:10.3233/JAD-150520,
author = {Kathleen C. Fraser and Jed A. Meltzer and Frank Rudzicz},
title ={Linguistic Features Identify Alzheimer’s Disease in Narrative Speech},
journal = {Journal of Alzheimer’s Disease},
volume = {49},
number = {2},
pages = {407-422},
year = {2016},
doi = {10.3233/JAD-150520},
    note ={PMID: 26484921},
URL = {https://journals.sagepub.com/doi/abs/10.3233/JAD-150520},
eprint = {https://journals.sagepub.com/doi/pdf/10.3233/JAD-150520}
}

@inproceedings{luz21_interspeech,
  title     = {{Detecting Cognitive Decline Using Speech Only: The ADReSSo Challenge}},
  author    = {Saturnino Luz and Fasih Haider and Sofia de la Fuente and Davida Fromm and Brian MacWhinney},
  year      = {2021},
  booktitle = {{Interspeech 2021}},
  pages     = {3780--3784},
  doi       = {10.21437/Interspeech.2021-1220},
  issn      = {2958-1796},
}

@inproceedings{luz24_interspeech,
  title     = {{Connected Speech-Based Cognitive Assessment in Chinese and English}},
  author    = {Saturnino Luz and Sofia {De La Fuente Garcia} and Fasih Haider and Davida Fromm and Brian MacWhinney and Alyssa Lanzi and Ya-Ning Chang and Chia-Ju Chou and Yi-Chien Liu},
  year      = {2024},
  booktitle = {{Interspeech 2024}},
  pages     = {947--951},
  doi       = {10.21437/Interspeech.2024-1807},
  issn      = {2958-1796},
}

@inproceedings{karlekar-etal-2018-detecting,
    title = "Detecting Linguistic Characteristics of {A}lzheimer{'}s Dementia by Interpreting Neural Models",
    author = "Karlekar, Sweta  and
      Niu, Tong  and
      Bansal, Mohit",
    editor = "Walker, Marilyn  and
      Ji, Heng  and
      Stent, Amanda",
    booktitle = "Proceedings of the 2018 Conference of the North {A}merican Chapter of the Association for Computational Linguistics: Human Language Technologies, Volume 2 (Short Papers)",
    month = jun,
    year = "2018",
    address = "New Orleans, Louisiana",
    publisher = "Association for Computational Linguistics",
    url = "https://aclanthology.org/N18-2110/",
    doi = "10.18653/v1/N18-2110",
    pages = "701--707"
}

@inproceedings{balagopalan20_interspeech,
  title     = {{To BERT or not to BERT: Comparing Speech and Language-Based Approaches for Alzheimer’s Disease Detection}},
  author    = {Aparna Balagopalan and Benjamin Eyre and Frank Rudzicz and Jekaterina Novikova},
  year      = {2020},
  booktitle = {{Interspeech 2020}},
  pages     = {2167--2171},
  doi       = {10.21437/Interspeech.2020-2557},
  issn      = {2958-1796},
}

@ARTICLE{9769980,
  author={Ilias, Loukas and Askounis, Dimitris},
  journal={IEEE Journal of Biomedical and Health Informatics}, 
  title={Explainable Identification of Dementia From Transcripts Using Transformer Networks}, 
  year={2022},
  volume={26},
  number={8},
  pages={4153-4164},
  doi={10.1109/JBHI.2022.3172479}}

@ARTICLE{10767227,
  author={Botelho, Catarina and Abad, Alberto and Schultz, Tanja and Trancoso, Isabel},
  journal={IEEE Access}, 
  title={Speech as a Biomarker for Disease Detection}, 
  year={2024},
  volume={12},
  number={},
  pages={184487-184508},
  doi={10.1109/ACCESS.2024.3506433}}

@inproceedings{yuan20_interspeech,
  title     = {{Disfluencies and Fine-Tuning Pre-Trained Language Models for Detection of Alzheimer’s Disease}},
  author    = {Jiahong Yuan and Yuchen Bian and Xingyu Cai and Jiaji Huang and Zheng Ye and Kenneth Church},
  year      = {2020},
  booktitle = {{Interspeech 2020}},
  pages     = {2162--2166},
  doi       = {10.21437/Interspeech.2020-2516},
  issn      = {2958-1796},
}

@ARTICLE{10.3389/fcomp.2020.624488,
AUTHOR={Yuan, Jiahong  and Cai, Xingyu  and Bian, Yuchen  and Ye, Zheng  and Church, Kenneth },
TITLE={Pauses for Detection of Alzheimer’s Disease},
JOURNAL={Frontiers in Computer Science},
VOLUME={Volume 2 - 2020},
YEAR={2021},
URL={https://www.frontiersin.org/journals/computer-science/articles/10.3389/fcomp.2020.624488},
DOI={10.3389/fcomp.2020.624488},
ISSN={2624-9898}}

@inproceedings{cummins20_interspeech,
  title     = {{A Comparison of Acoustic and Linguistics Methodologies for Alzheimer’s Dementia Recognition}},
  author    = {Nicholas Cummins and Yilin Pan and Zhao Ren and Julian Fritsch and Venkata Srikanth Nallanthighal and Heidi Christensen and Daniel Blackburn and Björn W. Schuller and Mathew Magimai-Doss and Helmer Strik and Aki Härmä},
  year      = {2020},
  booktitle = {{Interspeech 2020}},
  pages     = {2182--2186},
  doi       = {10.21437/Interspeech.2020-2635},
  issn      = {2958-1796},
}

@inproceedings{rohanian21_interspeech,
  title     = {{Alzheimer’s Dementia Recognition Using Acoustic, Lexical, Disfluency and Speech Pause Features Robust to Noisy Inputs}},
  author    = {Morteza Rohanian and Julian Hough and Matthew Purver},
  year      = {2021},
  booktitle = {{Interspeech 2021}},
  pages     = {3820--3824},
  doi       = {10.21437/Interspeech.2021-1633},
  issn      = {2958-1796},
}

@ARTICLE{8910399,
  author={Haider, Fasih and de la Fuente, Sofia and Luz, Saturnino},
  journal={IEEE Journal of Selected Topics in Signal Processing}, 
  title={An Assessment of Paralinguistic Acoustic Features for Detection of Alzheimer's Dementia in Spontaneous Speech}, 
  year={2020},
  volume={14},
  number={2},
  pages={272-281},
  doi={10.1109/JSTSP.2019.2955022}}

@inproceedings{alkhamissi-etal-2025-llm,
    title = "The {LLM} Language Network: A Neuroscientific Approach for Identifying Causally Task-Relevant Units",
    author = "AlKhamissi, Badr  and
      Tuckute, Greta  and
      Bosselut, Antoine  and
      Schrimpf, Martin",
    editor = "Chiruzzo, Luis  and
      Ritter, Alan  and
      Wang, Lu",
    booktitle = "Proceedings of the 2025 Conference of the Nations of the Americas Chapter of the Association for Computational Linguistics: Human Language Technologies (Volume 1: Long Papers)",
    month = apr,
    year = "2025",
    address = "Albuquerque, New Mexico",
    publisher = "Association for Computational Linguistics",
    url = "https://aclanthology.org/2025.naacl-long.544/",
    doi = "10.18653/v1/2025.naacl-long.544",
    pages = "10887--10911",
    ISBN = "979-8-89176-189-6"
}

@article{shapiro1965analysis,
  title={An analysis of variance test for normality (complete samples)},
  author={Shapiro, Samuel Sanford and Wilk, Martin B},
  journal={Biometrika},
  volume={52},
  number={3-4},
  pages={591--611},
  year={1965},
  publisher={Oxford University Press}
}

@ARTICLE{10.3389/fnagi.2021.635945,
AUTHOR={Balagopalan, Aparna  and Eyre, Benjamin  and Robin, Jessica  and Rudzicz, Frank  and Novikova, Jekaterina },
TITLE={Comparing Pre-trained and Feature-Based Models for Prediction of Alzheimer's Disease Based on Speech},
JOURNAL={Frontiers in Aging Neuroscience},
VOLUME={Volume 13 - 2021},
YEAR={2021},
URL={https://www.frontiersin.org/journals/aging-neuroscience/articles/10.3389/fnagi.2021.635945},
DOI={10.3389/fnagi.2021.635945},
ISSN={1663-4365},}

@article{10.1371/journal.pdig.0000168,
    doi = {10.1371/journal.pdig.0000168},
    author = {Agbavor, Felix AND Liang, Hualou},
    journal = {PLOS Digital Health},
    publisher = {Public Library of Science},
    title = {Predicting dementia from spontaneous speech using large language models},
    year = {2022},
    month = {12},
    volume = {1},
    url = {https://doi.org/10.1371/journal.pdig.0000168},
    pages = {1-14},
    number = {12},

}

@article{https://doi.org/10.4218/etrij.2023-0356,
author = {Bang, Jeong-Uk and Han, Seung-Hoon and Kang, Byung-Ok},
title = {Alzheimer's disease recognition from spontaneous speech using large language models},
journal = {ETRI Journal},
volume = {46},
number = {1},
pages = {96-105},
doi = {https://doi.org/10.4218/etrij.2023-0356},
url = {https://onlinelibrary.wiley.com/doi/abs/10.4218/etrij.2023-0356},
eprint = {https://onlinelibrary.wiley.com/doi/pdf/10.4218/etrij.2023-0356},
year = {2024}
}

@article{doi:10.3233/JAD-200888,
author = {Sofia de la Fuente Garcia and Craig W. Ritchie and Saturnino Luz},
title ={Artificial Intelligence, Speech, and Language Processing Approaches to Monitoring Alzheimer’s Disease: A Systematic Review},
journal = {Journal of Alzheimer’s Disease},
volume = {78},
number = {4},
pages = {1547-1574},
year = {2020},
doi = {10.3233/JAD-200888},
    note ={PMID: 33185605},
URL = {https://journals.sagepub.com/doi/abs/10.3233/JAD-200888},
eprint = {https://journals.sagepub.com/doi/pdf/10.3233/JAD-200888},
}

@inproceedings{wu2025axbench,
title={AxBench: Steering {LLM}s? Even Simple Baselines Outperform Sparse Autoencoders},
author={Zhengxuan Wu and Aryaman Arora and Atticus Geiger and Zheng Wang and Jing Huang and Dan Jurafsky and Christopher D Manning and Christopher Potts},
booktitle={Forty-second International Conference on Machine Learning},
year={2025},
url={https://openreview.net/forum?id=K2CckZjNy0}
}

@InProceedings{pmlr-v267-kantamneni25a,
  title = 	 {Are Sparse Autoencoders Useful? {A} Case Study in Sparse Probing},
  author =       {Kantamneni, Subhash and Engels, Joshua and Rajamanoharan, Senthooran and Tegmark, Max and Nanda, Neel},
  booktitle = 	 {Proceedings of the 42nd International Conference on Machine Learning},
  pages = 	 {29018--29049},
  year = 	 {2025},
  editor = 	 {Singh, Aarti and Fazel, Maryam and Hsu, Daniel and Lacoste-Julien, Simon and Berkenkamp, Felix and Maharaj, Tegan and Wagstaff, Kiri and Zhu, Jerry},
  volume = 	 {267},
  series = 	 {Proceedings of Machine Learning Research},
  month = 	 {13--19 Jul},
  publisher =    {PMLR},
  url = 	 {https://proceedings.mlr.press/v267/kantamneni25a.html}
}

@inproceedings{mondal-etal-2025-language,
    title = "Language-Specific Neurons Do Not Facilitate Cross-Lingual Transfer",
    author = "Mondal, Soumen Kumar  and
      Sen, Sayambhu  and
      Singhania, Abhishek  and
      Jyothi, Preethi",
    editor = "Drozd, Aleksandr  and
      Sedoc, Jo{\~a}o  and
      Tafreshi, Shabnam  and
      Akula, Arjun  and
      Shu, Raphael",
    booktitle = "The Sixth Workshop on Insights from Negative Results in NLP",
    month = may,
    year = "2025",
    address = "Albuquerque, New Mexico",
    publisher = "Association for Computational Linguistics",
    url = "https://aclanthology.org/2025.insights-1.6/",
    doi = "10.18653/v1/2025.insights-1.6",
    pages = "46--62",
    ISBN = "979-8-89176-240-4"
}

@inproceedings{gurgurov-etal-2025-language,
    title = "Language Arithmetics: Towards Systematic Language Neuron Identification and Manipulation",
    author = "Gurgurov, Daniil  and
      Trinley, Katharina  and
      Al Ghussin, Yusser  and
      Baeumel, Tanja  and
      van Genabith, Josef  and
      Ostermann, Simon",
    editor = "Inui, Kentaro  and
      Sakti, Sakriani  and
      Wang, Haofen  and
      Wong, Derek F.  and
      Bhattacharyya, Pushpak  and
      Banerjee, Biplab  and
      Ekbal, Asif  and
      Chakraborty, Tanmoy  and
      Singh, Dhirendra Pratap",
    booktitle = "Proceedings of the 14th International Joint Conference on Natural Language Processing and the 4th Conference of the Asia-Pacific Chapter of the Association for Computational Linguistics",
    month = dec,
    year = "2025",
    address = "Mumbai, India",
    publisher = "The Asian Federation of Natural Language Processing and The Association for Computational Linguistics",
    url = "https://aclanthology.org/2025.ijcnlp-long.156/",
    doi = "10.18653/v1/2025.ijcnlp-long.156",
    pages = "2911--2937",
    ISBN = "979-8-89176-298-5"
}

@article{CROISILE19961,
title = {Comparative Study of Oral and Written Picture Description in Patients with Alzheimer's Disease},
journal = {Brain and Language},
volume = {53},
number = {1},
pages = {1-19},
year = {1996},
issn = {0093-934X},
doi = {https://doi.org/10.1006/brln.1996.0033},
url = {https://www.sciencedirect.com/science/article/pii/S0093934X96900334},
author = {Bernard Croisile and Bernadette Ska and Marie-Josee Brabant and Annick Duchene and Yves Lepage and Gilbert Aimard and Marc Trillet}
}

@inproceedings{park25d_interspeech,
  title     = {{Reasoning-Based Approach with Chain-of-Thought for Alzheimer’s Detection Using Speech and Large Language Models}},
  author    = {Chanwoo Park and Anna Seo Gyeong Choi and Sunghye Cho and Chanwoo Kim},
  year      = {2025},
  booktitle = {{Interspeech 2025}},
  pages     = {2185--2189},
  doi       = {10.21437/Interspeech.2025-1226},
  issn      = {2958-1796},
}

@article{https://doi.org/10.1016/j.jalz.2011.03.005,
author = {McKhann, Guy M. and Knopman, David S. and Chertkow, Howard and Hyman, Bradley T. and Jack Jr., Clifford R. and Kawas, Claudia H. and Klunk, William E. and Koroshetz, Walter J. and Manly, Jennifer J. and Mayeux, Richard and Mohs, Richard C. and Morris, John C. and Rossor, Martin N. and Scheltens, Philip and Carrillo, Maria C. and Thies, Bill and Weintraub, Sandra and Phelps, Creighton H.},
title = {The diagnosis of dementia due to Alzheimer's disease: Recommendations from the National Institute on Aging-Alzheimer's Association workgroups on diagnostic guidelines for Alzheimer's disease},
journal = {Alzheimer's \& Dementia},
volume = {7},
number = {3},
pages = {263-269},
doi = {https://doi.org/10.1016/j.jalz.2011.03.005},
url = {https://alz-journals.onlinelibrary.wiley.com/doi/abs/10.1016/j.jalz.2011.03.005},
eprint = {https://alz-journals.onlinelibrary.wiley.com/doi/pdf/10.1016/j.jalz.2011.03.005},
year = {2011}
}

@article{https://doi.org/10.1016/j.jalz.2011.03.008,
author = {Albert, Marilyn S. and DeKosky, Steven T. and Dickson, Dennis and Dubois, Bruno and Feldman, Howard H. and Fox, Nick C. and Gamst, Anthony and Holtzman, David M. and Jagust, William J. and Petersen, Ronald C. and Snyder, Peter J. and Carrillo, Maria C. and Thies, Bill and Phelps, Creighton H.},
title = {The diagnosis of mild cognitive impairment due to Alzheimer's disease: Recommendations from the National Institute on Aging-Alzheimer's Association workgroups on diagnostic guidelines for Alzheimer's disease},
journal = {Alzheimer's \& Dementia},
volume = {7},
number = {3},
pages = {270-279},
doi = {https://doi.org/10.1016/j.jalz.2011.03.008},
url = {https://alz-journals.onlinelibrary.wiley.com/doi/abs/10.1016/j.jalz.2011.03.008},
eprint = {https://alz-journals.onlinelibrary.wiley.com/doi/pdf/10.1016/j.jalz.2011.03.008},
year = {2011}
}

@article{https://doi.org/10.1111/j.1365-2796.2004.01388.x,
author = {Petersen, R. C.},
title = {Mild cognitive impairment as a diagnostic entity},
journal = {Journal of Internal Medicine},
volume = {256},
number = {3},
pages = {183-194},
doi = {https://doi.org/10.1111/j.1365-2796.2004.01388.x},
url = {https://onlinelibrary.wiley.com/doi/abs/10.1111/j.1365-2796.2004.01388.x},
eprint = {https://onlinelibrary.wiley.com/doi/pdf/10.1111/j.1365-2796.2004.01388.x},
year = {2004}
}

@article{https://doi.org/10.1002/alz.71345,
author={{Alzheimer's Association}},
title = {2026 Alzheimer's disease facts and figures},
journal = {Alzheimer's \& Dementia},
volume = {22},
number = {4},
pages = {e71345},
doi = {https://doi.org/10.1002/alz.71345},
url = {https://alz-journals.onlinelibrary.wiley.com/doi/abs/10.1002/alz.71345},
eprint = {https://alz-journals.onlinelibrary.wiley.com/doi/pdf/10.1002/alz.71345},
year = {2026}
}

@article{scheltens2021alzheimer,
  title={Alzheimer's disease},
  author={Scheltens, Philip and De Strooper, Bart and Kivipelto, Miia and Holstege, Henne and Ch{\'e}telat, Gael and Teunissen, Charlotte E and Cummings, Jeffrey and Van der Flier, Wiesje M},
  journal={The Lancet},
  volume={397},
  number={10284},
  pages={1577--1590},
  year={2021},
  publisher={Elsevier}
}

@article{nichols2022estimation,
  title={Estimation of the global prevalence of dementia in 2019 and forecasted prevalence in 2050: an analysis for the Global Burden of Disease Study 2019},
  author={Nichols, Emma and Steinmetz, Jaimie D and Vollset, Stein Emil and Fukutaki, Kai and Chalek, Julian and Abd-Allah, Foad and Abdoli, Amir and Abualhasan, Ahmed and Abu-Gharbieh, Eman and Akram, Tayyaba Tayyaba and others},
  journal={The Lancet Public Health},
  volume={7},
  number={2},
  pages={e105--e125},
  year={2022},
  publisher={Elsevier}
}

@article{https://doi.org/10.1111/j.2517-6161.1995.tb02031.x,
author = {Benjamini, Yoav and Hochberg, Yosef},
title = {Controlling the False Discovery Rate: A Practical and Powerful Approach to Multiple Testing},
journal = {Journal of the Royal Statistical Society: Series B (Methodological)},
volume = {57},
number = {1},
pages = {289-300},
doi = {https://doi.org/10.1111/j.2517-6161.1995.tb02031.x},
url = {https://rss.onlinelibrary.wiley.com/doi/abs/10.1111/j.2517-6161.1995.tb02031.x},
eprint = {https://rss.onlinelibrary.wiley.com/doi/pdf/10.1111/j.2517-6161.1995.tb02031.x},
year = {1995}
}

\appendix

\section{Reproducibility Details}
\label{appendix:reproducibility}

\subsection{Models}
\label{appendix:models}

Table~\ref{tab:checkpoints} lists the checkpoint of each model. All weights are loaded from local copies of these checkpoints.

\begin{table}[h]
\centering
\small
\resizebox{\columnwidth}{!}{%
\begin{tabular}{@{}ll@{}}
\toprule
Model & Hugging Face checkpoint \\ \midrule
Qwen2.5-Omni-7B & \texttt{Qwen/Qwen2.5-Omni-7B} \\
Qwen3-Omni-30B & \texttt{Qwen/Qwen3-Omni-30B-A3B-Instruct} \\
Kimi-Audio & \texttt{moonshotai/Kimi-Audio-7B-Instruct} \\
MiniCPM-o-4.5 & \texttt{openbmb/MiniCPM-o-4\_5} \\
Gemma-4-12B & \texttt{google/gemma-4-12B-it} \\
Voxtral-Small-24B & \texttt{mistralai/Voxtral-Small-24B-2507} \\
Audio-Flamingo-Next & \texttt{nvidia/audio-flamingo-next-hf} \\
MOSS-Audio-8B & \texttt{OpenMOSS-Team/MOSS-Audio-8B-Instruct} \\ \bottomrule
\end{tabular}%
}
\caption{Model names used in this paper and the Hugging Face checkpoints they refer to.}
\label{tab:checkpoints}
\end{table}

\paragraph{Audio input.}
All models receive the same waveform: 16~kHz mono PCM, decoded and downmixed without level normalization, passed as float32 arrays. PROCESS-2 is distributed loudness-normalized; we apply no normalization to either corpus. In \texttt{T} the model receives no audio input at all, rather than silent audio.
We checked each model's effective audio-length limit in its installed processing code. Participant-only Pitt audio is 11--220~s long (median 47~s) and PROCESS-2 recordings are at most 261~s, so no limit binds.
Two limits needed intervention:
\begin{itemize}[left=0pt]
\item Qwen3-Omni-30B's shipped preprocessor configuration lacks the chunk-length field, so its feature extractor silently falls back to 30~s and cuts every longer input there. We restore the intended 300~s.
\item Gemma-4-12B truncates each audio clip at 30~s, so it receives consecutive clips of at most 30~s. For Pitt the clips are cut at utterance boundaries: 2 of 532 cuts fall inside an utterance longer than 30~s. For PROCESS-2 they are cut at detected silences, with a quietest-frame fallback, so that no stretch exceeds 25~s. Every model therefore hears the same speech. The audio span for Gemma-4-12B then includes the few clip-delimiter tokens.
\end{itemize}

\paragraph{Span detection.}
For Hugging Face processor models, the audio and transcript spans are located in the input token IDs: the audio span runs from the first to the last audio placeholder token, and the transcript span covers the tokens of the quoted transcript string. For Kimi-Audio and MiniCPM-o-4.5, whose chat interfaces build the input internally, the input IDs are captured by a hook on the language model.
For the two models without Pitt masks, the audio span of Voxtral-Small-24B ends at the last token that holds real audio, before its zero-padding to a multiple of 30~s, and that of MOSS-Audio-8B includes the time-marker digits inserted every 2~s.
The token-matched audio span is validated by setting its sample size to the full audio span: it then reproduces the standard audio-span statistics bit for bit, for both Kimi-Audio and the Qwen3-Omni-30B MoE path.

\subsection{Implementation Details}
\label{appendix:impl}

\paragraph{Token spans.}
For the two joint conditions, each model wrapper reports the token ranges of the audio block and the transcript block in the prompt. Activation statistics are then recorded over three spans: all prompt tokens (\texttt{all}), the audio tokens (\texttt{audio}) and the transcript tokens (\texttt{text}). The transcript is about 5$\times$ shorter than the audio in tokens. To check that this difference does not drive the results, we add a token-matched audio span (\texttt{audio\_tm}): for each item, a seeded random subset of its audio positions with as many positions as the item has transcript tokens.

\begin{table}[t]
\centering
\small
\resizebox{\columnwidth}{!}{%
\begin{tabular}{@{}lrrrr@{}}
\toprule
Model & Layers & MLP width & Neurons & Mask ($r{=}0.5\%$) \\ \midrule
Qwen2.5-Omni-7B & 28 & 18{,}944 & 530{,}432 & 2{,}652 \\
Qwen3-Omni-30B & 48 & $128\times768$ & 4{,}718{,}592 & 23{,}592 \\
Kimi-Audio & 28 & 18{,}944 & 530{,}432 & 2{,}652 \\
MiniCPM-o-4.5 & 36 & 12{,}288 & 442{,}368 & 2{,}211 \\
Gemma-4-12B & 48 & 15{,}360 & 737{,}280 & 3{,}686 \\ \bottomrule
\end{tabular}%
}
\caption{Decoder MLP sizes of the models that reach identification on Pitt (Audio-Flamingo-Next, Voxtral-Small-24B and MOSS-Audio-8B do not meet the analysis-eligibility criteria, \S\ref{subsec:eligibility}). For Qwen3-Omni-30B a neuron is one unit of one expert.}
\label{tab:models}
\end{table}

\paragraph{Mixture-of-experts neurons.}
For the sparse mixture-of-experts decoder (Qwen3-Omni-30B) we treat each unit of each expert as a neuron: neuron $e\cdot W+j$ of a layer is unit $j$ of expert $e$, where $W$ is the expert width. We hook each expert's gate nonlinearity and recover token positions by recomputing the router's top-8 selection in a pre-hook on the MoE block, so that spans work exactly as for dense models. A token that is not routed to expert $e$ counts as activation $0$ for all of $e$'s units, so every neuron is measured on the same token set; per-item, per-expert routed-token counts are stored, so routing-conditional statistics can be derived. A routed expert only runs on the tokens assigned to it, so a shift reaches those tokens only, and we report ablation alone for this model.

\paragraph{Analysis eligibility by model and condition.}
The twelve Pitt cells that advance to identification are listed in Table~\ref{tab:pools}. The eligibility criteria are applied separately to each input condition (Appendix~\ref{appendix:pilot}). On Pitt, some Qwen cells do not advance because of a strong option-letter preference. Audio-Flamingo-Next has no CI recording that meets the both-orders confidence criterion in any condition (0 of the 67 CI recordings of the evaluation split; its identification activations were not logged), while Voxtral-Small-24B and MOSS-Audio-8B remain outside the predefined Pitt UAR criterion (Pitt \texttt{T} UAR 0.548 and 0.500, below 0.65; the other-condition intervals include 0.5, with UAR 0.50--0.52; Table~\ref{tab:pilot_full}); MOSS-Audio-8B predicts CI for every Pitt evaluation recording. On PROCESS-2, the UAR criterion is met by \texttt{T} for Gemma-4-12B and Voxtral-Small-24B, \texttt{A} for Kimi-Audio, MiniCPM-o-4.5, Voxtral-Small-24B and MOSS-Audio-8B, and \texttt{AT} for Kimi-Audio, Qwen2.5-Omni-7B and Voxtral-Small-24B (Table~\ref{tab:pilot_full_p2}). Gemma-4-12B \texttt{T} and Voxtral-Small-24B \texttt{T} also meet the balanced-pool criterion (Table~\ref{tab:pools}). The pool criterion is defined on identification recordings; for cells whose identification activations were not logged it was screened on the evaluation split instead. The remaining cells have insufficient eligible recordings for the fixed $N=15$ per class: for example, Voxtral-Small-24B has 12 NC recordings in \texttt{AT} and 6 CI recordings in \texttt{A}; Kimi-Audio \texttt{A}/\texttt{AT}, MiniCPM-o-4.5 \texttt{A} and Qwen2.5-Omni-7B \texttt{AT}, whose activations were not logged, have at most four CI recordings meeting the confidence criterion among the 20 CI cases of the evaluation split. Identification is therefore restricted to cells satisfying the predefined reliability and pool criteria.

\paragraph{Eligibility on the identification split.}
The UAR criteria were applied on the evaluation split. Table~\ref{tab:iden_gate} applies the same criteria to the identification split, where activations were logged (297 Pitt recordings; intervals resample participants). All 12 Pitt cells that reach identification also pass there: every interval excludes 0.5, and every cell has at least 16 eligible recordings per class (Table~\ref{tab:pools}). The four further Pitt cells with logged activations (Qwen2.5-Omni-7B \texttt{T}, \texttt{A} and \texttt{AT-chat}, Qwen3-Omni-30B \texttt{T}) meet the UAR criterion on both splits and are excluded by the pool criterion alone (1--11 eligible CI recordings). The remaining 16 Pitt cells were ruled out on the evaluation split before any activation was logged, 14 by the UAR criterion and two (Audio-Flamingo-Next \texttt{AT} and \texttt{AT-chat}) by the pool criterion screened on that split, so their identification-split results are unknown.

\begin{table}[h]
\centering
\small
\resizebox{\ifdim\width>\linewidth\linewidth\else\width\fi}{!}{%
\begin{tabular}{@{}ll cc@{}}
\toprule
& & eval. & identification \\
Model & Cond. & UAR & UAR [95\% CI] \\ \midrule
Gemma-4-12B      & \texttt{AT}      & 0.67 & 0.663 [0.608, 0.715] \\
             & \texttt{AT-chat} & 0.68 & 0.639 [0.582, 0.695] \\
Kimi-Audio   & \texttt{A}       & 0.67 & 0.634 [0.573, 0.690] \\
             & \texttt{AT}      & 0.64 & 0.615 [0.554, 0.673] \\
             & \texttt{AT-chat} & 0.64 & 0.589 [0.528, 0.652] \\
MiniCPM-o-4.5 & \texttt{A}      & 0.66 & 0.635 [0.580, 0.688] \\
             & \texttt{AT}      & 0.63 & 0.736 [0.681, 0.789] \\
             & \texttt{AT-chat} & 0.66 & 0.682 [0.621, 0.739] \\
Qwen2.5-Omni-7B & \texttt{AT}      & 0.74 & 0.707 [0.652, 0.763] \\
Qwen3-Omni-30B   & \texttt{A}       & 0.73 & 0.691 [0.634, 0.747] \\
             & \texttt{AT}      & 0.75 & 0.751 [0.694, 0.806] \\
             & \texttt{AT-chat} & 0.75 & 0.683 [0.620, 0.740] \\ \bottomrule
\end{tabular}%
}
\caption{Analysis eligibility recomputed on the Pitt identification split for the 12 cells that reach identification. ``eval.'': UAR on the evaluation split (Table~\ref{tab:pilot_full}). The eligible identification recordings of each cell are in Table~\ref{tab:pools}.}
\label{tab:iden_gate}
\end{table}

Applying the eligibility criteria on the identification split alone would therefore retain every cell we report and could differ only by admitting some of the 16 unchecked cells. Such cells would add masks to the family of the diagonal ablation, not remove any. A bound that holds whatever the $p$-values of those masks would have been is the Benjamini--Hochberg correction of the conjunction test over all 32 Pitt model--condition cells (64 masks, against 24 in the reported family) with every unobserved mask counted as a non-rejection, which multiplies each reported $q$ by $64/24$: the four primary masks of Section~\ref{subsec:ablation} stay below 0.05 under it ($q\le0.006$), as does Kimi-Audio \texttt{AT} NC ($q=0.032$). Their significance thus does not depend on the eligibility decision. The count of models with a passing mask remains conditional on these criteria: three of the five models that reach identification, out of eight studied.

On PROCESS-2, Gemma-4-12B \texttt{T} and Voxtral-Small-24B \texttt{T} would miss the \texttt{T} threshold on the identification split (UAR 0.649, interval [0.563, 0.727], and 0.636, [0.546, 0.718]; 0.709 and 0.741 on the evaluation split). Their masks give null results, so no positive result depends on these two cells.

\paragraph{Number of random masks.}
The number of eligible recordings ranges from 16 to 86 per class and cell on Pitt and from 19 to 45 on PROCESS-2. Identification, analysis and statistics run on CPU.

\subsection{Data Preparation}
\label{appendix:data}

\paragraph{Choice of corpora.}
The ADReSS challenge \cite{luz2020adress} and its successors ADReSSo \cite{luz21_interspeech} and the multilingual ADReSS-M \cite{luz2023multilingual} are the common benchmarks for this task. We use the original Pitt recordings instead of these benchmark releases because our input conditions need unprocessed audio together with timed manual transcripts, and PROCESS-2 because it adds an independently collected corpus with manual transcripts and an MCI group.

\paragraph{Pitt.}
We use the original audio and the CHAT transcripts of the Cookie Theft task. Control maps to NC and probable or possible Alzheimer's disease to CI; MCI, vascular dementia and other diagnoses are held out and used only as a transfer test.
We parse the CHAT files and keep the participant tier (\texttt{*PAR}).
An utterance is kept only if it has a timing bullet, so audio and text cover the same speech. Its start is clipped to the end of the previous kept utterance, which removes overlap, and it is dropped if nothing remains.
The kept segments are cut from the original recordings and concatenated. For each segment we store its position in the concatenated audio and its raw CHAT text, and the transcripts are rendered from these segments at input time.
Seven Cookie Theft recordings are 8~kHz MP3s (all others are 44.1~kHz). All seven belong to dementia or MCI participants, which would make audio bandwidth a label shortcut. They are still recorded during activation logging but are excluded from aggregation and evaluation.
Three recordings lack a diagnosis in their header and are excluded.

\paragraph{PROCESS-2.}
PROCESS-2 contains remote cognitive assessments of 400 participants, recorded on their own devices. We use its Cookie Theft task with the manual transcripts, and map HC to NC and dementia to CI, holding out MCI as for Pitt.
PROCESS-2 recordings are not cut, since their transcripts carry no timings, so the models receive the full recording. We exclude the 17 recordings whose transcript contains another speaker's turn or a note about another speaker (14 MCI, 2 HC, 1 dementia).
The transcripts come in two styles, one marking pauses and fillers and one in plain prose. Table~\ref{tab:process2_style} shows how transcription style relates to diagnosis. The plain-prose style is concentrated in HC, and style alone separates HC from dementia at $\approx 0.63$ UAR. No plain-style HC participant has an MMSE score, which suggests a separate recruitment stream, and fillers are written out far less often in that style. Our clean rendering removes the explicit markup that differs between the styles (Appendix~\ref{appendix:render}), the partition is stratified by style, and every PROCESS-2 result is also reported within each style. Because no annotated transcript rendering is defined for PROCESS-2, experiments on this corpus use only the \texttt{T}, \texttt{A}, and \texttt{AT} conditions.

\begin{table}[h]
\centering
\small
\resizebox{\ifdim\width>\linewidth\linewidth\else\width\fi}{!}{%
\begin{tabular}{@{}lrrr@{}}
\toprule
& HC & MCI & Dementia \\ \midrule
Labeled style ($n$) & 99 & 124 & 38 \\
Plain style ($n$)    & 101 & 26 & 12 \\
Plain style (\%)     & 50.5 & 17.3 & 24.0 \\ \midrule
Fillers / 100 words, labeled & 3.9 & 3.7 & 2.8 \\
Fillers / 100 words, plain    & 1.4 & 2.8 & 3.7 \\
Median words, labeled        & 163 & 127 & 120 \\
Median words, plain           & 139 & 115 & 117 \\ \bottomrule
\end{tabular}%
}
\caption{PROCESS-2 Cookie Theft transcripts by transcription style (all 400 participants, before exclusions).}
\label{tab:process2_style}
\end{table}

\paragraph{Partitions.}
Both corpora use participant-disjoint identification (60\%) and evaluation (40\%) splits, stratified by class and, for PROCESS-2, by transcription style (seed 42). Evaluation uses one recording per participant; identification may use multiple visits from the same identification participant. The stratification is over participants (Pitt: 59 NC and 97 CI in identification, 40 NC and 67 CI in evaluation), so the class shares of recordings differ between the splits where the classes differ in visits per participant.
Pitt contains 297 identification recordings (152 NC, 145 CI) and 107 evaluation recordings (40 NC, 67 CI); its 297 identification recordings come from 156 participants. PROCESS-2 contains 148 identification recordings (119 NC, 29 CI) and 99 evaluation recordings (79 NC, 20 CI), together with 136 held-out MCI recordings. Held-out diagnoses use one recording per participant and exclude identification participants. The gender control relabels the same recordings.

\paragraph{Interpretation and scope.}
Pitt's utterance timings come from automatic forced alignment. PROCESS-2 has no timings, so its audio cannot be cut to the participant's utterances and keeps the pauses and background noise between them; recordings whose transcripts show another speaker are excluded, and a listening check of a sample found no spoken instructions from the assessment platform, but other voices in the remaining recordings cannot be ruled out entirely. Its transcription style is associated with diagnosis; our rendering and stratification reduce but cannot fully exclude this confound, especially in the audio. Transcripts are manual throughout; automatic transcripts, which a deployed system would use, were not evaluated.

\subsection{Transcript Renderings}
\label{appendix:render}

\paragraph{Clean (Pitt).}
All CHAT codes are removed: pauses, fillers, fragments, retracing markers, unintelligible-speech markers, events, error codes and replacement targets. The produced form is kept, not the target.
Repeated words stay, because they were spoken.
Prolongation marks and morphological parentheses are resolved, e.g.\ ``wa:ter'' $\rightarrow$ ``water'' and ``(be)cause'' $\rightarrow$ ``because''.

\paragraph{Annotated (Pitt, convention \texttt{v1}).}
The same words plus the fillers and word fragments, with the codes in Table~\ref{tab:render} rendered as readable text; everything else is removed as in the clean rendering.
The transcript is preceded by a fixed legend:
\emph{``Transcript notation: (short pause), (pause), (long pause) and (pause Ns) mark silences; uh/um are filled pauses; a word ending in `-' was broken off; [repeats], [self-correction], [reformulation] and [false start] follow the words the speaker went back over; (unintelligible) marks speech the transcriber could not understand.''}

\begin{table}[h]
\centering
\small
\resizebox{\ifdim\width>\linewidth\linewidth\else\width\fi}{!}{%
\begin{tabular}{@{}ll@{}}
\toprule
CHAT & Rendered as \\ \midrule
\texttt{(.)}, \texttt{(..)}, \texttt{(...)} & (short pause), (pause), (long pause) \\
timed pause \texttt{(2.5)} & (pause 2.5s) \\
filler \texttt{\&-uh} & uh \\
fragment \texttt{\&+wa} & wa- \\
\texttt{[/]}, \texttt{[//]}, \texttt{[///]} & [repeats], [self-correction], [reformulation] \\
\texttt{[/-]}, \texttt{[/?]} & [false start], [unclear retracing] \\
\texttt{xxx}, \texttt{yyy} & (unintelligible) \\
events \texttt{\&=laughs}, error codes \texttt{[*]} & removed \\ \bottomrule
\end{tabular}%
}
\caption{Annotated rendering (convention \texttt{v1}).}
\label{tab:render}
\end{table}

\paragraph{Clean (PROCESS-2).}
Only participant turns are kept.
All parenthetical and bracketed notes are removed: timed pauses, non-verbal events, uncertain words.
Fillers are removed, using the same filler list as for CHAT.
Curly quotes are normalized, and everything is lower-cased except the pronoun ``I''. Without this last step, removing a sentence-initial filler would leave a lower-case sentence start in one style only.
Word fragments are unmarked in these transcripts and stay.

\subsection{Prompts}
\label{appendix:prompt}

\begin{lstlisting}[style=promptbox, caption={Prompt skeleton (\texttt{AT} shown). The option order is reversed in the second run of every item.}, label={lst:cicn-prompt}]
The following comes from an older adult describing a picture of a kitchen scene.
Recording of the speaker: <audio>
Transcript of the speaker:
"{transcript}"
Which option better describes the speaker? Both options are equally likely.
A: The speaker is cognitively healthy.
B: The speaker is cognitively impaired.
Answer with only the option letter.
\end{lstlisting}

Listing~\ref{lst:cicn-prompt} shows the \texttt{AT} prompt. The other conditions differ only in the blocks they contain:
\begin{itemize}[left=0pt]
\item \texttt{T} starts the transcript block directly after the context sentence;
\item \texttt{A} ends the audio block with the question;
\item \texttt{AT-chat} inserts the legend line (Appendix~\ref{appendix:render}) between ``Transcript of the speaker:'' and the quoted transcript.
\end{itemize}
The gender task replaces the two option texts with ``The speaker is male.'' and ``The speaker is female.''
For each model the segments are rendered with the model's own chat template, with the audio at the position of \texttt{<audio>}.

\subsection{Deterministic Inference}
\label{appendix:determinism}

\paragraph{Prediction from first-step logits.}
Predictions do not depend on the generated text (\S\ref{subsec:probe}). We read the first-step logits before any logits processor (temperature, top-$k$/top-$p$, repetition penalty), so decoding settings cannot change the probabilities.
Generation is capped at 8 new tokens, and the generated text is logged only for inspection.

\paragraph{Decoding.}
Every model decodes greedily (one beam, no sampling). Several released checkpoints ship generation defaults that sample or use beam search, so we override them explicitly rather than rely on the defaults.
For the Hugging Face models, the first-step logits come from \texttt{generate(..., output\_logits=True)}. MiniCPM-o-4.5 and Kimi-Audio use chat interfaces that do not return logits, so we capture the first-step output of their language-model head with a forward hook.

\paragraph{Numerical determinism.}
We fix all random seeds, enable PyTorch's deterministic algorithms and run one item at a time on a single NVIDIA A100 80GB GPU, in the precision each model's inference code uses; results are bit-identical only on the same GPU type, so every run uses A100s. Running the full inference path twice as independent jobs gave identical logits, predictions and generated text for every model.

\section{Statistical Procedure}
\label{appendix:stats}

\paragraph{Test.}
For each metric we compare the real mask's value $x$ against the values $y_1,\dots,y_K$ of its $K$ matched random masks with a prediction-interval $t$-test,
\begin{equation}
t=\frac{x-\bar y}{s_y\sqrt{1+1/K}},\qquad \text{df}=K-1,
\end{equation}
where $\bar y$ and $s_y$ are the mean and standard deviation of the random-mask values. The test is two-sided; its $p$ is the probability that a new random mask would be at least as extreme as the real one.
We check the normality of the random-mask values with a Shapiro--Wilk test \cite{shapiro1965analysis}; for both co-primary metrics normality is rejected in at most a few cells per family.

\paragraph{Conjunction.}
Each mask is one hypothesis, tested by the intersection--union test on the two co-primary metrics: $p_{\text{conj}}=\max(p_{\Delta P},p_{\Delta\ell})$, which needs no further correction between the two metrics. A mask is counted as passing only if its effect also lies in the predicted direction.

\paragraph{Distribution-free checks.}
The $t$-test assumes normal random-mask values and ignores the sampling of speakers, and with $K=20$ a Shapiro--Wilk non-rejection is only weak evidence of normality. We therefore add three descriptive checks that need neither assumption; the effect size relative to the random-mask distribution remains the most interpretable result.
First, the rank $p$ of each real mask among its random masks. Its floor is $1/21$ at $K=20$, which cannot resolve the observed rejection pattern under a correction over a family, so the test uses the prediction interval.
Second, a family-level count: if ConAct selection were no better than random selection, a real mask would lie beyond all $K$ of its random masks with probability $1/(K+1)$, so about one of 24 masks would. On the Pitt diagonal 9 of 24 do on each co-primary metric, in the transfer to \texttt{T} 10 ($\Delta\ell$) and 9 ($\Delta P$) of 24, and on PROCESS-2 7 of 16. The masks of one model are not independent, so we attach no $p$-value to these counts and report them descriptively. On the diagonal the nine masks are the seven that pass the test, Kimi-Audio \texttt{A} NC and Qwen3-Omni-30B \texttt{A} NC.
Third, a bootstrap over evaluation recordings (2000 resamples, the same recordings for a real mask and its random masks). The four primary masks stay beyond every random mask in 100\% of resamples; Qwen3-Omni-30B's \texttt{AT} CI mask does so in 91\% ($\Delta\ell$) and 77\% ($\Delta P$). This shows stability over evaluation recordings for the given sample of random masks, not generalization across identification pools, prompt variants or stronger null masks.
We do not test the bootstrap difference between a real mask and the \emph{mean} of its random masks. That mean is known far more precisely than a single random mask varies, so its interval excludes zero for 21 of the 24 diagonal masks, including masks that several of their own random masks exceed.

\paragraph{Families.}
The Benjamini--Hochberg correction runs over the masks of one experiment family, pooling models. A family is defined by task, intervention and design (diagonal, transfer, span, cross-task, cross-corpus), and Table~\ref{tab:overview} lists them. A steering $\alpha$ sweep is one family. Analyses with a single statistic apply the same correction to its $p$-values.

\paragraph{Notation in tables.}
Tables that compare real masks with matched controls use a common notation so that captions need only define table-specific quantities. Random-control columns and parenthesized values give the control mean (and the standard deviation where shown); $K$, the null, and the correction family are listed in Table~\ref{tab:overview}. Unless that table specifies a single-statistic analysis, $p$ is the uncorrected value for one co-primary metric and $q$ is the Benjamini--Hochberg value of their conjunction. Bold or $^{\ast}$ marks $q<0.05$ with the effect in the required direction. $^{\S}$ is descriptive: the real value lies beyond all matched controls on the statistic shown; it is not an additional inferential significance marker.

\begin{table*}[h]
\centering
\small
\resizebox{\ifdim\width>\linewidth\linewidth\else\width\fi}{!}{%
\begin{tabular}{@{}p{0.3\textwidth} p{0.27\textwidth} r p{0.27\textwidth}@{}}
\toprule
Analysis & Null & $K$ & Family \\ \midrule
Ablation, own condition (\S\ref{subsec:ablation}) & matched random masks & 20 & 24 masks \\
Ablation, own condition (\S\ref{subsec:ablation}) & sets matched on activation statistics & 20 & 8 masks \\
Ablation, transfer to \texttt{T} (\S\ref{subsec:attribution}) & matched random masks under \texttt{T} & 20 & 24 masks \\
Class-mean shift (\S\ref{subsec:ablation}) & norm-matched random masks & 20 & 21 mask--$\alpha$ cells \\
CICN masks on gender (\S\ref{subsec:ablation}) & matched random masks & 20 & cross-task, $|\Delta\ell|$ only \\
Span masks (App.~\ref{appendix:overlap}) & matched random masks & 20 & 36 masks \\
Swapped transcripts (App.~\ref{appendix:overlap}) & random masks, same swap & 20 & 3 masks \\
Position-restricted ablation (\S\ref{subsec:positions}) & random masks, same positions & 20 & 4 cells per position set \\
Transcript positions under \texttt{T} (\S\ref{subsec:positions}) & random masks, same positions & 20 & 4 masks \\
Token contrasts (App.~\ref{appendix:tokens}) & matched random masks & 20 & per model, condition and mask \\
Pitt masks on PROCESS-2 (\S\ref{subsec:crosscorpus}) & random masks drawn on Pitt & 20 & 16 masks; 8 \texttt{AT-chat} masks \\
Held-out diagnoses (\S\ref{subsec:heldout}) & matched random masks & 20 & 8 (Pitt); 6 and 2 (PROCESS-2) \\
Offset or discrimination (\S\ref{subsec:bias}) & matched random masks & 20 & 8 masks, 3 statistics \\
Mask and pool size (App.~\ref{appendix:sensitivity}) & matched random masks & 20 & 8 masks per variant \\
Prompt variants (App.~\ref{appendix:prompt_robust}) & matched random masks, same prompt & 20 & 4 masks per variant \\
Positions under the prompt variants (App.~\ref{appendix:position_prompts}) & random masks, same positions and prompt & 20 & 3 masks per position set and variant \\
Readout (\S\ref{subsec:exploit}) & readouts on random masks & 20 & per design \\
Readout with a unit budget (\S\ref{subsec:exploit}) & layer-matched random draws & 20 & 7 cells $\times$ 6 budgets \\ \bottomrule
\end{tabular}%
}
\caption{Overview of the analyses: the null each is compared with, the number $K$ of random masks per real mask, and the family over which the Benjamini--Hochberg correction runs. Unless stated, each mask enters with the conjunction of its two co-primary metrics. Details and caveats are in the sections given.}
\label{tab:overview}
\end{table*}

\paragraph{Condition contrasts.}
For the unmodified model, the UAR difference between two input conditions is tested on the evaluation recordings both conditions share (all of them). Recordings are resampled jointly for both conditions (1000 resamples); the 95\% interval is the percentile interval and the two-sided $p$ is twice the smaller tail mass at 0, floored at $1/1000$. The Benjamini--Hochberg correction runs over all contrasts of one corpus (Pitt: 8 models $\times$ 3 contrasts; PROCESS-2: 8 $\times$ 2). Here the recordings are the unit of analysis, because the question is how the model behaves on the population of speakers, not whether one set of neurons differs from another.

\paragraph{Unit of inference.}
A paired test over items (the per-item effect of the real mask minus the mean per-item effect of its random masks) answers a different question and cannot establish mask specificity: applied leave-one-out to random masks only, it declared 75--95\% of them significant, because inference is deterministic and any two masks differ at the item level. A specificity test asks whether \emph{this} set of neurons differs from other sets of the same size and layout, so the unit of analysis is the mask.

\section{Baseline Performance and Identification Pools}
\label{appendix:additional}

\subsection{Baseline Performance}
\label{appendix:pilot}

\begin{table}[h]
\centering
\small
\resizebox{\columnwidth}{!}{%
\begin{tabular}{@{}ll rcr rr@{}}
\toprule
Model & Cond. & UAR & 95\% CI & F1 & $P(\text{CI})$ & pred.\ CI \\ \midrule
Qwen2.5-Omni-7B & \texttt{T}       & 0.664 & [0.580, 0.751] & 0.634 & 0.464 & 0.430 \\
             & \texttt{A}       & 0.587 & [0.527, 0.644] & 0.471 & 0.342 & 0.159 \\
             & \texttt{AT}      & 0.736 & [0.646, 0.817] & 0.715 & 0.492 & 0.495 \\
             & \texttt{AT-chat} & 0.582 & [0.542, 0.629] & 0.435 & 0.377 & 0.103 \\ \midrule
Qwen3-Omni-30B   & \texttt{T}       & 0.736 & [0.668, 0.806] & 0.682 & 0.343 & 0.346 \\
             & \texttt{A}       & 0.726 & [0.636, 0.808] & 0.699 & 0.432 & 0.458 \\
             & \texttt{AT}      & 0.746 & [0.658, 0.824] & 0.732 & 0.522 & 0.533 \\
             & \texttt{AT-chat} & 0.749 & [0.673, 0.819] & 0.710 & 0.383 & 0.411 \\ \midrule
Kimi-Audio   & \texttt{T}       & 0.628 & [0.544, 0.717] & 0.629 & 0.628 & 0.785 \\
             & \texttt{A}       & 0.673 & [0.581, 0.768] & 0.672 & 0.559 & 0.617 \\
             & \texttt{AT}      & 0.636 & [0.543, 0.729] & 0.637 & 0.620 & 0.645 \\
             & \texttt{AT-chat} & 0.638 & [0.547, 0.737] & 0.627 & 0.534 & 0.523 \\ \midrule
MiniCPM-o-4.5 & \texttt{T}      & 0.536 & [0.442, 0.624] & 0.533 & 0.673 & 0.720 \\
             & \texttt{A}       & 0.658 & [0.566, 0.755] & 0.663 & 0.666 & 0.748 \\
             & \texttt{AT}      & 0.630 & [0.546, 0.720] & 0.634 & 0.652 & 0.738 \\
             & \texttt{AT-chat} & 0.660 & [0.570, 0.751] & 0.665 & 0.649 & 0.701 \\ \midrule
Gemma-4-12B      & \texttt{T}       & 0.570 & [0.504, 0.645] & 0.543 & 0.865 & 0.888 \\
             & \texttt{A}       & 0.599 & [0.500, 0.688] & 0.589 & 0.554 & 0.523 \\
             & \texttt{AT}      & 0.668 & [0.581, 0.765] & 0.673 & 0.650 & 0.710 \\
             & \texttt{AT-chat} & 0.684 & [0.592, 0.769] & 0.653 & 0.472 & 0.430 \\ \midrule
Audio-Flamingo-Next      & \texttt{T}       & 0.490 & [0.446, 0.529] & 0.294 & 0.392 & 0.037 \\
             & \texttt{A}       & 0.517 & [0.456, 0.580] & 0.381 & 0.424 & 0.121 \\
             & \texttt{AT}      & 0.579 & [0.501, 0.657] & 0.493 & 0.444 & 0.224 \\
             & \texttt{AT-chat} & 0.589 & [0.522, 0.655] & 0.487 & 0.446 & 0.187 \\ \midrule
Voxtral-Small-24B & \texttt{T}      & 0.548 & [0.497, 0.614] & 0.496 & 0.870 & 0.935 \\
             & \texttt{A}       & 0.518 & [0.485, 0.562] & 0.432 & 0.739 & 0.972 \\
             & \texttt{AT}      & 0.513 & [0.500, 0.542] & 0.412 & 0.840 & 0.991 \\
             & \texttt{AT-chat} & 0.513 & [0.500, 0.542] & 0.412 & 0.752 & 0.991 \\ \midrule
MOSS-Audio-8B   & \texttt{T}       & 0.500 & [0.500, 0.500] & 0.385 & 0.934 & 1.000 \\
             & \texttt{A}       & 0.500 & [0.500, 0.500] & 0.385 & 0.959 & 1.000 \\
             & \texttt{AT}      & 0.500 & [0.500, 0.500] & 0.385 & 0.966 & 1.000 \\
             & \texttt{AT-chat} & 0.500 & [0.500, 0.500] & 0.385 & 0.952 & 1.000 \\ \bottomrule
\end{tabular}%
}
\caption{Unmodified Pitt results. ``pred.\ CI'' is the fraction of items predicted CI.}
\label{tab:pilot_full}
\end{table}

\begin{table}[h]
\centering
\small
\resizebox{\columnwidth}{!}{%
\begin{tabular}{@{}ll rcr rr@{}}
\toprule
Model & Cond. & UAR & 95\% CI & F1 & $P(\text{CI})$ & pred.\ CI \\ \midrule
Qwen2.5-Omni-7B  & \texttt{T}  & 0.549 & [0.468, 0.652] & 0.553 & 0.330 & 0.121 \\
              & \texttt{A}  & 0.568 & [0.480, 0.665] & 0.576 & 0.310 & 0.091 \\
              & \texttt{AT} & 0.606 & [0.506, 0.719] & 0.618 & 0.369 & 0.131 \\ \midrule
Qwen3-Omni-30B    & \texttt{T}  & 0.487 & [0.467, 0.500] & 0.438 & 0.077 & 0.020 \\
              & \texttt{A}  & 0.525 & [0.500, 0.583] & 0.494 & 0.086 & 0.010 \\
              & \texttt{AT} & 0.562 & [0.494, 0.661] & 0.565 & 0.166 & 0.051 \\ \midrule
Kimi-Audio    & \texttt{T}  & 0.531 & [0.451, 0.621] & 0.529 & 0.290 & 0.101 \\
              & \texttt{A}  & 0.650 & [0.546, 0.766] & 0.675 & 0.254 & 0.111 \\
              & \texttt{AT} & 0.661 & [0.544, 0.786] & 0.649 & 0.310 & 0.242 \\ \midrule
MiniCPM-o-4.5 & \texttt{T}  & 0.606 & [0.496, 0.728] & 0.618 & 0.184 & 0.131 \\
              & \texttt{A}  & 0.612 & [0.504, 0.731] & 0.618 & 0.247 & 0.172 \\
              & \texttt{AT} & 0.575 & [0.495, 0.675] & 0.584 & 0.132 & 0.081 \\ \midrule
Gemma-4-12B       & \texttt{T}  & 0.709 & [0.634, 0.780] & 0.551 & 0.604 & 0.616 \\
              & \texttt{A}  & 0.530 & [0.434, 0.628] & 0.531 & 0.404 & 0.152 \\
              & \texttt{AT} & 0.510 & [0.395, 0.639] & 0.488 & 0.384 & 0.384 \\ \midrule
Audio-Flamingo-Next       & \texttt{T}  & 0.500 & [0.500, 0.500] & 0.444 & 0.300 & 0.000 \\
              & \texttt{A}  & 0.525 & [0.500, 0.579] & 0.494 & 0.351 & 0.010 \\
              & \texttt{AT} & 0.525 & [0.500, 0.579] & 0.494 & 0.357 & 0.010 \\ \midrule
Voxtral-Small-24B & \texttt{T}  & 0.741 & [0.643, 0.830] & 0.636 & 0.516 & 0.465 \\
              & \texttt{A}  & 0.653 & [0.541, 0.761] & 0.558 & 0.510 & 0.505 \\
              & \texttt{AT} & 0.576 & [0.505, 0.641] & 0.353 & 0.635 & 0.828 \\ \midrule
MOSS-Audio-8B    & \texttt{T}  & 0.570 & [0.535, 0.612] & 0.307 & 0.725 & 0.889 \\
              & \texttt{A}  & 0.544 & [0.514, 0.576] & 0.260 & 0.793 & 0.929 \\
              & \texttt{AT} & 0.519 & [0.500, 0.544] & 0.209 & 0.842 & 0.970 \\ \bottomrule
\end{tabular}%
}
\caption{Unmodified PROCESS-2 results, pooled over transcription styles.}
\label{tab:pilot_full_p2}
\end{table}

\begin{table}[h]
\centering
\small
\resizebox{\ifdim\width>\linewidth\linewidth\else\width\fi}{!}{%
\begin{tabular}{@{}l cc cc cc@{}}
\toprule
& \multicolumn{2}{c}{\texttt{T}} & \multicolumn{2}{c}{\texttt{A}} & \multicolumn{2}{c}{\texttt{AT}} \\ \cmidrule(lr){2-3}\cmidrule(lr){4-5}\cmidrule(lr){6-7}
Model & lab. & plain & lab. & plain & lab. & plain \\ \midrule
Qwen2.5-Omni-7B   & 0.59 & 0.51 & 0.56 & 0.58 & 0.67 & 0.44 \\
Qwen3-Omni-30B     & 0.47 & 0.50 & 0.50 & 0.60 & 0.54 & 0.60 \\
Kimi-Audio     & 0.53 & 0.55 & 0.66 & 0.59 & 0.70 & 0.53 \\
MiniCPM-o-4.5  & 0.62 & 0.56 & 0.67 & 0.44 & 0.61 & 0.48 \\
Gemma-4-12B        & 0.71 & 0.73 & 0.57 & 0.43 & 0.47 & 0.61 \\
Audio-Flamingo-Next        & 0.50 & 0.50 & 0.53 & 0.50 & 0.53 & 0.50 \\
Voxtral-Small-24B  & 0.68 & 0.85 & 0.64 & 0.69 & 0.56 & 0.61 \\
MOSS-Audio-8B     & 0.54 & 0.60 & 0.55 & 0.54 & 0.51 & 0.53 \\ \bottomrule
\end{tabular}%
}
\caption{Unmodified PROCESS-2 UAR within transcription style (lab.: labeled; plain: plain prose; Table~\ref{tab:process2_style}). Labeled style: 54 evaluation recordings (39 NC / 15 CI); plain style: 45 (40 NC / 5 CI). With 5 CI recordings, each plain-style value moves by 0.10 per CI recording.}
\label{tab:pilot_strata}
\end{table}

\begin{table}[h]
\centering
\small
\resizebox{\columnwidth}{!}{%
\begin{tabular}{@{}l rrr rr@{}}
\toprule
& \multicolumn{3}{c}{Pitt} & \multicolumn{2}{c}{PROCESS-2} \\ \cmidrule(lr){2-4}\cmidrule(lr){5-6}
Model & \texttt{AT}$-$\texttt{T} & \texttt{AT}$-$\texttt{A} & \texttt{chat}$-$\texttt{AT} & \texttt{AT}$-$\texttt{T} & \texttt{AT}$-$\texttt{A} \\ \midrule
Qwen2.5-Omni-7B  & +0.07 & \textbf{+0.15}$^{\ast}$ & \textbf{$-$0.15}$^{\ast}$ & +0.06 & +0.04 \\
Qwen3-Omni-30B    & +0.01 & +0.02 & +0.00 & +0.08 & +0.04 \\
Kimi-Audio    & +0.01 & $-$0.04 & +0.00 & +0.13 & +0.01 \\
MiniCPM-o-4.5 & +0.09 & $-$0.03 & +0.03 & $-$0.03 & $-$0.04 \\
Gemma-4-12B       & +0.10 & +0.07 & +0.02 & \textbf{$-$0.20}$^{\ast}$ & $-$0.02 \\
Audio-Flamingo-Next       & +0.09 & +0.06 & +0.01 & +0.03 & 0.00 \\
Voxtral-Small-24B & $-$0.04 & $-$0.01 & 0.00 & \textbf{$-$0.17}$^{\ast}$ & $-$0.08 \\
MOSS-Audio-8B    & 0.00 & 0.00 & 0.00 & \textbf{$-$0.05}$^{\ast}$ & $-$0.03 \\ \bottomrule
\end{tabular}%
}
\caption{Paired UAR differences between input conditions of the unmodified model (\texttt{chat} = \texttt{AT-chat}). $^{\ast}$ marks a significant paired-bootstrap contrast (Appendix~\ref{appendix:stats}).}
\label{tab:modality}
\end{table}

\paragraph{Audio-induced decision shifts.}
Joint input changes the operating point rather than behaving as a simple accuracy upper bound (Table~\ref{tab:modality}). On PROCESS-2, adding audio significantly lowers UAR for Gemma-4-12B and Voxtral-Small-24B, but shifts their class priors in opposite directions; on Pitt, Qwen2.5-Omni-7B likewise changes substantially as transcript and rendered disfluency information are added. Non-participant speech and measured recording-quality variables do not account for the PROCESS-2 shifts (Appendix~\ref{appendix:audio_quality}).

\subsection{Recording Properties and the Effect of Audio}
\label{appendix:audio_quality}

For each evaluation recording we compute five properties of the audio the models receive (16\,kHz mono, 25\,ms frames with 10\,ms hop). The \emph{energy dynamic range}, a proxy for the signal-to-noise ratio, is the 95th minus the 10th percentile of frame energy, and the \emph{noise floor} is the 10th percentile. The \emph{speech fraction} is the share of frames above the midpoint between the two. The \emph{cutoff} is the highest frequency whose long-term level is within 60\,dB of the spectral peak, so a device or codec low-pass appears as a low cutoff. \emph{Duration} is the length of the audio.
For each model we correlate each property with the per-item change from \texttt{T} to \texttt{AT} in the first-step log-odds of the correct class (Spearman; Benjamini--Hochberg over the 40 model--property tests of a corpus).

The PROCESS-2 recordings are generally clean and show no bandwidth shortcut analogous to the excluded Pitt recordings; a small cutoff difference between transcription styles suggests some residual recording-setup variation.

Duration is the recording property most consistently associated with the audio-induced decision shift on PROCESS-2, with weaker associations on Pitt. The PROCESS-2 modality effect remains within both duration halves, so duration alone does not explain it.

\paragraph{Interpretation and scope.}
Neither non-participant speech nor class differences in recording length account for the observed shifts; MOSS-Audio-8B shows the same qualitative pattern more weakly in a near-constant prediction regime.

\subsection{Identification Pools}
\label{appendix:pools}

Table~\ref{tab:pools} gives, for every cell that reaches identification, the number of eligible recordings per class and the number of tokens pooled.

\begin{table}[h]
\centering
\small
\resizebox{\columnwidth}{!}{%
\begin{tabular}{@{}ll rr rr@{}}
\toprule
& & \multicolumn{2}{c}{eligible / total} & \multicolumn{2}{c}{tokens pooled ($N=15$)} \\ \cmidrule(lr){3-4}\cmidrule(lr){5-6}
Model & Cond. & NC & CI & NC & CI \\ \midrule
\multicolumn{6}{@{}l}{\textit{Pitt}} \\
Qwen2.5-Omni-7B & \texttt{AT}      & 24 / 152 & 16 / 145 & 42{,}988 & 61{,}434 \\
Qwen3-Omni-30B   & \texttt{A}       & 64 / 152 & 20 / 145 & 21{,}960 & 23{,}438 \\
             & \texttt{AT}      & 47 / 152 & 46 / 145 & 23{,}286 & 28{,}064 \\
             & \texttt{AT-chat} & 73 / 152 & 19 / 145 & 25{,}008 & 36{,}058 \\
Kimi-Audio   & \texttt{A}       & 57 / 152 & 52 / 145 & 18{,}630 & 22{,}650 \\
             & \texttt{AT}      & 48 / 152 & 77 / 145 & 20{,}248 & 28{,}438 \\
             & \texttt{AT-chat} & 63 / 152 & 50 / 145 & 21{,}006 & 30{,}400 \\
MiniCPM-o-4.5 & \texttt{A}      & 23 / 152 & 67 / 145 & 20{,}922 & 20{,}804 \\
             & \texttt{AT}      & 66 / 152 & 81 / 145 & 26{,}726 & 23{,}506 \\
             & \texttt{AT-chat} & 58 / 152 & 84 / 145 & 24{,}134 & 24{,}656 \\
Gemma-4-12B      & \texttt{AT}      & 37 / 152 & 86 / 145 & 45{,}502 & 42{,}778 \\
             & \texttt{AT-chat} & 82 / 152 & 58 / 145 & 44{,}642 & 58{,}660 \\ \midrule
\multicolumn{6}{@{}l}{\textit{PROCESS-2}} \\
Gemma-4-12B       & \texttt{T}      & 33 / 119 & 23 / 29 & 8{,}744 & 6{,}822 \\
Voxtral-Small-24B & \texttt{T}      & 45 / 119 & 19 / 29 & 8{,}490 & 5{,}376 \\ \bottomrule
\end{tabular}%
}
\caption{Identification recordings that the unmodified model classifies correctly under both option orders with margin $\ge0.1$ (``eligible''), and the prompt tokens pooled from the 15 used per class (both orders, span \texttt{all}).}
\label{tab:pools}
\end{table}

\section{Concentrated Causal Control}
\label{appendix:causal_control}

\subsection{Diagonal Ablation: Summary of the Strongest Masks}
\label{appendix:ablation_full}

Table~\ref{tab:robust} summarizes the stronger robustness checks for the masks highlighted in Figure~\ref{fig:ablation}: activity-matched controls (Appendix~\ref{appendix:controlmasks}) and the two prompt variants (Appendix~\ref{appendix:prompt_robust}). Uniform-random results are already shown in Figure~\ref{fig:ablation}.

\begin{table}[h]
\centering
\small
\resizebox{\columnwidth}{!}{%
\begin{tabular}{@{}lll r rrr rr@{}}
\toprule
& & & & \multicolumn{3}{c}{Activity-matched sets} & \multicolumn{2}{c}{Prompt variants, $q$} \\ \cmidrule(lr){5-7} \cmidrule(lr){8-9}
Model & Cond. & Mask & $\Delta\ell$ real & $\Delta\ell$ sets & $p$ & $q$ & paraphrase & numeric labels \\ \midrule
Qwen3-Omni-30B   & \texttt{AT}      & NC & +1.77 & $-$0.16{\scriptsize$\pm$0.76} & 0.030 & \textbf{0.040} & \textbf{$<$0.001} & \textbf{$<$0.001} \\
MiniCPM-o-4.5    & \texttt{A}       & CI & $-$2.51 & $-$0.27{\scriptsize$\pm$0.95} & 0.033 & \textbf{0.040} & \textbf{$<$0.001} & \textbf{0.006} \\
MiniCPM-o-4.5    & \texttt{AT}      & CI & $-$2.63 & $-$0.64{\scriptsize$\pm$0.86} & 0.035 & \textbf{0.040} & \textbf{$<$0.001} & \textbf{0.002} \\
Kimi-Audio       & \texttt{AT-chat} & NC & +1.60 & +0.05{\scriptsize$\pm$0.60} & 0.033 & \textbf{0.040} & \textbf{0.006} & \textbf{0.005} \\ \midrule
Kimi-Audio       & \texttt{AT}      & NC & +1.46 & $-$0.05{\scriptsize$\pm$0.56} & 0.020 & \textbf{0.040} & -- & -- \\
MiniCPM-o-4.5    & \texttt{AT-chat} & CI & $-$1.35 & +0.10{\scriptsize$\pm$0.49} & 0.013 & \textbf{0.040} & -- & -- \\
Qwen3-Omni-30B   & \texttt{AT}      & CI & $-$0.83 & +1.60{\scriptsize$\pm$0.83} & 0.010 & 0.040$^{\dagger}$ & -- & -- \\
Kimi-Audio       & \texttt{A}       & NC & +0.79 & +0.15{\scriptsize$\pm$0.41} & 0.15 & 0.15 & -- & -- \\ \bottomrule
\end{tabular}%
}
\caption{Robustness of the eight strongest Pitt CICN masks under activity-matched and prompt-variant controls. The real-mask $\Delta\ell$ is retained for effect-size context; its uniform-random comparison is shown in Figure~\ref{fig:ablation}. $\Delta P(\text{CI})$ also enters each significance test. $p$ is the uncorrected conjunction value; the equal $q$ values follow from the Benjamini--Hochberg step-up. The first four rows are the primary masks. --: not evaluated under that prompt variant. $^{\dagger}$ indicates an activity-matched comparison with a larger effect in the opposite direction, which is not counted as passing.}
\label{tab:robust}
\end{table}

\subsection{Activity-Matched and Alternative Selection Controls}
\label{appendix:controlmasks}

\paragraph{Activity-matched controls.}
Activity-matched sets provide a stronger null than uniform random masks because their intervention effects are more variable. Against 20 such sets per mask, six of the eight strongest Pitt masks pass on both co-primary metrics in the predicted direction, including all four primary masks ($q=0.040$; Table~\ref{tab:robust}). Kimi-Audio \texttt{A} NC does not pass. For Qwen3-Omni-30B \texttt{AT} CI, most activity-matched sets produce larger shifts than the CICN mask, in the opposite direction.

\paragraph{Alternative selection contrasts.}

ConAct selects neurons with extreme class gaps from confidently correct recordings, so we repeat selection under several control contrasts for the eight strongest Pitt masks. Independent random masks have negligible overlap, whereas masks from disjoint confident pools can overlap substantially, establishing the relevant stability scale.
\begin{itemize}
\item \emph{Shuffled labels.} The true labels produce larger aggregate ConAct gaps than all permutations, although shuffled masks still share some high-selectivity neurons with CICNs.
\item \emph{Recording properties.} Masks selected by duration, speech rate, disfluency rate, or information-unit count overlap CICNs no more than the shuffled-label controls.
\item \emph{Decision margin.} High-versus-low confidence masks show little overlap with CICNs.
\item \emph{Unfiltered pools.} Diagnosis-based masks from unfiltered recordings partially recover the CICN sets, whereas masks from wrong or low-margin recordings overlap less and scarcely with the opposite-class mask.
\end{itemize}
These controls indicate that the diagnosis contrast, rather than confidence filtering or any single recording property, accounts for most of the selected-neuron identity. Ablating recording-property masks supports the same conclusion most clearly for Kimi-Audio, where they reproduce only a minority of the CICN effect or move the decision oppositely. MiniCPM-o-4.5 and especially Qwen3-Omni-30B show stronger property-selected effects, however, so the direction of the CICN shift is not unique to the CI contrast in every model. Activity-matched sets likewise retain part of the class contrast, and in several cells they approach or exceed the real mask. The recording-property comparison remains descriptive because its random masks are matched to the CICN masks rather than separately to each property mask.

\paragraph{Choice of null.}
The matched random masks are the null for the question we ask: whether causal leverage over the decision is concentrated in a small neuron set or spread evenly over the decoder's MLP neurons. Uniform draws with the same layer counts show what an evenly spread effect looks like.
Sets that also reproduce the selection, such as neurons matched on activation statistics or ConAct on relabeled pools, address a different question, namely which ingredient of the selection confers the leverage, and they are not effect-free references: ConAct selects on activation frequency, so such sets share neurons and part of the class gap with the CICN masks, as shown above.
Our claim is accordingly specificity relative to arbitrary neurons of the same number and depth, which for six masks also holds against equally active neurons at the 5\% false discovery rate; we do not attribute the effect to the class contrast alone.

\subsection{Sensitivity to Mask Size and Pool Size}
\label{appendix:sensitivity}

Table~\ref{tab:sensitivity} repeats the diagonal ablation of the eight masks with the largest own-condition effects in the main setting (the seven that pass and Kimi-Audio \texttt{A} NC; Figure~\ref{fig:ablation}) under four variants:
\begin{itemize}
\item smaller identification pools, $N=5$ and $N=10$, which are nested in the $N=15$ pool;
\item a smaller mask, $r=0.1\%$, and a larger one, $r=1\%$.
\end{itemize}
Each variant is tested against $K=20$ matched random masks drawn for it and forms its own family, separate from the main-result correction.

\begin{table}[h]
\centering
\small
\resizebox{\columnwidth}{!}{%
\begin{tabular}{@{}lll rr rr rr rr rr@{}}
\toprule
& & & \multicolumn{2}{c}{main} & \multicolumn{2}{c}{$N=5$} & \multicolumn{2}{c}{$N=10$} & \multicolumn{2}{c}{$r=0.1\%$} & \multicolumn{2}{c}{$r=1\%$} \\
\cmidrule(lr){4-5}\cmidrule(lr){6-7}\cmidrule(lr){8-9}\cmidrule(lr){10-11}\cmidrule(lr){12-13}
Model & Cond. & Mask & $\Delta P$ & flip & $\Delta P$ & flip & $\Delta P$ & flip & $\Delta P$ & flip & $\Delta P$ & flip \\ \midrule
Kimi-Audio & \texttt{A}       & NC & +0.139 & 0.50$^{\S}$ & +0.125 & 0.63 & +0.105 & 0.42 & +0.011 & 0.17 & \textbf{+0.220} & 0.79$^{\S}$ \\
Kimi-Audio & \texttt{AT}      & NC & \textbf{+0.228} & 0.95$^{\S}$ & +0.164 & 0.57 & +0.132 & 0.67$^{\S}$ & +0.046 & 0.33$^{\S}$ & +0.262 & 1.00$^{\S}$ \\
Kimi-Audio & \texttt{AT-chat} & NC & \textbf{+0.267} & 0.88$^{\S}$ & +0.250 & 0.88$^{\S}$ & +0.207 & 0.77$^{\S}$ & \textbf{+0.152} & 0.58$^{\S}$ & +0.064 & 0.35 \\
MiniCPM-o-4.5  & \texttt{A}       & CI & \textbf{$-$0.212} & 0.38$^{\S}$ & $-$0.105 & 0.12$^{\S}$ & \textbf{$-$0.189} & 0.36$^{\S}$ & \textbf{$-$0.105} & 0.14$^{\S}$ & \textbf{$-$0.203} & 0.41$^{\S}$ \\
MiniCPM-o-4.5  & \texttt{AT}      & CI & \textbf{$-$0.253} & 0.55$^{\S}$ & \textbf{$-$0.355} & 0.68$^{\S}$ & $-$0.133 & 0.23 & $-$0.087 & 0.16 & \textbf{$-$0.268} & 0.68$^{\S}$ \\
MiniCPM-o-4.5  & \texttt{AT-chat} & CI & \textbf{$-$0.136} & 0.16$^{\S}$ & +0.029 & 0.00 & +0.110 & 0.00 & \textbf{$-$0.101} & 0.13$^{\S}$ & \textbf{$-$0.216} & 0.40$^{\S}$ \\
Qwen3-Omni-30B & \texttt{AT}      & NC & \textbf{+0.141} & 0.35$^{\S}$ & \textbf{+0.138} & 0.45$^{\S}$ & \textbf{+0.129} & 0.45$^{\S}$ & +0.032 & 0.10 & \textbf{+0.290} & 0.84$^{\S}$ \\
Qwen3-Omni-30B & \texttt{AT}      & CI & \textbf{$-$0.065} & 0.13$^{\S}$ & +0.035 & 0.02 & $-$0.053 & 0.08 & $-$0.025 & 0.04$^{\S}$ & $-$0.064 & 0.17$^{\S}$ \\ \bottomrule
\end{tabular}%
}
\caption{Sensitivity of Pitt ablation effects to identification-pool size and CICN mask size for the eight strongest masks. The main setting is $N=15$, $r=0.5\%$; each variant uses its own matched random controls. $\Delta P=\Delta P(\text{CI})$; flip denotes $\text{flip}_{\text{own}}$.}
\label{tab:sensitivity}
\end{table}

Across the 32 sensitivity cells, 29 preserve the direction of the main effect; the few reversals occur in masks that are also least stable across corpora (Table~\ref{tab:xcorpus}). Smaller masks and smaller identification pools generally weaken the effects, whereas the 1\% masks exceed the main 0.5\% setting for five of the eight masks and match it within 0.01 for two (Table~\ref{tab:sensitivity}). Kimi-Audio \texttt{AT-chat} NC is the exception: its 1\% mask contains the 0.5\% mask, yet the effect falls from $+0.267$ to $+0.064$ and no longer passes. The effects of nested masks are therefore not additive, and for this mask the neurons ranked between 0.5\% and 1\% offset most of the effect of those ranked above them.

Twelve variants survive correction, with robustness higher to mask-size changes than to smaller identification pools. The main conclusion is therefore set-level: a sufficiently large selected mask can exert reproducible causal leverage, while the exact effect, and by implication the identity of individual selected neurons, depends more strongly on the recordings used for identification.

\paragraph{Scope of the sensitivity analysis.}
Only the mask size and the pool size are varied here; the selector (ConAct) was not varied. Prompt variants are in Appendix~\ref{appendix:prompt_robust}.

\subsection{Class Contrasts Adjusted for Recording Properties}
\label{appendix:recording_contrasts}

Over all 297 identification recordings, CI recordings contain more silence and slower speech than NC recordings (standardized mean differences 0.33 and 0.47) and barely differ in duration (0.03); within the 15-per-class pools the differences can be larger (up to 0.96, 1.07 and 1.38 standard deviations for duration, silence and speech rate). The class gap that ConAct selects on could therefore be a gap in these properties, above all for masks selected from speech alone. On the identification split we regress each mask neuron's per-recording activation probability on the class, with and without three standardized covariates: duration of the participant's audio, proportion of silence (from frame energies) and speech rate (words per second). Table~\ref{tab:residual_gap} gives the class coefficient averaged over the mask, for the 30 pool recordings the selection saw and for all 297 identification recordings.

\begin{table}[h]
\centering
\small
\resizebox{\columnwidth}{!}{%
\begin{tabular}{@{}lll rrr rrr@{}}
\toprule
& & & \multicolumn{3}{c}{pool (30)} & \multicolumn{3}{c}{all identification (297)} \\ \cmidrule(lr){4-6}\cmidrule(lr){7-9}
Model & Cond. & Mask & raw & adjusted & kept & raw & adjusted & kept \\ \midrule
Kimi-Audio & \texttt{A}       & NC & 0.127 & 0.086 & 68\% & 0.051 & 0.033 & 65\% \\
Kimi-Audio & \texttt{AT}      & NC & 0.092 & 0.080 & 86\% & 0.046 & 0.042 & 91\% \\
Kimi-Audio & \texttt{AT-chat} & NC & 0.094 & 0.069 & 73\% & 0.039 & 0.032 & 82\% \\
MiniCPM-o-4.5  & \texttt{A}       & CI & 0.139 & 0.080 & 58\% & 0.056 & 0.035 & 61\% \\
MiniCPM-o-4.5  & \texttt{AT}      & CI & 0.119 & 0.078 & 66\% & 0.054 & 0.035 & 64\% \\
MiniCPM-o-4.5  & \texttt{AT-chat} & CI & 0.078 & 0.066 & 85\% & 0.047 & 0.036 & 77\% \\
Qwen3-Omni-30B & \texttt{AT}      & NC & 0.035 & 0.029 & 82\% & 0.019 & 0.014 & 75\% \\
Qwen3-Omni-30B & \texttt{AT}      & CI & 0.048 & 0.042 & 87\% & 0.022 & 0.018 & 83\% \\ \bottomrule
\end{tabular}%
}
\caption{Class gap in CICN activation probability (mask's class minus the other class, averaged over mask neurons) before and after adjustment for duration, silence proportion, and speech rate, shown for the selection pool and all identification recordings. ``kept'' is the adjusted/raw gap ratio.}
\label{tab:residual_gap}
\end{table}

After adjustment, the masks retain most of their class gap (58--91\% across settings; Table~\ref{tab:residual_gap}), and nearly all selected neurons keep the direction of their class gap: 97--100\% per mask in the pool and 93--100\% over all identification recordings. Duration, silence, and speech rate therefore explain a meaningful but minority share of the selection contrast. Because these covariates are themselves impairment-related, the retained fraction is conservative; the intervention controls above separately test whether recording properties can reproduce the causal effect.

\subsection{Prompt Variants}
\label{appendix:prompt_robust}

Every mask is identified under the prompt of Listing~\ref{lst:cicn-prompt}. To test whether the ablation effects depend on that wording or on the option letters, we ablate the four primary masks under two further prompts, without re-identifying them:
\begin{itemize}[left=0pt]
\item \emph{Paraphrase} (Listing~\ref{lst:cicn-prompt-para}): every sentence and both option texts are reworded with the same meaning;
\item \emph{Numeric labels}: the original prompt with the options labeled ``1'' and ``2'' and the last line ``Answer with only the option number.'' The decision is read from the first-step logits of the two digits.
\end{itemize}
Each mask is compared with its 20 matched random masks ablated under the same prompt, and with the unmodified model under that prompt. Each variant forms its own family (4 masks).

\begin{lstlisting}[style=promptbox, caption={Paraphrased prompt (\texttt{AT} shown).}, label={lst:cicn-prompt-para}]
Below, an elderly person talks about what is happening in a picture of a kitchen.
Audio of the person speaking: <audio>
What the person said:
"{transcript}"
Which of the two statements fits this person better? Each is as probable as the other.
A: This person has normal cognition.
B: This person has a cognitive impairment.
Reply with just the letter of your choice.
\end{lstlisting}

All four masks retain their predicted-direction effects under both prompt variants (Table~\ref{tab:robust}), with magnitudes comparable to the original prompt.

Positional robustness under these variants is evaluated separately in Appendix~\ref{appendix:position_prompts}.

\paragraph{Interpretation and scope.}
The variants change wording and answer tokens but retain the same binary task. Unmodified accuracy also varies across prompts; in particular, Kimi-Audio \texttt{AT-chat} is unreliable under the paraphrase, so that result demonstrates decision control rather than improvement of a reliable classifier.

\subsection{Speaker-Attribute Control: Full Results}
\label{appendix:attribute}

\begin{table}[h]
\centering
\small
\resizebox{\columnwidth}{!}{%
\begin{tabular}{@{}l rr rr@{}}
\toprule
& \multicolumn{2}{c}{CICN mask $\rightarrow$ gender} & \multicolumn{2}{c}{gender mask $\rightarrow$ CI} \\ \cmidrule(lr){2-3}\cmidrule(lr){4-5}
CICN cell (own $|\Delta\ell|$ on CI) & real & random & male & female \\ \midrule
Qwen3 \texttt{AT} NC (1.87)        & 0.62 & 0.49{\scriptsize$\pm$0.08} & 0.87 {\scriptsize(0.65$\pm$0.09)} & 0.97 {\scriptsize(0.70$\pm$0.15)} \\
Qwen3 \texttt{AT} CI (1.10)        & \textbf{1.61}$^{\ast}$ & 0.48{\scriptsize$\pm$0.05} & \multicolumn{2}{c}{\scriptsize(same cell as above)} \\
Kimi \texttt{AT} NC (1.46)         & 0.21 & 0.22{\scriptsize$\pm$0.08} & \textbf{0.89}$^{\ast}$ {\scriptsize(0.40$\pm$0.12)} & 0.35 {\scriptsize(0.41$\pm$0.17)} \\
Kimi \texttt{AT-chat} NC (1.60)    & 0.23 & 0.21{\scriptsize$\pm$0.05} & 0.89 {\scriptsize(0.42$\pm$0.21)} & 0.30 {\scriptsize(0.57$\pm$0.28)} \\
MiniCPM \texttt{A} CI (2.51)       & \textbf{1.35}$^{\ast}$ & 0.57{\scriptsize$\pm$0.27} & 0.42 {\scriptsize(0.52$\pm$0.25)} & 0.47 {\scriptsize(0.44$\pm$0.20)} \\
MiniCPM \texttt{AT} CI (2.63)      & 1.09 & 0.59{\scriptsize$\pm$0.31} & \textbf{1.95}$^{\ast}$ {\scriptsize(0.47$\pm$0.30)} & 0.84 {\scriptsize(0.57$\pm$0.27)} \\
MiniCPM \texttt{AT-chat} CI (1.36) & 0.72 & 0.56{\scriptsize$\pm$0.25} & 0.82 {\scriptsize(0.55$\pm$0.30)} & \textbf{1.74}$^{\ast}$ {\scriptsize(0.65$\pm$0.37)} \\ \bottomrule
\end{tabular}%
}
\caption{Cross-task ablation control measured by $|\Delta\ell|$. Left: CICN masks evaluated on gender; right: gender masks identified in the same model and condition and evaluated on CI. The own-task value in parentheses is the mean absolute per-item change, which is at least as large as the magnitude of the signed mean $\Delta\ell$ on the same recordings (Table~\ref{tab:robust}). $^{\ast}$ marks cells that pass the predefined cross-task test; others are n.s.}
\label{tab:attribute}
\end{table}

\paragraph{Baselines and positive control.}
For gender, whose decisions are near-certain, the primary metric is $|\Delta\ell|$.
Unmodified gender detection is near ceiling, so an unchanged outcome is not trivial. Gender masks identified with the same pipeline move gender log-odds far beyond matched random masks, confirming that the procedure can recover attribute-relevant neurons. Their direction is not consistently class-specific, however, indicating a generic decision drift analogous to that observed for CI.

\paragraph{CICNs leave gender mostly unchanged.}
Six of seven CICN masks leave gender decisions unchanged, and five also keep gender log-odds within their matched-random range (Table~\ref{tab:attribute}). MiniCPM-o-4.5's \texttt{A} CI mask shows a log-odds cross-effect without changing any gender decision. Qwen3-Omni-30B's \texttt{AT} CI mask is the main substantive exception, showing a measurable gender effect and above-chance overlap with the female mask.

\paragraph{Gender neurons and CI.}
The reverse direction is similarly mixed. Qwen3-Omni-30B \texttt{AT} and MiniCPM-o-4.5 \texttt{A} show the cleanest dissociation, whereas Kimi-Audio \texttt{AT} and MiniCPM-o-4.5 \texttt{AT}/\texttt{AT-chat} show cross-task effects. Because the corresponding gender and CI masks overlap little, these cross-effects are more consistent with interaction at a downstream decision stage than with shared neuron identity.

\paragraph{Attribute control under steering.}
Under the class-mean shift ($\alpha\in\{1,4,-4\}$, norm-matched random masks; Kimi-Audio and MiniCPM-o-4.5), CICN masks leave gender log-odds within the random range in all 15 mask--$\alpha$ cells, gender masks move gender in all 30, and gender masks leave CI log-odds within the random range in 27 of 30.

\paragraph{Interpretation and scope.}
Gender is nearly independent of diagnosis in our partition (CI-vs-NC odds ratio $\approx1.2$ in both the identification and the evaluation split). Gender decisions are near-certain (mean margin 0.79--0.96 vs.\ 0.17--0.35 for CI), so $\Delta P$ and flips saturate there; using $|\Delta\ell|$ instead makes effect sizes on the two tasks only roughly comparable. The control covers gender only, and specificity is relative to it.

\subsection{Steering: Full Results}
\label{appendix:steering}

The class-mean shift moves each masked coordinate along its class-mean difference, in the spirit of difference-of-means steering \cite{rimsky-etal-2024-steering, mean-centring, turner2024steeringlanguagemodelsactivation}:
\begin{equation}
\tilde g_{l,t,n}=g_{l,t,n}+\alpha\,\big(\mu^{(c)}_{l,n}-\mu^{(\bar c)}_{l,n}\big)\quad\text{for } n\in\mathcal{I}^{(c)}_l ,
\end{equation}
where $\bar c$ is the other class and $\mu$ is the mean gate activation over all positions of the identification pool. $\alpha>0$ pushes toward the mask's class and $\alpha<0$ away from it.
We use an additive shift because it directly follows the signed class-mean difference even when the two classes have similar activation magnitudes. The shift is applied at $\alpha\in\{0.5,1,2,4,8,-2,-4\}$.
A random mask's own class-mean gaps are much smaller than the real mask's, so we rescale them per layer to the $L_2$ norm of the real mask's gaps (\emph{norm-matched}): the random mask then differs from the real one only in which neurons are pushed, not in how hard.
For a class-$c$ mask, $\text{flip}_{\text{own}}$ counts correct $\bar c$ items flipping to $c$ when $\alpha>0$ and, as under ablation, correct class-$c$ items flipping to $\bar c$ when $\alpha<0$.

\begin{table}[h]
\centering
\small
\resizebox{\columnwidth}{!}{%
\begin{tabular}{@{}lr rrr rrr r rr@{}}
\toprule
& & \multicolumn{3}{c}{$\Delta P(\text{CI})$} & \multicolumn{3}{c}{$\Delta\ell$} & & \multicolumn{2}{c}{$\text{flip}_{\text{own}}$} \\ \cmidrule(lr){3-5}\cmidrule(lr){6-8}\cmidrule(lr){10-11}
Model / cond. / mask & $\alpha$ & real & norm-m. & $p$ & real & norm-m. & $p$ & $q$ & real & norm-m. \\ \midrule
Kimi \texttt{AT} NC      & $-4$ & +0.197 & $-$0.010 & 0.007 & +1.27 & $-$0.06 & 0.006 & 0.072 & 0.90$^{\S}$ & 0.13 \\
Kimi \texttt{AT} NC      & $+4$ & $-$0.148 & +0.020 & 0.024 & $-$0.91 & +0.12 & 0.027 & 0.094 & 0.20$^{\S}$ & 0.03 \\
Kimi \texttt{AT-chat} NC & $-4$ & +0.225 & +0.002 & 0.010 & +1.25 & +0.01 & 0.007 & 0.072 & 0.77$^{\S}$ & 0.14 \\
Kimi \texttt{AT-chat} NC & $+1$ & $-$0.059 & +0.002 & 0.013 & $-$0.31 & 0.00 & 0.013 & 0.072 & 0.14$^{\S}$ & 0.02 \\
Kimi \texttt{AT-chat} NC & $+4$ & $-$0.220 & +0.001 & 0.017 & $-$1.23 & $-$0.01 & 0.015 & 0.072 & 0.57$^{\S}$ & 0.09 \\
Kimi \texttt{A} NC       & $+4$ & $-$0.135 & +0.012 & 0.085 & $-$0.81 & +0.07 & 0.078 & 0.26 & 0.36 & 0.10 \\
MiniCPM \texttt{A} CI    & $-4$ & $-$0.072 & 0.000 & 0.17 & $-$0.97 & $-$0.10 & 0.18 & 0.29 & 0.12$^{\S}$ & 0.02 \\
MiniCPM \texttt{AT} CI   & $+4$ & +0.109 & $-$0.018 & 0.15 & +1.06 & $-$0.16 & 0.18 & 0.29 & 0.12 & 0.07 \\
MiniCPM \texttt{AT} CI   & $-4$ & $-$0.080 & +0.021 & 0.26 & $-$0.89 & +0.21 & 0.25 & 0.39 & 0.23 & 0.05 \\
Kimi \texttt{A} CI       & $+4$ & +0.065 & +0.029 & 0.70 & +0.37 & +0.19 & 0.77 & 0.90 & 0.33 & 0.24 \\ \bottomrule
\end{tabular}%
}
\caption{Steering against norm-matched random directions; ``norm-m.'' gives the matched-control mean.}
\label{tab:steer_random}
\end{table}

Shifting the same neurons along their class-mean difference steers Kimi-Audio and MiniCPM-o-4.5 decisions monotonically and reversibly with $\alpha$, producing substantial flips at stronger steering strengths; Qwen2.5-Omni-7B shows little response.

Norm-matched random pushes are a demanding baseline because arbitrary directions of the same norm are not neutral (Table~\ref{tab:steer_random}). Kimi-Audio shows the clearest directional control and several cells exceed every matched random push descriptively, but no mask--$\alpha$ cell survives the family correction.

\section{Decision Bias versus Diagnostic Discrimination}
\label{appendix:bias}

\paragraph{Offset and discrimination measures.}
If ablation only added a constant to the log-odds, it would change where the decision threshold sits and not how well the model separates the classes. Table~\ref{tab:bias} tests this on the evaluation recordings. With $\ell_0$ and $\ell_1$ the first-step CI-vs-NC log-odds of an item, averaged over the two option orders, without and with the mask ablated and $d=\ell_1-\ell_0$, we report the change in AUC, the Spearman correlation of $\ell_0$ and $\ell_1$, and the offset share $\bar d^{\,2}/\overline{d^2}$, the part of the mean squared per-item change that one additive constant accounts for; each is compared with the mask's 20 random masks.

\paragraph{Residual effects and offset correction.}
The residual beyond the intercept shift remains mask-structured: besides reordering items (Section~\ref{subsec:bias}), real masks compress the logit range relative to their matched random controls. The intercept nevertheless dominates: after subtracting the mean shift $\bar d$ from $\ell_1$, 90--97\% of primary-mask thresholded decisions match the unmodified model. Subtracting a constant cannot change the ranking, so AUC is unaffected by this correction; $\Delta$AUC in Table~\ref{tab:bias} is computed on $\ell_0$ and $\ell_1$ directly. The two AUC decreases exceed those of the matched random masks ($q=0.009$ and $0.003$), while a paired item bootstrap of $\Delta$AUC against zero gives intervals $[-0.106,-0.013]$ and $[-0.078,-0.005]$ ($q=0.086$ and $0.12$ over the eight masks).

\paragraph{Recording-level modulation.}
The per-recording intervention shift is not systematically modulated by duration, speech rate, filler rate, pause rate, or information-unit count.

\section{Control across Modalities, Corpora, and Diagnoses}
\label{appendix:transfer}

\subsection{Cross-Modal Transfer, Mask Overlap, and Span Controls}
\label{appendix:overlap}

We examine whether shared intervention effects across input conditions are accompanied by overlapping neuron sets and whether transfer persists when selection is restricted to audio positions or the modalities provide conflicting evidence.

\paragraph{Mask overlap and layer profiles.}
Table~\ref{tab:overlap} gives the overlap of masks across conditions and spans.

\begin{table}[h]
\centering
\small
\resizebox{\columnwidth}{!}{%
\begin{tabular}{@{}ll rrr@{}}
\toprule
Model & Mask & \texttt{A}--\texttt{AT} & \texttt{A}--\texttt{AT-chat} & \texttt{AT}--\texttt{AT-chat} \\ \midrule
Qwen3-Omni-30B & NC & 0.28 & 0.32 & 0.42 \\
           & CI & 0.19 & 0.57 & 0.29 \\
Kimi-Audio & NC & 0.19 & 0.20 & 0.30 \\
           & CI & 0.20 & 0.28 & 0.30 \\
MiniCPM-o-4.5 & NC & 0.33 & 0.24 & 0.21 \\
           & CI & 0.40 & 0.24 & 0.22 \\
Gemma-4-12B    & NC & -- & -- & 0.23 \\
           & CI & -- & -- & 0.29 \\ \midrule
\multicolumn{5}{@{}l}{\textit{Within \texttt{AT}} (Kimi-Audio / Qwen3-Omni-30B)} \\
& & audio--all & audio--text & audio\_tm--audio \\
& NC & 0.56 / 0.62 & 0.09 / 0.16 & 0.65 / 0.62 \\
& CI & 0.62 / 0.72 & 0.10 / 0.10 & 0.65 / 0.72 \\ \bottomrule
\end{tabular}%
}
\caption{Jaccard overlap between ConAct masks (span \texttt{all} unless stated). The Jaccard expected for independent masks with the same per-layer counts is 0.002--0.008 in every cell.}
\label{tab:overlap}
\end{table}

\begin{figure}[H]
\centering
\includegraphics[width=\columnwidth]{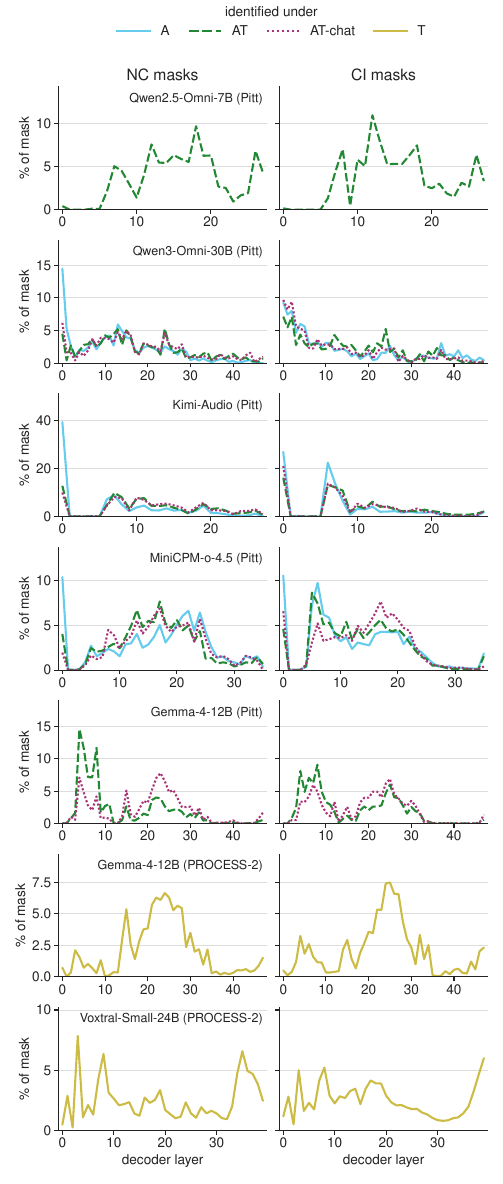}
\caption{Layer distribution of the ConAct masks (span \texttt{all}): share of a mask's neurons in each decoder layer, one line per identification condition. Rows are models and corpora; the $y$ axis differs between rows.}
\label{fig:layers}
\end{figure}

Figure~\ref{fig:layers} shows that mask depth profiles are broadly consistent across conditions within a model. Kimi-Audio, MiniCPM-o-4.5, and Qwen3-Omni-30B are comparatively front-loaded, whereas Qwen2.5-Omni-7B and Gemma-4-12B place little mass in their earliest layers. Gemma-4-12B also shows a corpus-dependent shift in its bimodal layer profile.

\begin{table}[h]
\centering
\small
\resizebox{\columnwidth}{!}{%
\begin{tabular}{@{}ll rrr@{}}
\toprule
& & \multicolumn{3}{c}{evaluated under} \\ \cmidrule(l){3-5}
Model (NC mask) & span & \texttt{T} & \texttt{A} & \texttt{AT} \\ \midrule
Kimi-Audio & audio          & +0.27$^{\ast}$ {\scriptsize(4.5)} & +0.29$^{\ast}$ {\scriptsize(3.3)} & +0.27$^{\ast}$ {\scriptsize(2.9)} \\
           & audio (tok.-matched) & +0.28$^{\ast}$ {\scriptsize(4.7)} & +0.34$^{\ast}$ {\scriptsize(5.4)} & +0.28$^{\ast}$ {\scriptsize(3.8)} \\
           & text           & $-$0.04 {\scriptsize($-$1.4)} & $-$0.07 {\scriptsize($-$1.3)} & $-$0.02 {\scriptsize($-$0.3)} \\ \midrule
Qwen3-Omni-30B & audio          & +0.31$^{\ast}$ {\scriptsize(9.1)} & +0.16$^{\ast}$ {\scriptsize(5.5)} & +0.17$^{\ast}$ {\scriptsize(6.1)} \\
           & audio (tok.-matched) & +0.19$^{\ast}$ {\scriptsize(4.1)} & +0.08 {\scriptsize(1.7)} & +0.09$^{\ast}$ {\scriptsize(2.6)} \\
           & text           & +0.14$^{\ast}$ {\scriptsize(3.8)} & +0.07 {\scriptsize(2.3)} & +0.09$^{\ast}$ {\scriptsize(2.9)} \\ \bottomrule
\end{tabular}%
}
\caption{Span dissociation within \texttt{AT}: $\Delta P(\text{CI})$ when ablating NC masks identified from the audio-token span, a token-matched subset of that span, or the transcript-token span; $z$ relative to matched random masks is shown in parentheses.}
\label{tab:span}
\end{table}

\paragraph{Span dissociation.}
The expected audio-versus-text asymmetry does not emerge (Table~\ref{tab:span}). The audio-span NC masks of Kimi-Audio and Qwen3-Omni-30B pass against 20 matched random masks under \texttt{A}, \texttt{AT}, and \texttt{T}, shifting transcript-only decisions about as much as speech-only decisions in Kimi-Audio and more in Qwen3-Omni-30B; Kimi-Audio's text-span mask is weak throughout, whereas Qwen3-Omni-30B retains smaller text-span effects under some conditions.

\paragraph{Interpretation and scope.}
The audio/text difference is not explained simply by token count: token-matched audio masks overlap strongly with full audio-span masks and largely preserve their causal behavior. Across the full span family, significant effects concentrate in audio and token-matched audio masks, with model-specific contributions from text-span masks. One Kimi-Audio control acts significantly in the opposite direction, reinforcing that span selection does not guarantee the predicted class direction. Among the masks not shown in Table~\ref{tab:transfer}, MiniCPM-o-4.5's \texttt{A} NC mask moves transcript-only decisions toward NC ($\Delta P(\text{CI})=-0.148$, $q=0.014$), the same direction as that model's CI masks. Two NC masks pass under \texttt{T} without passing in their own condition. For Qwen3-Omni-30B \texttt{A} NC the effect itself is larger under \texttt{T} ($\Delta P(\text{CI})$ $+0.100\rightarrow+0.176$, random-mask sd $0.030\rightarrow0.031$, $z$ $2.4\rightarrow4.9$); for Kimi-Audio \texttt{A} NC the effect is slightly larger and the random masks vary less ($+0.139\rightarrow+0.164$, sd $0.082\rightarrow0.055$, $z$ $2.1\rightarrow3.1$).

\paragraph{Swapped transcripts.}
When each transcript is replaced by one from the opposite class, all three joint-condition primary masks still pass against 20 matched random masks and retain effects comparable to the matched-transcript setting. These controls are ablated under the same transcript replacement. The result establishes robustness of the intervention to conflicting inputs, without by itself identifying how the model combines their evidence.

\subsection{Steering Transcript-Only Decisions}
\label{appendix:steer_transfer}

Kimi-Audio's audio-discovered NC masks also steer transcript-only decisions: eight of nine tested mask--$\alpha$ cells pass against norm-matched random pushes, with one \texttt{A}-mask exception. Specificity is clearer under \texttt{T} than in the masks' own conditions because the matched-random steering effects are less variable.

\subsection{Cross-Corpus Transfer to PROCESS-2}
\label{appendix:process2}

\begin{table}[h]
\centering
\small
\resizebox{\columnwidth}{!}{%
\begin{tabular}{@{}lll rrr rrr r r@{}}
\toprule
& & & \multicolumn{3}{c}{$\Delta P(\text{CI})$} & \multicolumn{3}{c}{$\Delta\ell$} & & \\ \cmidrule(lr){4-6} \cmidrule(lr){7-9}
Model & Cond. & Mask & real & random & $p$ & real & random & $p$ & $q$ & $\text{flip}_{\text{own}}$ ($n$) \\ \midrule
Kimi-Audio   & \texttt{AT} & NC & \textbf{+0.298} & +0.002 & $<$0.001 & \textbf{+1.72} & $-$0.01 & 0.001 & \textbf{0.006} & 0.65 (65) \\
MiniCPM-o-4.5    & \texttt{A}  & CI & \textbf{$-$0.143} & +0.018 & 0.001 & \textbf{$-$2.26} & +0.14 & $<$0.001 & \textbf{0.006} & -- (7) \\
Qwen3-Omni-30B   & \texttt{A}  & NC & \textbf{+0.112} & +0.019 & $<$0.001 & \textbf{+1.85} & +0.27 & $<$0.001 & \textbf{0.006} & 0.06 (79) \\
Qwen3-Omni-30B   & \texttt{AT} & NC & \textbf{+0.110} & +0.015 & 0.001 & \textbf{+1.91} & +0.21 & $<$0.001 & \textbf{0.006} & 0.14 (77) \\
Kimi-Audio   & \texttt{A}  & NC & \textbf{+0.202} & $-$0.008 & 0.002 & \textbf{+1.15} & $-$0.10 & 0.010 & \textbf{0.033} & 0.29 (75) \\
MiniCPM-o-4.5    & \texttt{AT} & CI & $-$0.067 & 0.000 & 0.069 & $-$1.26 & 0.00 & 0.037 & 0.16 & -- (4) \\
Qwen2.5-Omni-7B & \texttt{AT} & NC & +0.061 & +0.011 & 0.037 & +0.50 & +0.06 & 0.007 & 0.099 & 0.13 (72) \\
\multicolumn{3}{@{}l}{\textit{9 further cells}} & \multicolumn{8}{l}{$|\Delta P|\le 0.050$; $q\ge0.29$} \\ \midrule
\multicolumn{11}{@{}l}{\textit{Pitt \texttt{AT-chat} masks, ablated under PROCESS-2 \texttt{AT}}} \\
Kimi-Audio   & \texttt{AT-chat}$\rightarrow$\texttt{AT} & NC & \textbf{+0.299} & +0.040 & $<$0.001 & \textbf{+1.74} & +0.25 & $<$0.001 & \textbf{$<$0.001} & 0.65 (65) \\
Qwen3-Omni-30B   & \texttt{AT-chat}$\rightarrow$\texttt{AT} & NC & +0.075 & +0.019 & 0.028 & +1.42 & +0.21 & 0.003 & 0.10 & 0.06 (77) \\
\multicolumn{3}{@{}l}{\textit{6 further cells}} & \multicolumn{8}{l}{$q\ge0.10$} \\ \bottomrule
\end{tabular}%
}
\caption{Pitt CICN masks ablated on PROCESS-2 with Pitt-matched random controls. Upper part: \texttt{A} and \texttt{AT} masks; lower part: Pitt \texttt{AT-chat} masks evaluated under PROCESS-2 \texttt{AT}. $n$ is the denominator of $\text{flip}_{\text{own}}$; flip rates are omitted for $n<10$. The two blocks are separate correction families.}
\label{tab:xcorpus}
\end{table}

\paragraph{Pitt masks on PROCESS-2.}
Beyond the four Pitt-passing masks that also pass on PROCESS-2 (\S\ref{subsec:attribution}), two masks that were non-significant on Pitt pass there (Table~\ref{tab:xcorpus}). The class asymmetry also carries over. A transferred effect shows that the same neurons shift the same prompted, NC-biased decision on both corpora, not that the model uses the same evidence on both; the audio conditions also differ between corpora (Appendix~\ref{appendix:data}).

Baseline performance within each transcription style is in Table~\ref{tab:pilot_strata}. Because transcription style remains imbalanced within the eligible identification pools, every PROCESS-2 effect is also inspected separately within style.

\begin{table}[h]
\centering
\small
\resizebox{\columnwidth}{!}{%
\begin{tabular}{@{}lll rr rr@{}}
\toprule
& & & \multicolumn{2}{c}{$\Delta P(\text{CI})$} & \multicolumn{2}{c}{$\text{flip}_{\text{own}}$} \\ \cmidrule(lr){4-5}\cmidrule(lr){6-7}
Model & Cond. & Mask & lab. & plain & lab. & plain \\ \midrule
\multicolumn{7}{@{}l}{\textit{Pitt masks}} \\
Kimi-Audio   & \texttt{A}  & NC & +0.202 {\scriptsize($-$0.013)} & +0.203 {\scriptsize($-$0.002)} & 0.28 {\scriptsize(0.03)} & 0.31 {\scriptsize(0.04)} \\
Kimi-Audio   & \texttt{AT} & NC & +0.282 {\scriptsize($-$0.001)} & +0.317 {\scriptsize(+0.006)} & 0.65 {\scriptsize(0.06)} & 0.65 {\scriptsize(0.06)} \\
Kimi-Audio   & \texttt{AT-chat}$\rightarrow$\texttt{AT} & NC & +0.291 {\scriptsize(+0.037)} & +0.309 {\scriptsize(+0.043)} & 0.71 {\scriptsize(0.09)} & 0.59 {\scriptsize(0.11)} \\
MiniCPM-o-4.5    & \texttt{A}  & CI & $-$0.145 {\scriptsize(+0.015)} & $-$0.142 {\scriptsize(+0.022)} & -- & -- \\
MiniCPM-o-4.5    & \texttt{AT} & CI & $-$0.080 {\scriptsize(0.000)} & $-$0.051 {\scriptsize($-$0.001)} & -- & -- \\
Qwen3-Omni-30B   & \texttt{A}  & NC & +0.121 {\scriptsize(+0.022)} & +0.102 {\scriptsize(+0.016)} & 0.05 {\scriptsize(0.00)} & 0.08 {\scriptsize(0.00)} \\
Qwen3-Omni-30B   & \texttt{AT} & NC & +0.111 {\scriptsize(+0.024)} & +0.108 {\scriptsize(+0.004)} & 0.16 {\scriptsize(0.01)} & 0.13 {\scriptsize(0.02)} \\
Qwen2.5-Omni-7B & \texttt{AT} & NC & +0.063 {\scriptsize(+0.011)} & +0.058 {\scriptsize(+0.011)} & 0.05 {\scriptsize(0.02)} & 0.20 {\scriptsize(0.05)} \\ \midrule
\multicolumn{7}{@{}l}{\textit{PROCESS-2 masks}} \\
Gemma-4-12B       & \texttt{T} & NC & +0.137 {\scriptsize(+0.001)} & +0.139 {\scriptsize($-$0.013)} & 0.53 {\scriptsize(0.11)} & 0.44 {\scriptsize(0.10)} \\
Gemma-4-12B       & \texttt{T} & CI & $-$0.090 {\scriptsize(+0.014)} & $-$0.089 {\scriptsize($-$0.010)} & 0.14 {\scriptsize(0.06)} & -- \\
Voxtral-Small-24B & \texttt{T} & NC & $-$0.005 {\scriptsize(+0.028)} & +0.002 {\scriptsize(+0.033)} & 0.00 {\scriptsize(0.11)} & 0.07 {\scriptsize(0.17)} \\
Voxtral-Small-24B & \texttt{T} & CI & $-$0.017 {\scriptsize(+0.036)} & 0.000 {\scriptsize(+0.042)} & 0.00 {\scriptsize(0.00)} & -- \\ \bottomrule
\end{tabular}%
}
\caption{Within-style PROCESS-2 ablation: real-mask effect with the matched-random mean in parentheses (recording counts in Table~\ref{tab:pilot_strata}). ``--'' indicates fewer than 10 baseline-correct items of the mask's class, so no flip rate is reported. These values are descriptive and are not tested separately.}
\label{tab:p2_style}
\end{table}

No effect in Table~\ref{tab:p2_style} is confined to one transcription style: the Pitt masks that pass on the pooled recordings (Table~\ref{tab:xcorpus}) change $P(\text{CI})$ by similar amounts in both styles, and so does Gemma-4-12B's NC mask.

\paragraph{Interpretation and scope.}
PROCESS-2 is strongly NC-skewed at the model-output level, leaving too few correctly classified CI items for stable CI flip rates. Inference therefore rests on $\Delta P(\text{CI})$ and $\Delta\ell$ over all evaluation recordings.

\paragraph{Masks identified on PROCESS-2.}
Only Gemma-4-12B and Voxtral-Small-24B provide sufficient transcript-only identification pools. Re-selection yields a corpus-specific profile: Gemma-4-12B's NC mask has the largest and style-consistent effect but does not pass the specificity test, while Voxtral-Small-24B remains within its matched-random distribution.

\paragraph{Mask overlap between corpora.}
Gemma-4-12B provides the available cross-corpus overlap comparison (\texttt{T} on PROCESS-2; \texttt{AT} and \texttt{AT-chat} on Pitt). Same-class masks overlap by Jaccard 0.011--0.022, 3--6$\times$ the chance level (0.003--0.005), and their per-layer profiles correlate at Spearman $\rho=0.49$--$0.75$ (Figure~\ref{fig:layers}). Within Pitt, the two conditions overlap more strongly (0.23--0.29), indicating a coarse structural alignment across the combined corpus-and-condition change rather than neuron-for-neuron identity.

\subsection{Held-Out Diagnoses}
\label{appendix:heldout}

\begin{table}[h]
\centering
\small
\resizebox{\columnwidth}{!}{%
\begin{tabular}{@{}lll rrr rrr r@{}}
\toprule
& & & \multicolumn{3}{c}{$\Delta P(\text{CI})$} & \multicolumn{3}{c}{$\Delta\ell$} & \\ \cmidrule(lr){4-6}\cmidrule(lr){7-9}
Model & Cond. & Mask & real & random & $p$ & real & random & $p$ & $q$ \\ \midrule
\multicolumn{10}{@{}l}{\textit{Pitt held-out} (16 MCI, 4 vascular, 3 other)} \\
Kimi-Audio & \texttt{A}       & NC & \textbf{+0.200} & $-$0.027{\scriptsize$\pm$0.085} & 0.018 & \textbf{+1.09} & $-$0.17 & 0.026 & \textbf{0.034} \\
Kimi-Audio & \texttt{AT}      & NC & \textbf{+0.301} & $-$0.002{\scriptsize$\pm$0.068} & $<$0.001 & \textbf{+1.84} & $-$0.03 & $<$0.001 & \textbf{0.001} \\
Kimi-Audio & \texttt{AT-chat} & NC & \textbf{+0.290} & +0.021{\scriptsize$\pm$0.056} & $<$0.001 & \textbf{+1.75} & +0.11 & $<$0.001 & \textbf{0.001} \\
MiniCPM-o-4.5  & \texttt{A}       & CI & \textbf{$-$0.203} & +0.016{\scriptsize$\pm$0.054} & 0.001 & \textbf{$-$2.20} & +0.17 & 0.002 & \textbf{0.003} \\
MiniCPM-o-4.5  & \texttt{AT}      & CI & \textbf{$-$0.140} & +0.002{\scriptsize$\pm$0.049} & 0.010 & \textbf{$-$1.82} & +0.01 & 0.004 & \textbf{0.016} \\
MiniCPM-o-4.5  & \texttt{AT-chat} & CI & \textbf{$-$0.083} & +0.020{\scriptsize$\pm$0.045} & 0.037 & \textbf{$-$0.93} & +0.19 & 0.039 & \textbf{0.044} \\
Qwen3-Omni-30B & \texttt{AT}      & NC & \textbf{+0.169} & +0.008{\scriptsize$\pm$0.026} & $<$0.001 & \textbf{+2.33} & +0.08 & $<$0.001 & \textbf{$<$0.001} \\
Qwen3-Omni-30B & \texttt{AT}      & CI & $-$0.030 & +0.015{\scriptsize$\pm$0.033} & 0.20 & $-$0.52 & +0.19 & 0.078 & 0.20 \\ \midrule
\multicolumn{10}{@{}l}{\textit{PROCESS-2 MCI} (136 recordings), Pitt masks} \\
Kimi-Audio & \texttt{A}       & NC & \textbf{+0.190} & $-$0.021{\scriptsize$\pm$0.068} & 0.007 & \textbf{+1.09} & $-$0.14 & 0.013 & \textbf{0.016} \\
Kimi-Audio & \texttt{AT}      & NC & \textbf{+0.272} & $-$0.004{\scriptsize$\pm$0.063} & $<$0.001 & \textbf{+1.57} & $-$0.04 & 0.001 & \textbf{0.001} \\
MiniCPM-o-4.5  & \texttt{A}       & CI & \textbf{$-$0.184} & +0.019{\scriptsize$\pm$0.044} & $<$0.001 & \textbf{$-$2.53} & +0.19 & $<$0.001 & \textbf{0.001} \\
MiniCPM-o-4.5  & \texttt{AT}      & CI & \textbf{$-$0.119} & 0.000{\scriptsize$\pm$0.042} & 0.013 & \textbf{$-$1.72} & $-$0.01 & 0.004 & \textbf{0.016} \\
Qwen3-Omni-30B & \texttt{AT}      & NC & \textbf{+0.117} & +0.026{\scriptsize$\pm$0.022} & 0.001 & \textbf{+1.98} & +0.34 & $<$0.001 & \textbf{0.001} \\
Qwen3-Omni-30B & \texttt{AT}      & CI & +0.003 & +0.025{\scriptsize$\pm$0.024} & 0.38 & $-$0.09 & +0.34 & 0.25 & 0.38 \\
\multicolumn{10}{@{}l}{\textit{Pitt \texttt{AT-chat} masks under PROCESS-2 \texttt{AT}}} \\
Kimi-Audio & \texttt{AT-chat}$\rightarrow$\texttt{AT} & NC & \textbf{+0.284} & +0.038{\scriptsize$\pm$0.044} & $<$0.001 & \textbf{+1.69} & +0.22 & $<$0.001 & \textbf{$<$0.001} \\
MiniCPM-o-4.5  & \texttt{AT-chat}$\rightarrow$\texttt{AT} & CI & \textbf{$-$0.078} & +0.016{\scriptsize$\pm$0.035} & 0.016 & \textbf{$-$1.24} & +0.15 & 0.009 & \textbf{0.016} \\ \bottomrule
\end{tabular}%
}
\caption{CICN ablation on held-out diagnoses with matched random controls. The eight strongest Pitt masks are used; on PROCESS-2, \texttt{AT-chat} masks are evaluated under \texttt{AT}. Flip rates are undefined because these recordings have no NC/CI evaluation label. Each block is a separate correction family.}
\label{tab:heldout_abl}
\end{table}

Unmodified models place Pitt MCI speakers closer to NC, whereas PROCESS-2 MCI speakers occupy a more intermediate regime. On the held-out intervention tests, seven of eight Pitt masks and five of six PROCESS-2 masks pass against their matched random controls, as do both transferred \texttt{AT-chat} masks; Qwen3-Omni-30B's \texttt{AT} CI mask is the consistent exception (Table~\ref{tab:heldout_abl}).

\paragraph{Interpretation and scope.}
Most Pitt held-out recordings are MCI, and restricting the analysis to those MCI cases yields effects similar to the full held-out set. Thus the transfer is not driven by Pitt's small number of non-MCI held-out diagnoses.

\section{Localizing the Intervention Effects}
\label{appendix:localization}

\subsection{Position-Restricted Ablation}
\label{appendix:positions}

\begin{table*}[t]
\centering
\small
\resizebox{\textwidth}{!}{%
\begin{tabular}{@{}l r rrr rrr rrr@{}}
\toprule
& Kimi-Audio \texttt{AT} NC & \multicolumn{3}{c}{Kimi-Audio \texttt{AT-chat} NC} & \multicolumn{3}{c}{MiniCPM-o-4.5 \texttt{A} CI} & \multicolumn{3}{c}{MiniCPM-o-4.5 \texttt{AT} CI} \\ \cmidrule(lr){2-2}\cmidrule(lr){3-5}\cmidrule(lr){6-8}\cmidrule(lr){9-11}
Ablated positions & orig. & orig. & para. & num. & orig. & para. & num. & orig. & para. & num. \\ \midrule
all (Figure~\ref{fig:ablation})     & +0.228$^{\ast}$ & +0.267$^{\ast}$ & -- & -- & $-$0.212$^{\ast}$ & -- & -- & $-$0.253$^{\ast}$ & -- & -- \\
all but the last                   & +0.222$^{\ast}$ & +0.252$^{\ast}$ & +0.179$^{\ast}$ & +0.223$^{\ast}$ & $-$0.217$^{\ast}$ & $-$0.324$^{\ast}$ & $-$0.135$^{\ast}$ & $-$0.261$^{\ast}$ & $-$0.330$^{\ast}$ & $-$0.218$^{\ast}$ \\
\quad audio span                   & +0.069$^{\ast}$ & +0.038 & +0.030 & +0.041 & $-$0.100$^{\ast}$ & $-$0.109$^{\ast}$ & $-$0.085$^{\ast}$ & $-$0.084$^{\ast}$ & $-$0.083$^{\ast}$ & $-$0.083$^{\ast}$ \\
\quad transcript span              & +0.022$^{\ast}$ & +0.016 & -- & -- & -- & -- & -- & $-$0.034$^{\ast}$ & -- & -- \\
\quad template                     & +0.152$^{\ast}$ & +0.227$^{\ast}$ & -- & -- & $-$0.123$^{\ast}$ & -- & -- & $-$0.161$^{\ast}$ & -- & -- \\
\quad\quad before the content      & +0.042$^{\ast}$ & +0.027 & -- & -- & $-$0.005 & -- & -- & $-$0.012 & -- & -- \\
\quad\quad after the content       & +0.117 & +0.166$^{\ast}$ & +0.156$^{\ast}$ & +0.135 & $-$0.112$^{\ast}$ & $-$0.189$^{\ast}$ & $-$0.036 & $-$0.157$^{\ast}$ & $-$0.187$^{\ast}$ & $-$0.076 \\
last position only                 & +0.009 & +0.014 & 0.000 & $-$0.005 & +0.009 & +0.003 & +0.009 & +0.017 & +0.001 & +0.011 \\ \midrule
random masks ($|\text{mean}|$, any row) & $\le$0.016 & $\le$0.023 & -- & -- & $\le$0.018 & -- & -- & $\le$0.003 & -- & -- \\ \bottomrule
\end{tabular}%
}
\caption{Position-restricted ablation on Pitt under the original prompt (orig.) and the two prompt variants (para.: paraphrase; num.: numeric option labels; Appendix~\ref{appendix:prompt_robust}): $\Delta P(\text{CI})$ when the mask is ablated only at the listed prefill positions. ``Template'' is every position outside the audio and transcript spans except the last; ``before the content'' cannot attend to the recording. $^{\ast}$ marks cells that pass the predefined test for that position set and prompt. --: not run (the variants cover three masks and four position sets; MiniCPM-o-4.5 \texttt{A} has no transcript span); the random-mask row is given for the original prompt. The shares quoted in Appendix~\ref{appendix:position_prompts} are computed on $\Delta\ell$.}
\label{tab:positions}
\end{table*}

\paragraph{Transcript positions under \texttt{T}.}
Under \texttt{T}, we restrict ablation to transcript positions for four masks: Kimi-Audio \texttt{A} and \texttt{AT} NC and MiniCPM-o-4.5 \texttt{A} and \texttt{AT} CI. The effects equal 11--38\% of the corresponding all-position $\Delta\ell$. Both speech-only masks exceed their 20 matched random controls on both co-primary metrics at transcript positions. These ratios compare separate interventions and should not be interpreted as additive mediation shares.

\paragraph{Interpretation and scope.}
The four masks in Table~\ref{tab:positions} are not the four primary masks of Section~\ref{subsec:ablation}. In a sparse mixture-of-experts layer each expert receives only the tokens routed to it, without their sequence positions. Our activation logging recovers these positions by recomputing the routing (Appendix~\ref{appendix:impl}), but our intervention code does not, so position-restricted ablation is not implemented for this model; this is a limit of the implementation, not of the method. Qwen3-Omni-30B \texttt{AT} NC is therefore left out, and Kimi-Audio \texttt{AT} NC, the strongest remaining dense-model mask that passes in its own condition, takes its place. The localization results thus cover Kimi-Audio and MiniCPM-o-4.5 only.
Positions between the audio and transcript belong to the template but to neither the pre- nor post-content segments. The after-content segment is short, which increases its matched-random variance. Kimi-Audio \texttt{AT} NC also passes when ablated only before the content ($+0.042$, $q=0.001$), about 19\% of its all-but-last effect on either co-primary metric, so part of this mask's influence is set by the prompt frame before the recording is seen. Kimi-Audio's \texttt{AT-chat} NC effect is too small to pass when audio and transcript spans are isolated separately.

\subsection{Positional Robustness across Prompt Variants}
\label{appendix:position_prompts}

The variants are defined in Appendix~\ref{appendix:prompt_robust}.

We repeated the main position-restricted ablations under both variants for Kimi-Audio \texttt{AT-chat} NC and MiniCPM-o-4.5 \texttt{A}/\texttt{AT} CI, each against matched random masks under the same prompt. The all-but-last intervention remains robust for all three masks, whereas last-position ablation remains null. Audio-span effects pass for both MiniCPM-o-4.5 masks under all three prompts; Kimi-Audio's is not significant under any of them ($+0.030$ to $+0.041$). The after-content concentration is preserved under the paraphrase but weakens under numeric option labels (Table~\ref{tab:positions}).

\paragraph{What changes with the answer format.}
The paraphrase rewords the question and both option texts, and the after-content positions keep their share: they carry 53--79\% of the all-but-last $\Delta\ell$ and pass for all three masks ($q\le0.036$), against 48--61\% under the original prompt. The numeric-label prompt keeps the original option texts and changes only the labels and the final instruction, yet the after-content share falls to 24--56\% and no mask passes there ($q=0.096$--$0.28$). For the two MiniCPM-o-4.5 masks the after-content effect is about half of its original size or less, while the audio-span effect is nearly unchanged, so under numeric labels the audio span carries a larger effect than the after-content positions for both. For Kimi-Audio the after-content effect remains the largest localized one but no longer separates from its random masks.
Two conclusions therefore have different scope. That the masks act before the final position holds under all three prompts. That their influence is concentrated at the question and options holds for letter-labeled options, whatever the option wording, and is weaker with numeric labels. The shares compare separate interventions and are not an additive decomposition.

\subsection{Answer-Aligned Readout across Depth}
\label{appendix:logit_lens}

\begin{figure}[h]
\centering
\includegraphics[width=\columnwidth]{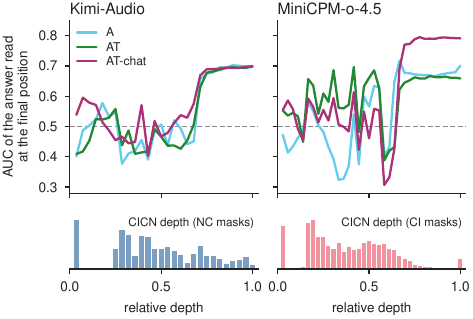}
\caption{Logit lens on Pitt. Top: AUC of the option-logit readout from each layer's residual stream at the final prompt position, one line per input condition. Bottom: depth distribution of the CICN masks used in the position-restricted ablations (blue: NC masks; red: CI masks).}
\label{fig:lens}
\end{figure}

As a complement to the position-restricted ablations, a logit lens shows at which depth the answer becomes readable. We read the CI-vs-NC log-odds of the option letters out of the residual stream at the final prompt position after every decoder layer, through the final normalization and the unembedding, and compute its AUC against the labels (Figure~\ref{fig:lens}).

The readout's diagnostic AUC remains near or below chance through much of the network and reaches the model's answer-level AUC only around two-thirds depth, while 77--92\% of CICNs lie before that point. Most selected neurons therefore sit earlier than the layers at which the final option readout becomes strongly diagnostic, which is consistent with the upstream role indicated by the positional interventions (\S\ref{subsec:positions}). MiniCPM-o-4.5 exhibits a reversed intermediate readout. The logit lens does not establish when diagnostic information first becomes available internally.

\subsection{Token-Level Correlates of CICN Activity}
\label{appendix:tokens}

\paragraph{CI-direction score.}
At prompt position $t$, the score of a mask $\mathcal{I}^{(c)}$ is
\begin{equation}
s_t=\pm\frac{1}{|\mathcal{I}^{(c)}|}\sum_{(l,n)\in\mathcal{I}^{(c)}}\operatorname{sign}\!\big(\mu^{(c)}_{l,n}-\mu^{(\bar c)}_{l,n}\big)\,\frac{a_{l,n,t}-m_{l,n}}{\sigma_{l,n}},
\end{equation}
where $m_{l,n}$ and $\sigma_{l,n}$ are the class-balanced mean and pooled standard deviation of the neuron over the identification pool, and the leading sign is $+$ for CI masks and $-$ for NC masks.

\paragraph{Token categories and scope.}
Each transcript token gets a category: the rendered pause, filler, word-fragment, retracing and unintelligible-speech markers of the annotated transcript (\texttt{AT-chat} only); words naming Cookie Theft information units \cite{CROISILE19961} (\emph{boy}, \emph{cookie}, \emph{sink}, \emph{stool}, \emph{water}, \ldots); other content words; function words (a stop-word list, including contractions of its words such as \emph{she's} and \emph{don't}); and punctuation. Audio tokens are labeled silence or speech from the energy of 25-ms frames (silent if 30~dB below the recording's 90th-percentile frame level), mapped linearly onto the audio span; a token counts as silence if at least 80\% of its time is silent and as speech if at most 20\% is. The information-unit category comes from a fixed word list rather than annotated content units, and the audio categories from frame energy rather than annotated pauses. The token analysis covers Kimi-Audio and MiniCPM-o-4.5 on Pitt.

\paragraph{Procedure.}
For every evaluation recording we compute the CI-direction score $s_t$ of a mask at each prompt position and average it per token category and recording.
Categories are compared within recordings and between classes, for the real mask and its 20 matched random masks, with Benjamini--Hochberg correction over the contrasts of each model, condition and mask.

\begin{figure*}[h]
\centering
\includegraphics[width=\textwidth]{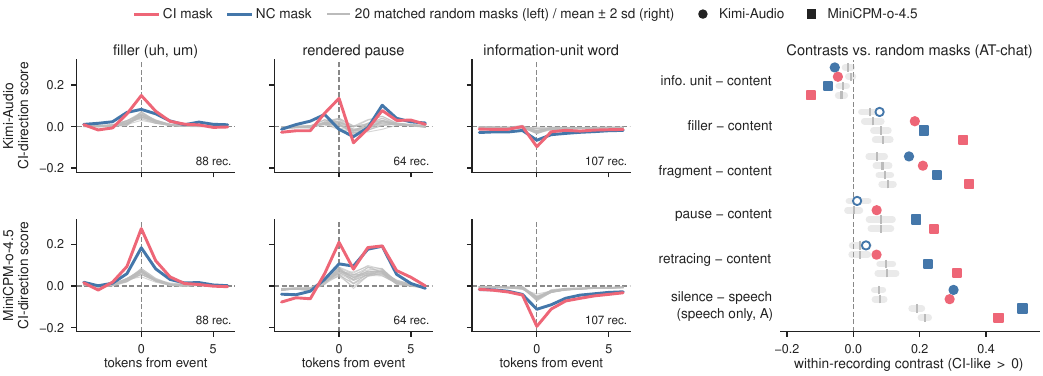}
\caption{Token-level correlates of CICN activity on Pitt. Left: CI-direction score (positive: CI-like) around fillers, rendered pauses, and Cookie Theft information-unit words, relative to each recording's transcript mean; thin gray lines are matched random masks and ``rec.'' counts recordings containing the event. Right: within-recording token-category contrasts, with matched-random mean $\pm$ 2 sd; filled markers pass the predefined token-level test. Values, including \texttt{AT}, are in Table~\ref{tab:tokens_full}.}
\label{fig:tokens}
\end{figure*}

\paragraph{What CICN activity correlates with.}
In Kimi-Audio and MiniCPM-o-4.5, the masks are more NC-like on words that name the picture's content \cite{CROISILE19961, doi:10.3233/JAD-150520} and more CI-like on fillers, word fragments, pauses, retracings and silence, each beyond all 20 random masks with one exception (Figure~\ref{fig:tokens}, Table~\ref{tab:tokens_full}).
The markers are not the source of the masks' class difference: removing all marker, silence and information-unit tokens leaves 95--101\% of the class gap, a recording-level state carried by ordinary tokens.
These analyses characterize associations between the selected activations and input cues; they do not establish that those cues mediate the dominant decision-bias shift.

\begin{table}[h]
\centering
\small
\resizebox{\columnwidth}{!}{%
\begin{tabular}{@{}ll rrrr@{}}
\toprule
& & \multicolumn{2}{c}{Kimi-Audio} & \multicolumn{2}{c}{MiniCPM-o-4.5} \\ \cmidrule(lr){3-4}\cmidrule(lr){5-6}
Contrast & Mask & \texttt{AT} & \texttt{AT-chat} & \texttt{AT} & \texttt{AT-chat} \\ \midrule
info.\ unit $-$ content & CI & $-$0.055$^{\ast}$ {\scriptsize($-$0.008)} & $-$0.047$^{\ast}$ {\scriptsize($-$0.008)} & $-$0.128$^{\ast}$ {\scriptsize($-$0.039)} & $-$0.128$^{\ast}$ {\scriptsize($-$0.036)} \\
                        & NC & $-$0.039$^{\ast}$ {\scriptsize($-$0.010)} & $-$0.056$^{\ast}$ {\scriptsize($-$0.015)} & $-$0.076$^{\ast}$ {\scriptsize($-$0.036)} & $-$0.076$^{\ast}$ {\scriptsize($-$0.032)} \\
filler $-$ content      & CI & -- & +0.185$^{\ast}$ {\scriptsize(+0.058)} & -- & +0.332$^{\ast}$ {\scriptsize(+0.088)} \\
fragment $-$ content    & CI & -- & +0.210$^{\ast}$ {\scriptsize(+0.088)} & -- & +0.350$^{\ast}$ {\scriptsize(+0.103)} \\
pause $-$ content       & CI & -- & +0.070$^{\ast}$ {\scriptsize(+0.001)} & -- & +0.244$^{\ast}$ {\scriptsize(+0.083)} \\
retracing $-$ content   & CI & -- & +0.070$^{\ast}$ {\scriptsize(+0.019)} & -- & +0.313$^{\ast}$ {\scriptsize(+0.101)} \\
silence $-$ speech      & CI & $-$0.034$^{\ddagger}$ {\scriptsize($-$0.005)} & +0.186$^{\ast}$ {\scriptsize(+0.043)} & +0.296$^{\ast}$ {\scriptsize(+0.157)} & +0.167$^{\ast}$ {\scriptsize(+0.073)} \\
                        & NC & +0.048$^{\ast}$ {\scriptsize($-$0.001)} & +0.144$^{\ast}$ {\scriptsize(+0.044)} & +0.315$^{\ast}$ {\scriptsize(+0.130)} & +0.245$^{\ast}$ {\scriptsize(+0.085)} \\ \bottomrule
\end{tabular}%
}
\caption{Token-level within-recording differences in CI-direction score on Pitt (positive: more CI-like); matched-random means are in parentheses. Pause, filler, fragment, and retracing markers occur only in \texttt{AT-chat}. $^{\ast}$ marks contrasts that pass the predefined token-level test; $^{\ddagger}$ marks the same criterion with an effect opposite to the expected sign.}
\label{tab:tokens_full}
\end{table}

\section{Distributed Diagnostic Information}
\label{appendix:exploit}

\begin{figure*}[h]
\centering
\includegraphics[width=0.7\textwidth]{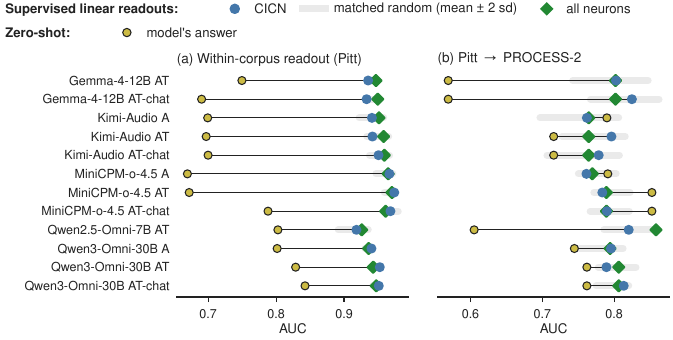}
\caption{Predictive utility of CI-contrast neurons: whole-mask readouts. Blue, gray and green are supervised linear readouts trained on Pitt (CICNs, matched-random neurons with mean $\pm$ 2 sd over 20 masks, and all neurons); yellow is the model's zero-shot answer, joined to the CICN readout by a line. (a) Within-corpus readout: CICN readouts reach 0.92--0.97 AUC on Pitt, against 0.67--0.84 for the model's own answer. (b) Cross-corpus generalization: the same readouts evaluated on PROCESS-2, with rows in the same order as (a).}
\label{fig:exploit}
\end{figure*}

\paragraph{Readout construction.}
Readouts are trained on the full identification split, including the 30 recordings used for mask selection, and tested once on the evaluation split (Pitt: 297 training recordings, 152 NC and 145 CI; 107 test recordings, 40 NC and 67 CI). The two splits share no participant, so no participant used for neuron selection or readout training appears in the test set, and no readout parameter or hyperparameter is fitted or tuned on the evaluation split. 
Analysis cells were selected using evaluation-based eligibility screening (\S\ref{subsec:eligibility}) and, for the unit-budget analysis, evaluation-split ablation effects.
Features are each neuron's mean gate activation over the prompt, averaged over option orders and standardized with the mean and standard deviation of the training set, which are applied unchanged to the test set. The standardization uses no labels and no evaluation recording, and is computed once on the training set.
The classifier is a class-balanced ridge regression on $\pm1$ targets, solved in the dual. Its penalty is $\lambda=c\,\bar k$, where $\bar k$ is the mean diagonal of the centered training kernel and $c$ is searched over 13 values from $10^{-3}$ to $10^{3}$ in half-decade steps. We choose $c$ by participant-grouped, class-stratified 5-fold cross-validation on the training set, maximizing the mean fold AUC, and refit on the whole training set. AUC needs no threshold; where a decision is reported, the threshold maximizes UAR on the out-of-fold training scores.
CICN readouts use the union of a cell's NC and CI masks. Each of the 20 random readouts uses the union of the corresponding matched random masks and goes through the same procedure, as does the all-neuron readout.

\paragraph{Within-corpus readouts.}
Within Pitt, CICN readouts outperform the model's own answer in all twelve cells (Figure~\ref{fig:exploit}a).

\paragraph{Reference baselines on the same split.}
To calibrate these values, we train class-balanced logistic regressions, the standard classifier for these sparse features, on the same training recordings, with the same grouped cross-validation, and test them on the same 107 recordings. Recording length does not separate the classes: participant-only duration gives 0.45 AUC, word count 0.51, and duration, word count and speech rate together 0.62. A TF-IDF model on word uni- and bigrams of the clean transcript reaches 0.91 (95\% bootstrap interval [0.85, 0.96]), and 0.92 on the annotated transcript. The readouts (0.92--0.97) are therefore in the range of a lexical baseline (0.91); we do not test the differences between them. Their high AUC reflects how separable Pitt Cookie Theft descriptions are with supervised training, and none of the length baselines reproduces it. Supervised detectors in earlier work range from LLMs used as embedding extractors, prompted classifiers, or fine-tuned models \cite{10.1371/journal.pdig.0000168, https://doi.org/10.4218/etrij.2023-0356, park25d_interspeech} to paralinguistic feature sets \cite{8910399}, self-supervised speech encoders \cite{luz21_interspeech}, and fusions of separately built acoustic and linguistic models \cite{cummins20_interspeech}; we use these simple references only for calibration. The comparison of interest is between neuron sets under one fixed procedure.

\paragraph{Cross-corpus readouts.}
On PROCESS-2, the Pitt-trained CICN readouts reach 0.76--0.82 AUC and exceed the model's answer in eight of twelve cells, without a consistent advantage over matched-random or all-neuron readouts (Figure~\ref{fig:exploit}b).

\paragraph{Recording and transcription controls.}
We train readouts to predict a confound instead of the diagnosis: a recording-quality proxy on Pitt (the within-class median split of the energy dynamic range, Appendix~\ref{appendix:audio_quality}) and transcription style on PROCESS-2. Confound directions are estimated on the training recordings after the class means are removed, so that they do not carry the diagnosis.
This proxy is decodable from all neurons at 0.90--0.95 AUC in the 14 Pitt cells with audio. Projecting the confound directions out of the features lowers this to 0.59--0.63, while the CI readout AUC changes by $-0.011$ to $+0.014$ within Pitt and by $-0.033$ to $+0.001$ on PROCESS-2.
Transcription style is decodable at 0.91 and 0.88 AUC in the two PROCESS-2 transcript-only cells (Gemma-4-12B and Voxtral-Small-24B); the projection lowers it to 0.56 and 0.57, with CI readout changes of at most 0.002 AUC.
The CI readouts thus depend little on the linearly decodable part of either confound.

\paragraph{Readout with a unit budget.}
Because whole-mask readouts saturate (\S\ref{subsec:exploit_method}), we repeat the analysis at fixed budgets using the top-ranked CICNs against layer-matched random draws. The seven cells are those that contain the eight masks with the largest own-condition effects (Appendix~\ref{appendix:sensitivity}): Kimi-Audio and MiniCPM-o-4.5 under \texttt{A}, \texttt{AT} and \texttt{AT-chat}, and Qwen3-Omni-30B under \texttt{AT}. A budget of $k$ units takes the $k/2$ neurons with the highest ConAct score from each of the cell's two masks, for $k=8$, 32, 128, 512 and 2{,}048, and the whole masks; each of the 20 random readouts draws the same per-layer counts from the matched random masks.
The 30 recordings used to select the masks are excluded from training to reduce selection leakage (267 remain). Other visits of the same participants stay in the training set; the test set shares no participant with either. Removing every recording of those participants as well (231--240 remain) or including the 30 recordings produces essentially the same curve (mean difference $+0.116$ and $+0.114$ at 8 units, $+0.036$ and $+0.031$ at 32).

At 8 and 32 units, the top-ranked CICNs outperform the mean of the random draws in all seven cells, by 0.04--0.18 and 0.003--0.12 AUC (Figure~\ref{fig:concentration}). No individual budget comparison survives correction in the reported training set (smallest $q=0.063$; two of 42 do when all recordings of the pool participants are removed) because tiny random subsets are highly variable, so the small-budget advantage is exploratory, descriptive evidence of rank efficiency.

\paragraph{Transcript-only readouts.}
The two transcript-only Pitt cells with logged activations have no CICN masks (Appendix~\ref{appendix:impl}), so only the all-neuron readout is available there. It reaches 0.93 AUC for Qwen2.5-Omni-7B and 0.95 for Qwen3-Omni-30B, against 0.75 and 0.78 for the models' own answers. The strong in-corpus decodability is therefore not specific to an audio shortcut.


\section{License}
DementiaBank Pitt was obtained through DementiaBank membership and PROCESS-2 under its data use agreement; both were used in accordance with their terms of use and intended research purposes. Neither audio nor transcripts are redistributed, and only aggregate results are reported. Pre-trained model checkpoints were used in accordance with their respective licenses.


\end{document}